\documentclass[a4paper,12pt]{article}
\usepackage[T1,T2A]{fontenc}
\usepackage[utf8]{inputenc}
\usepackage[ukrainian,english]{babel}

\usepackage[a4paper]{geometry}
\usepackage{amsmath}

\usepackage{graphicx}
\usepackage[english]{babel}
\usepackage{authblk}

\title{{On fast charged particles scattering on zigzag nanotube}}

\usepackage{lipsum}

\usepackage{multicol}
\usepackage{subcaption}
\usepackage{amssymb}
\usepackage{placeins}

\author[1]{V.D.~Omelchenko \thanks{koriukina@kipt.kharkov.ua; ORCID: 0000-0002-5113-3028 }}
\affil[1]{National Science Center ``Kharkiv Institute of Physics and Technology'', Kharkiv, Ukraine}

\begin{document}
\maketitle

\begin{abstract}
A fast charged particle scattering on a single-wall carbon nanotube of zigzag type was considered. The differential cross sections of scattering on nanotubes of different spatial orientation with respect to the incident particles were obtained. The eikonal approximation of quantum electrodynamics and the continuous potential approximation were used.
\end{abstract}

\section{Introduction}
Studying the process of scattering of fast charged particles on nanotubes \cite{Iij} is important and interesting problem from at least two points of view. On the one hand, nanotubes can serve as channels through which positive particles can be guided toward certain directions \cite{GrSh, FomN}. This property provides opportunities for beam control. On the other hand, nanotubes can be examined with high-energy particles. One of the instruments for such an examination is the rainbow scattering which is sensitive to the structure of the target \cite{Petrovic04, Petrovic13}. 

In this paper, we are considering a fast charged particle scattering on a zigzag carbon nanotube with different orientations with respect to the incident particle. As description of the scattering process, we find the differential scattering cross sections in the eikonal approximation of quantum electrodynamics \cite{AIA96}. The eikonal approximation was chosen for its benefits: contrary to the classical approach, it accounts for quantum nature of the incident particle, and unlike frequently used Born approximation, it has a wider applicability region.  

\section{Problem formulation}
\begin{figure}[!t]
      \centering
	   \begin{subfigure}{0.49\linewidth}
		\includegraphics[width=\textwidth]{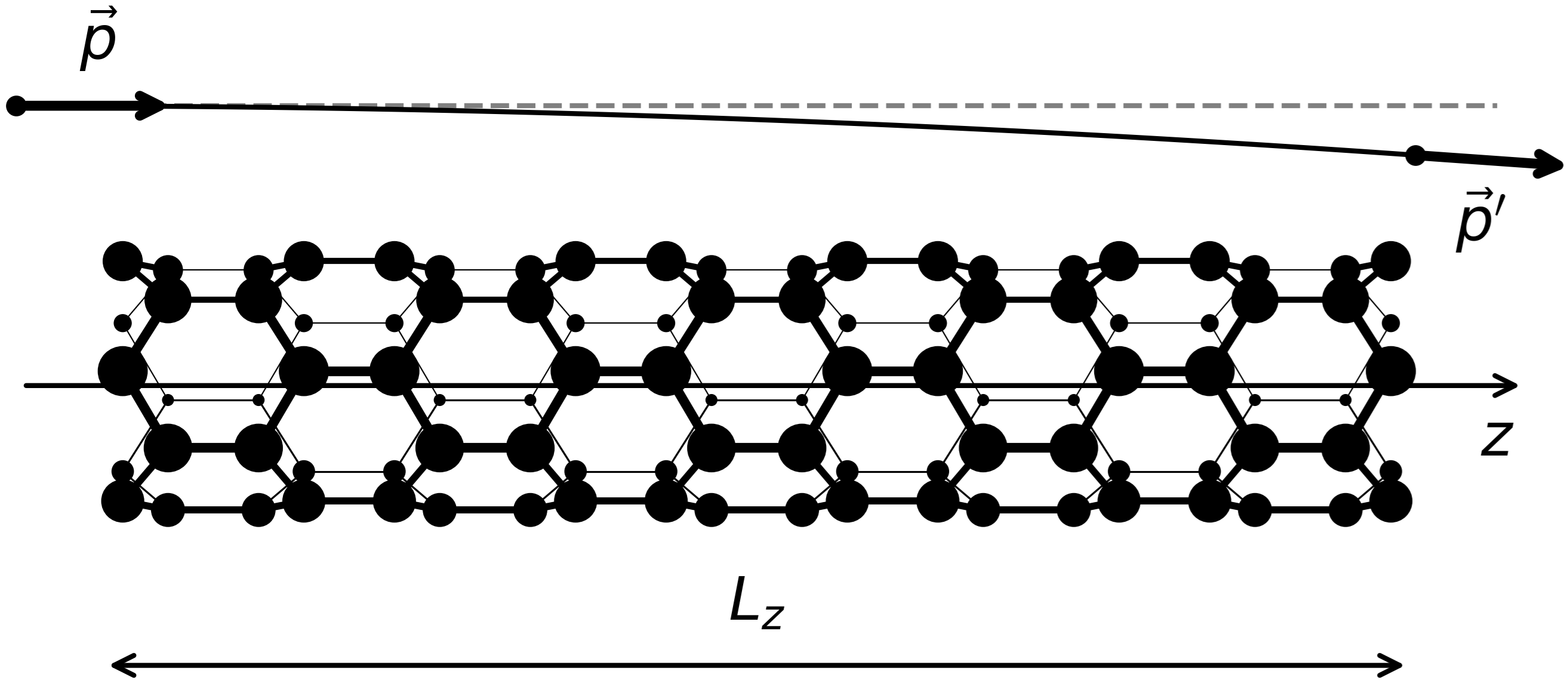}
		\caption{straight nanotube}
		\label{fig:nt_ss}
	   \end{subfigure}
	     \begin{subfigure}{0.49\linewidth}
		 \includegraphics[width=\textwidth]{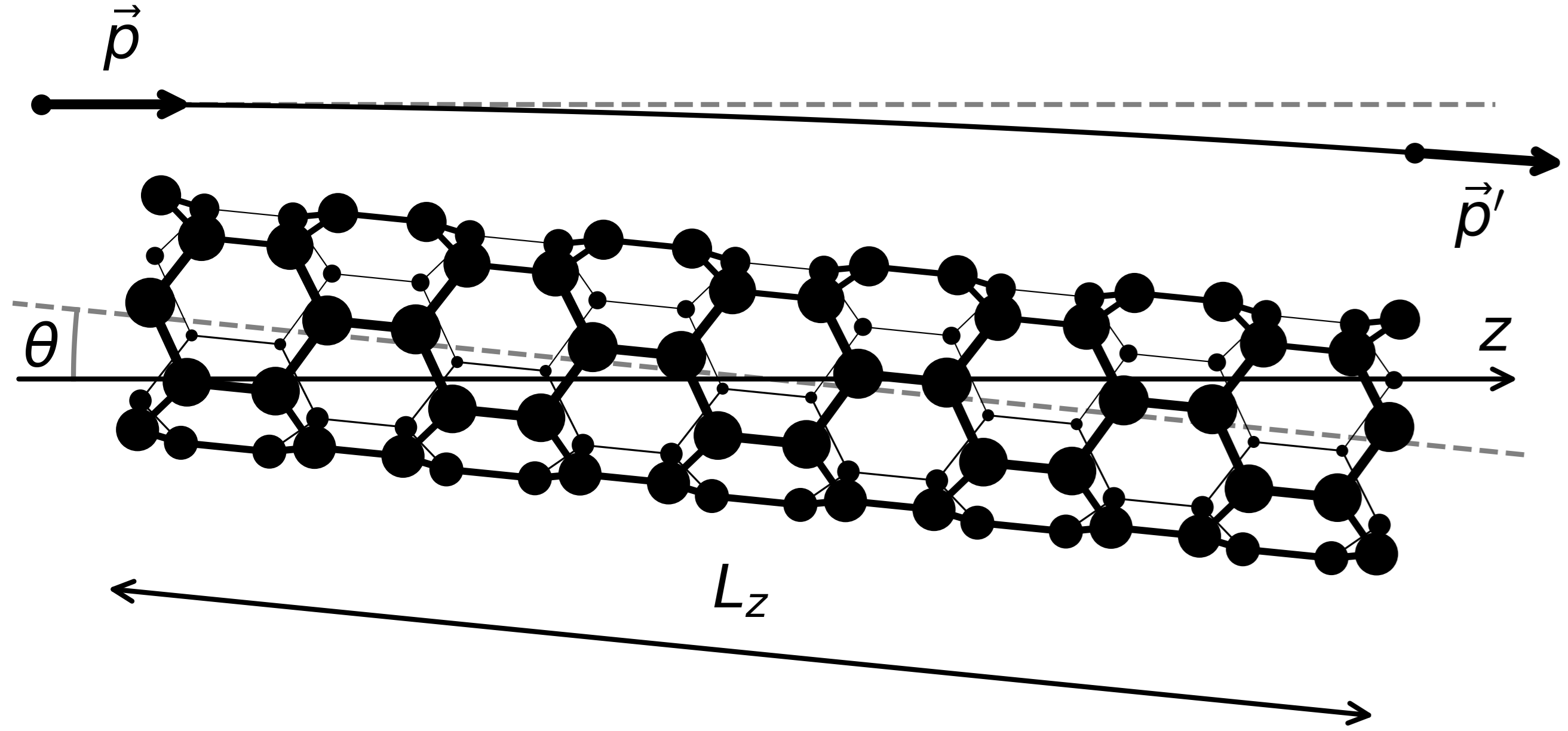}
		\caption{tilted nanotube}
		 \label{fig:nt_ts}
	      \end{subfigure}
\caption{Fast charged particle scattering on zigzag nanotubes}
 \label{fig_nt_s}
\end{figure}

\begin{figure}[!t]
      \centering
		\includegraphics[width=0.5\textwidth]{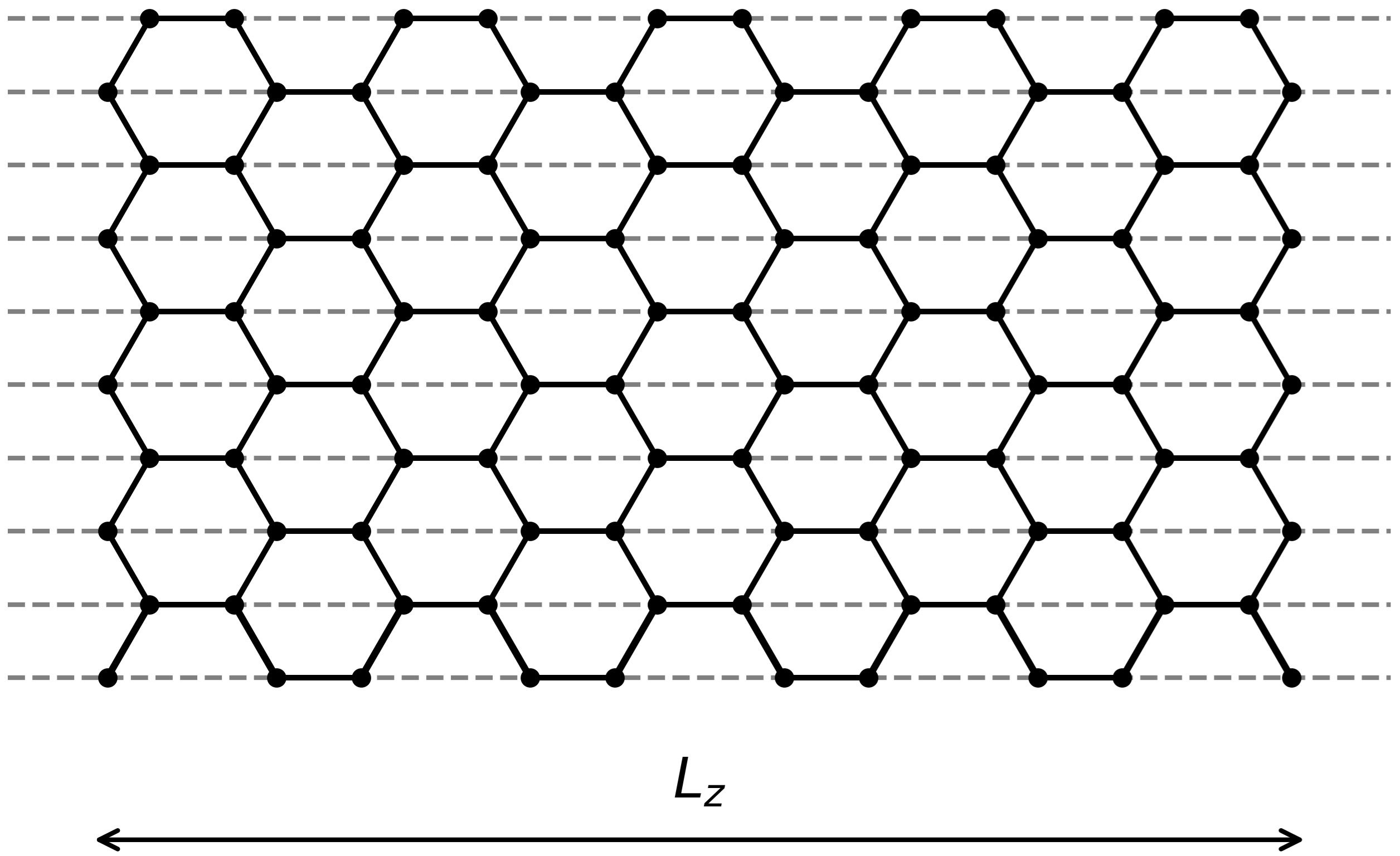}
\caption{"Unfolded" zigzag nanotube: dashed lines denote atomic strings}
 \label{fig_nt_st}
\end{figure}

Let us consider a fast charged particle scattering on a $(n,0)$ single-wall zigzag nanotube of a certain spatial orientation. Let us denote the coordinates as $\vec{r}=(\vec{\rho},z)$, $\vec{\rho}=(x,y)$. Let the initial momentum of the particle $\vec{p}$ be parallel to $z$-axis. Let there be a straight alignment of the nanotube -- when its axis is parallel to $z$-axis, and tilted  -- when there is an angle $\theta$ between the nanotube axis and $z$-axis in the $(z,x)$-plane (Fig. \ref{fig_nt_s}). Let us consider nanotubes as sets of $N_s=2n$ atomic strings (Fig. \ref{fig_nt_st}). The nanotube and its strings have the same length $L_z$, which for even or odd number $N_z$ of atoms in the string is:
\begin{eqnarray}\label{eq0_1}
L_z=\begin{cases}
a_g\left(\frac{3}{2}N_z-1\right), \ N_z=2m; \\
\frac{3}{2}a_g(N_z-1), \ N_z=2m+1;
\end{cases}
\end{eqnarray}
where $a_g=1.42 \ \text{\AA}$ is the length of the bond between carbon atoms.

\noindent For a straight nanotube, these atomic strings have positions $\vec{\rho}_j=(x_j,y_j)$ in the vertices of a regular $2n$-polygon. The edges of this polygon are of length $a_g \cos\frac{\pi}{6}$. Let us also associate the index $k$ with atom's number along each string, so that $z_k$ is atom's position along the axis of the corresponding string. Let $z_k \in \left[-L_z/2, L_z/2 \right]$. For a tilted nanotube, the positions of individual atoms in the string are $\vec{\rho}_{jk}=(\tilde{x}_{jk}, y_j)$, $\tilde{x}_{jk}=x_j \cos\theta+z_k \sin \theta$, which for small enough $\theta$ are approximately
\begin{eqnarray}\label{eq0}
\tilde{x}_{jk}=x_j+z_k \sin \theta. 
\end{eqnarray}
The $j$-th tilted string projection to the $x$-axis is approximately
\begin{eqnarray}\label{eq0_2}
x \in \left[x_j -\frac{L_z}{2} |\sin\theta|, x_j +\frac{L_z}{2} |\sin\theta| \right].
\end{eqnarray}

Let us also account for the thermal vibrations of atoms in the nanotube by introducing each atom's position displacement $\vec{u}_{jk}=(u_{x, jk}, u_{y, jk})$ in the $(x,y)$-plane. We assume that the law of distribution of these values is Gaussian:  
\begin{eqnarray}\label{eq0_3}
f(\vec{u}_{jk})=\frac{\exp \left(- \ \frac{u_{jk}^2}{2 \langle {u^2_x} \rangle }\right)}{ 2 \pi \langle {u^2_x} \rangle },
\end{eqnarray}
where $\langle {u^2_x} \rangle$ is mean square value of thermal displacements of atoms both in $x$ and $y$ directions, ${u}_{jk}^2=u_{x, jk}^2+ u_{y, jk}^2$.

In the eikonal approximation, the distribution of atoms in the target along $z$-axis (direction of initial momentum $\vec{p}$) does not influence the differential scattering cross section. So, we do not specify atoms positions and displacements in $z$-axis direction.

\section{Differential scattering cross section}
We will use basic ideas of the eikonal approximation \cite{AIA96} and Glauber's approach \cite{Glauber} to obtain the differential scattering cross section for this case, analogously to what we did in \cite{PerPl1}. We start with general formula for the scattering amplitude in the eikonal approximation \cite{AIA96}:
\begin{eqnarray}\label{eq1} 
a(\vec{q}_{\perp})=\frac{i}{2\pi} \int_{\mathbb{R}^2} d^2 \rho \ e^{i \vec{q}_{\perp} \vec{\rho}} \left\{ 1- \exp\left[ i \chi^{(N)}_0 (\vec{\rho}) \right] \right\}, 
\end{eqnarray}
where $\int_{\mathbb{R}^M}d^2 \rho ...$ denotes the integral over the entire real $M$-dimensional volume, 
\begin{eqnarray}\label{eq2}
\chi^{(N)}_0(\vec{\rho})=-e\int_{-\infty}^{\infty} dz U^{(N)}(\vec{\rho},z)
\end{eqnarray}
and $U^{(N)}(\vec{r})=\sum_{m=1}^N u(\vec{r}-\vec{r}_m)$ is total target's potential being a sum of individual potentials $u(\vec{r}-\vec{r}_m)$ of all $N$ atoms of the target, where each of them has position $\vec{r}_m$. We note that $N=N_s N_z$ in our case. So, it follows that
\begin{eqnarray}\label{eq3}
\chi^{(N)}_0 (\vec{\rho})=\sum_{m=1}^N \chi_0 (\vec{\rho}-\vec{\rho}_m),
\end{eqnarray}
where 
\begin{eqnarray}\label{eq3_1}
\chi_0 (\vec{\rho}-\vec{\rho}_m)=-e\int_{-\infty}^{\infty} dz u(\vec{\rho}-\vec{\rho}_m,z).\end{eqnarray}
The differential scattering cross section \cite{AIA96}
\begin{eqnarray}\label{eq4}
\frac{d\sigma}{d^2q_{\perp}}= |{a}(\vec{q}_{\perp})|^2
\end{eqnarray}
with amplitude \eqref{eq1} is difficult to calculate due to complicated form of the real target potential. So, we need to average the differential cross section over positions of atoms in the target:
\begin{eqnarray}\label{eq5}
\frac{d\sigma}{d^2q_{\perp}}=\frac{1}{4\pi^2} \int_{\mathbb{R}^4}d^2 \rho d^2\rho' e^{i\vec{q}_{\perp}(\vec{\rho}-\vec{\rho}')} \langle 1-e^{i \chi^{(N)}_0(\vec{\rho})} -e^{-i \chi^{(N)}_0(\vec{\rho}')}+ e^{i[\chi^{(N)}_0(\vec{\rho})-\chi^{(N)}_0(\vec{\rho}')]} \rangle,
\end{eqnarray}
where brackets $\langle ... \rangle$ denote averaging. 
Let us consider the term $\langle e^{i[\chi^{(N)}_0(\vec{\rho})-\chi^{(N)}_0(\vec{\rho}')]} \rangle$ for the case of a straight nanotube:
\begin{eqnarray}\label{eq6}
\langle e^{i[\chi^{(N)}_0(\vec{\rho})-\chi^{(N)}_0(\vec{\rho}')]} \rangle = \left(\prod_{j=1}^{N_s} \prod_{k=1}^{N_z} \int_{\mathbb{R}^2} d^2u_{jk} f(\vec{u}_{jk}) \right) \times \nonumber \\ 
\times \prod_{j=1}^{N_s} \prod_{k=1}^{N_z} \exp \left(i[\chi_0(\vec{\rho}-\vec{\rho}_j-\vec{u}_{jk})-\chi_0(\vec{\rho}'-\vec{\rho}_j-\vec{u}_{jk})] \right),
\end{eqnarray}
Considering that $f(\vec{u}_{jk})$ is the same for all atoms since all atoms' displacements are distributed according to the same law, we can write \eqref{eq6} in the form 
\begin{eqnarray}\label{eq7}
\langle e^{i[\chi^{(N)}_0(\vec{\rho})-\chi^{(N)}_0(\vec{\rho}')]} \rangle =  \prod_{j=1}^{N_s} \left\{ \int_{\mathbb{R}^2} d^2u f(\vec{u})  \exp \left(i[\chi_0(\vec{\rho}-\vec{\rho}_j-\vec{u})] - \right.  \right. \nonumber \\  
\left. \left. -\chi_0(\vec{\rho}'-\vec{\rho}_j-\vec{u})] \right) \right\}^{N_z},
\end{eqnarray}
where $\vec{u}$ is a value with distribution law \eqref{eq0_3}.

Using Glauber's idea \cite{Glauber} we can rewrite it as \cite{PerPl1, Eik}
\begin{eqnarray}\label{eq8}
\langle e^{i[\chi^{(N)}_0(\vec{\rho})-\chi^{(N)}_0(\vec{\rho}')]} \rangle =  \exp \left\{N_z \left[i \sum_{j=1}^{N_s} \langle \chi_{(j)}-\chi'_{(j)} \rangle - \right. \right.  \nonumber \\  
\left. \left. -\frac{1}{2} \sum_{j=1}^{N_s} \langle \left(\chi_{(j)}-\chi'_{(j)} \right)^2 \rangle + \frac{1}{2} \sum_{j=1}^{N_s}\langle \chi_{(j)}-\chi'_{(j)} \rangle ^2 + ... \right]\right\},
\end{eqnarray}
where $\chi_{(j)}=\chi_0(\vec{\rho}-\vec{\rho}_j-\vec{u})$, $\chi'_{(j)}=\chi_0(\vec{\rho}'-\vec{\rho}_j-\vec{u})$.
Expression $\langle e^{i\chi^{(N)}_0(\vec{\rho})} \rangle$ is obtained using the same idea, but with $\chi'_{(j)}=0$. Expression  $\langle e^{-i\chi^{(N)}_0(\vec{\rho}')} \rangle$ is complex conjugate of  $\langle e^{i\chi^{(N)}_0(\vec{\rho})} \rangle$ with $\vec{\rho}=\vec{\rho}'$.
If we consider \eqref{eq5} up to linear potential terms:  $\sum_{j=1}^{N_s} \langle \chi_{(j)}-\chi'_{(j)} \rangle$, $\sum_{j=1}^{N_s} \langle \chi_{(j)} \rangle$, $\sum_{j=1}^{N_s} \langle \chi'_{(j)} \rangle$, the differential scattering cross section can be written as  
\begin{eqnarray}\label{eq9}
\frac{d\sigma}{d^2q_{\perp}}=\frac{1}{4\pi^2} \Big| \int_{\mathbb{R}^2} d^2 \rho \ e^{i \vec{q}_{\perp} \vec{\rho}} \left\{ 1- \exp \left[i N_z \sum_{j=1}^{N_s} \langle \chi_{(j)} \rangle \right] \right\} \Big|^2.
\end{eqnarray}
The value $\langle \chi_{(j)} \rangle$ is the same for all atoms in the straight string so we can consider $N_z \langle \chi_{(j)} \rangle=\tilde{\chi}_{(j)}^{(s)}$, where $\tilde{\chi}_{(j)}^{(s)}=\tilde{\chi}_0^{(s)}(\vec{\rho}-\vec{\rho}_j)$ is $\chi_0$-function for $j$-th straight string of atoms. The function $\tilde{\chi}_{(j)}^{(s)}=N_z \langle \chi_{(j)} \rangle$ corresponds to the average potential of an atomic string in the continuous potential approximation \cite{Lind}. So, the scattering amplitude becomes  
\begin{eqnarray}\label{eq10}
a(\vec{q}_{\perp})=\frac{i}{2\pi} \int_{\mathbb{R}^2} d^2 \rho \ e^{i \vec{q}_{\perp} \vec{\rho}} \left\{ 1- \exp \left[i \sum_{j=1}^{N_s} \tilde{\chi}_{(j)}^{(s)} \right] \right\}.
\end{eqnarray}
In the works \cite{PerPl1, PerPl2}, we showed that for structures which can be considered isolated, the amplitude of scattering on a set of these structures can be expressed through structure factor and the amplitude of scattering on a single structure:
\begin{eqnarray}\label{eq11}
a(\vec{q}_{\perp})=S_{N_s} \ a^{(1)}(\vec{q}_{\perp}),
\end{eqnarray}
where structure factor $S_{N_s}$ is
\begin{eqnarray}\label{eq12}
S_{N_s}=\sum_{j=1}^{N_s} e^{i \vec{q}_{\perp} \vec{\rho}_j},
\end{eqnarray}
the amplitude of scattering on a single structure is
\begin{eqnarray}\label{eq13}
a^{(1)}(\vec{q}_{\perp})=\frac{i}{2\pi} \int_{\mathbb{R}^2} d^2 \rho \ e^{i \vec{q}_{\perp} \vec{\rho}} \left\{ 1- \exp \left[i \tilde{\chi}_{(0)}^{(s)} \right] \right\}
\end{eqnarray}and $\tilde{\chi}_{(0)}=\tilde{\chi}_0^{(s)}(\vec{\rho})$ corresponds to the $\chi_0$-function for a single structure -- for a single straight string, in our case -- which is located at the coordinate origin. We note that atomic strings should be considered isolated if the distance between neighboring strings is much greater than the distance at which their potentials significantly decrease. Then, the differential cross section of scattering on the set of these structures is also expressed through structure factor and the differential cross section of scattering on a single structure. 
\begin{eqnarray}\label{eq13_1}
\frac{d\sigma}{d^2q_{\perp}}= D_{N_s} \frac{d\sigma^{(1)}}{d^2q_{\perp}},
\end{eqnarray}
where $D_{N_s}= \Big|\sum_{j=1}^{N_s} e^{i \vec{q}_{\perp} \vec{\rho}_j} \Big|^2$ coincides with Laue-Bragg factor \cite{AIA96} and
\begin{eqnarray}\label{eq13_2}
\frac{d\sigma^{(1)}}{d^2q_{\perp}}=\Big| a^{(1)}(\vec{q}_{\perp}) \Big|^2
\end{eqnarray}
is the differential cross section of scattering on a single structure (atomic string in our case). Later we will discuss whether this approximation is applicable for this problem. Also, in the next section we will discuss how these calculations relate to the case of scattering on a tilted nanotube.

\section{Pontential-dependent functions}
To calculate the differential scattering cross sections both for straight and tilted nanotubes we need to know the corresponding $\chi_0$-functions for these nanotubes which centers are located at $(0,0)$ in the $(x,y)$-plane, which means $x_j=0$, $y_j=0$. So, index $j$ may be omitted in these calculations. To restore the dependence on $j$, one needs to make substitution $\vec{\rho} \rightarrow \vec{\rho} - \vec{\rho}_j$ in the final expression for $\chi_0$-function. 

To obtain $\tilde{\chi}_{(0)}$ for a straight atomic string, we need to calculate $\langle \chi_0 \rangle$  for a single atom and multiply it by amount  $N_z$ of atoms in the string.
Let us define each atom's potential as screened Coulomb potential \cite{AIA96}
\begin{eqnarray}\label{eq14}
u(\vec{r})=\frac{Ze}{r}\exp \left(-\frac{r}{R} \right),
\end{eqnarray}
where $R$ is the screening radius, $Ze$ is the charge of atom nucleus. 
Using the Fourier component of unaveraged $\chi_0$-function for a single atom (defined in \eqref{eq3_1}) \cite{AIA96}
\begin{eqnarray}\label{eq15}
\chi_{\vec{\mu}}=\frac{4 \pi Z\alpha}{\mu^2+R^{-2}}
\end{eqnarray}
where $\vec{\mu}=(\mu_x,\mu_y)$, $\mu^2=\mu_x^2+\mu_y^2$, we can write the expression for the averaged  $\langle \chi_0 \rangle$ as
\begin{eqnarray}\label{eq16}
\langle \chi_0 \rangle= 4 \pi Z\alpha \int_{\mathbb{R}^2} d^2u \ \frac{e^{- \ \frac{u^2}{2\langle {u^2_x} \rangle}}}{2 \pi \langle {u^2_x} \rangle} \int_{\mathbb{R}^2} \frac{d^2 \mu}{(2 \pi)^2} e^{i \mu_x(x-u_x)} e^{i \mu_y(y-u_y)} \frac{1}{\mu^2+R^{-2}}.
\end{eqnarray}
After integration over $u_x$, $u_y$ we obtain
\begin{eqnarray}\label{eq17}
\langle \chi_0 \rangle = \frac{Z\alpha}{\pi} \int_{\mathbb{R}^2}  d^2 \mu \ e^{i \mu_x x} e^{i \mu_y y} e^{- \ \frac{\mu^2_x \langle {u^2_x} \rangle}{2}} e^{- \ \frac{\mu^2_y \langle {u^2_x} \rangle}{2}} \frac{1}{\mu^2+R^{-2}}.
\end{eqnarray}
Spherical symmetry of the integral leads to
\begin{eqnarray}\label{eq17_1}
\langle \chi_0 \rangle = 2 Z\alpha \int_{0}^{\infty}  d \mu \ J_0(\mu \rho) \mu \ e^{- \ \frac{\mu^2 \langle {u^2_x} \rangle}{2}} \frac{1}{\mu^2+R^{-2}}
\end{eqnarray}
This integral can be exactly represented through series:
\begin{eqnarray}\label{eq18}
\int_{0}^{\infty}  d \mu \ J_0(\mu \rho) \mu \ \frac{ e^{- \ \frac{\mu^2 \langle {u^2_x} \rangle}{2}}}{\mu^2+R^{-2}}= 
\begin{cases}
\frac{1}{2} \ e^{ \frac{\langle {u^2_x} \rangle}{2R^2}} \sum_{k=0}^{\infty} \frac{(-1)^k}{k!} \left( \frac{\rho^2}{2\langle {u^2_x} \rangle} \right)^k E_{k+1} \left(\frac{\langle {u^2_x} \rangle}{2R^2} \right), \\
e^{ \frac{\langle {u^2_x} \rangle}{2R^2}} \left[ K_0 \left(\frac{\rho}{R} \right) -\frac{1}{2} \  \sum_{k=0}^{\infty} \frac{(-1)^k}{k!} \left( \frac{\langle {u^2_x} \rangle}{2R^2}  \right)^k E_{k+1} \left(  \frac{\rho^2}{2\langle {u^2_x} \rangle} \right) \right].
\end{cases}
\end{eqnarray}
where $E_{k}(\eta)$ is exponential integral function.

We note that both series in upper and down parts of \eqref{eq18} contain expressions of the form $\sum_{k=0}^{\infty} \frac{(-1)^k}{k!} \xi^k E_{k+1}(\eta)$, where $\xi \geq 0$, $\eta > 0$. For $\eta >0$, the following inequality holds: $E_{k+1}(\eta)<E_1(\eta)$ for $k \geq 1$. So we can estimate this series from above as $\sum_{k=0}^{\infty} \frac{(-1)^k}{k!} \xi^k E_{k+1}(\eta) \leq E_{1}(\eta) \sum_{k=0}^{\infty} \frac{(-1)^k}{k!} \xi^k$ which is equivalent to estimation $\sum_{k=0}^{\infty} \frac{(-1)^k}{k!} \xi^k E_{k+1}(\eta) \leq E_{1}(\eta) e^{-\xi}$. It means that both series in \eqref{eq18} converge for arguments of $E_k$ being greater than $0$. Considering that $\xi^k$ has smaller values for relatively small $\xi$ and $E_{k+1}(\eta)$ has smaller values for relatively large $\eta$, the "upper" definition of the integral suits better for $\rho \lesssim R$ and the "down" definition -- for $\rho \gtrsim R$. So, we obtain  
\begin{eqnarray}\label{eq19}
\langle \chi_0 \rangle = 2 Z\alpha \ e^{ \frac{\langle {u^2_x} \rangle}{2R^2}}
\begin{cases}
\frac{1}{2}  \sum_{k=0}^{\infty} \frac{(-1)^k}{k!} \left( \frac{\rho^2}{2\langle {u^2_x} \rangle} \right)^k E_{k+1} \left(\frac{\langle {u^2_x} \rangle}{2R^2} \right), \ \rho \lesssim R \\
K_0 \left(\frac{\rho}{R} \right) -\frac{1}{2} \  \sum_{k=0}^{\infty} \frac{(-1)^k}{k!} \left( \frac{\langle {u^2_x} \rangle}{2R^2}  \right)^k E_{k+1} \left(  \frac{\rho^2}{2\langle {u^2_x} \rangle} \right), \ \rho \gtrsim R.
\end{cases}
\end{eqnarray}
Formula \eqref{eq19} agrees well with idea that at large distances under condition $\langle {u^2_x} \rangle=0$, $\langle \chi_0 \rangle = 2 Z \alpha K_0(\rho/R)$ coincides with $\chi_0$ for a single atom's potential without averaging \cite{AIA96}. 
So, for a straight atomic string
\begin{eqnarray}\label{eq19_1}
\tilde{\chi}_0^{(s)} = A \ e^{ \frac{\langle {u^2_x} \rangle}{2R^2}}
\begin{cases}
\frac{1}{2}  \sum_{k=0}^{\infty} \frac{(-1)^k}{k!} \left( \frac{\rho^2}{2\langle {u^2_x} \rangle} \right)^k E_{k+1} \left(\frac{\langle {u^2_x} \rangle}{2R^2} \right), \ \rho \lesssim R \\
K_0 \left(\frac{\rho}{R} \right) -\frac{1}{2} \  \sum_{k=0}^{\infty} \frac{(-1)^k}{k!} \left( \frac{\langle {u^2_x} \rangle}{2R^2}  \right)^k E_{k+1} \left(  \frac{\rho^2}{2\langle {u^2_x} \rangle} \right), \ \rho \gtrsim R;
\end{cases}
\end{eqnarray}
where $A=2Z \alpha N_z$.

For the case of a tilted string $x$-component of atoms positions in the string are not the same like in the case of a straight string. Considering that $\chi_0$-function for a string is a sum of $\chi_0$-functions of all atoms in the string,
\begin{eqnarray}\label{eq20}
\tilde{\chi}_0^{(t)}= 4 \pi Z\alpha \sum_{k=1}^{N_z} \int_{\mathbb{R}^2} d^2u \ \frac{e^{- \ \frac{u^2}{2\langle {u^2_x} \rangle}}}{2 \pi \langle {u^2_x} \rangle} \int_{\mathbb{R}^2} \frac{d^2 \mu}{(2 \pi)^2} e^{i \mu_x(x-\tilde{x}_k-u_x)} e^{i \mu_y(y-u_y)} \frac{1}{\mu^2+R^{-2}}.
\end{eqnarray}

\begin{figure}[!t]
\begin{subfigure}{0.49\linewidth}
\includegraphics[width=\textwidth]{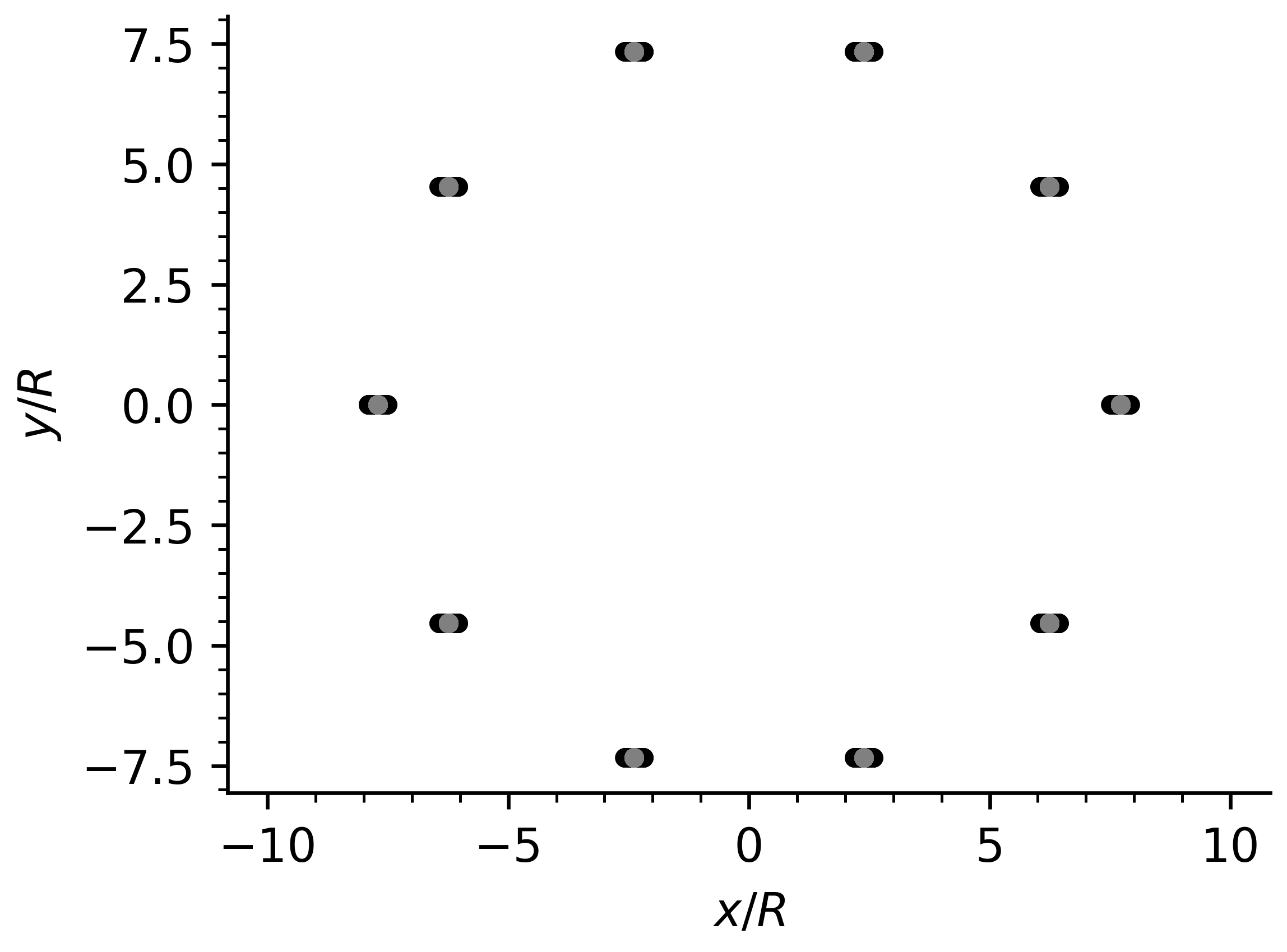}
\caption{$L_x=0.4R$}
\label{fig:lx2}
\end{subfigure}
\begin{subfigure}{0.49\linewidth}
\includegraphics[width=\textwidth]{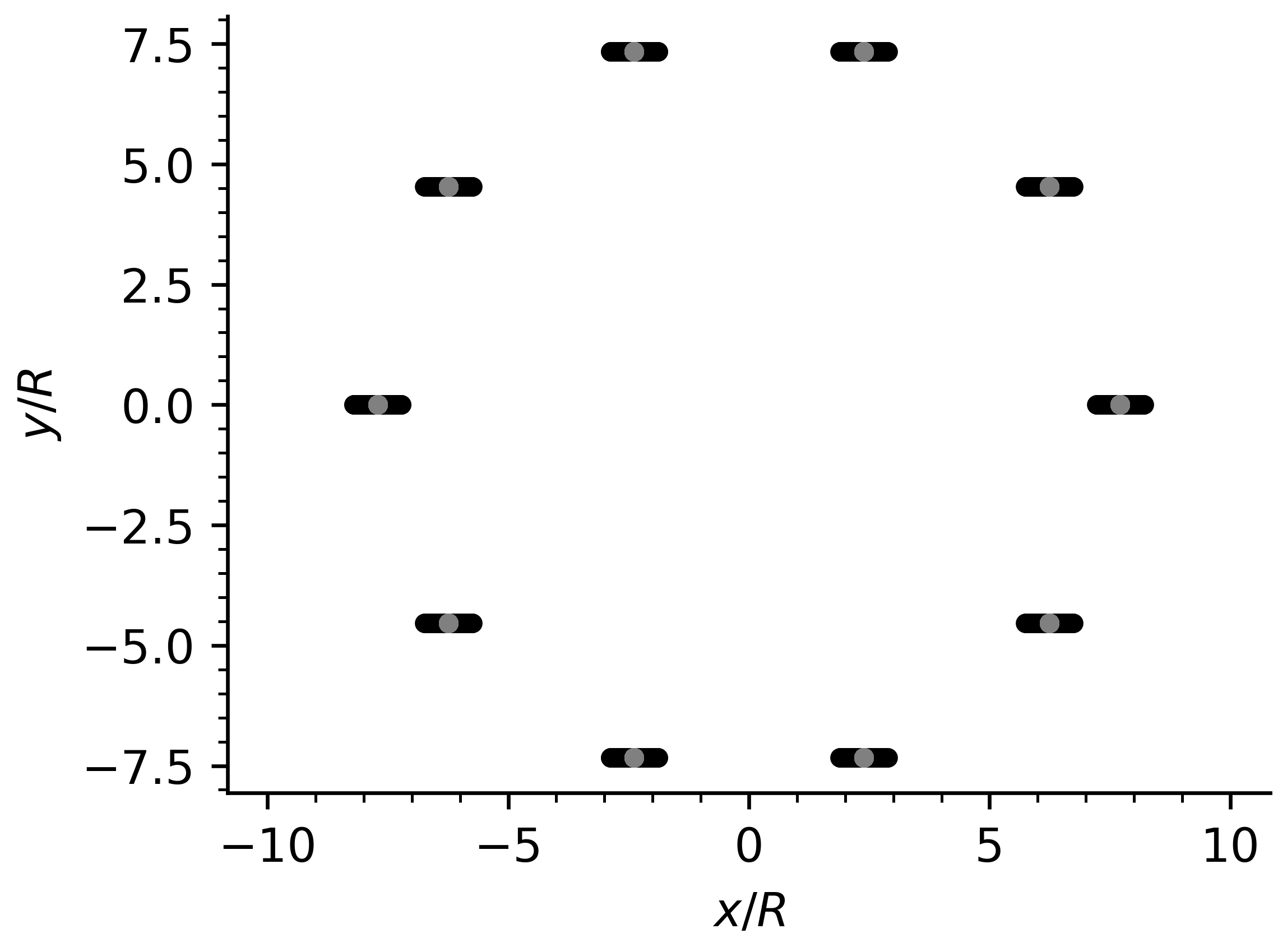}
\caption{$L_x=1.0R$}
\label{fig:lx8}
\end{subfigure}
\vfill
\begin{subfigure}{0.49\linewidth}
\includegraphics[width=\textwidth]{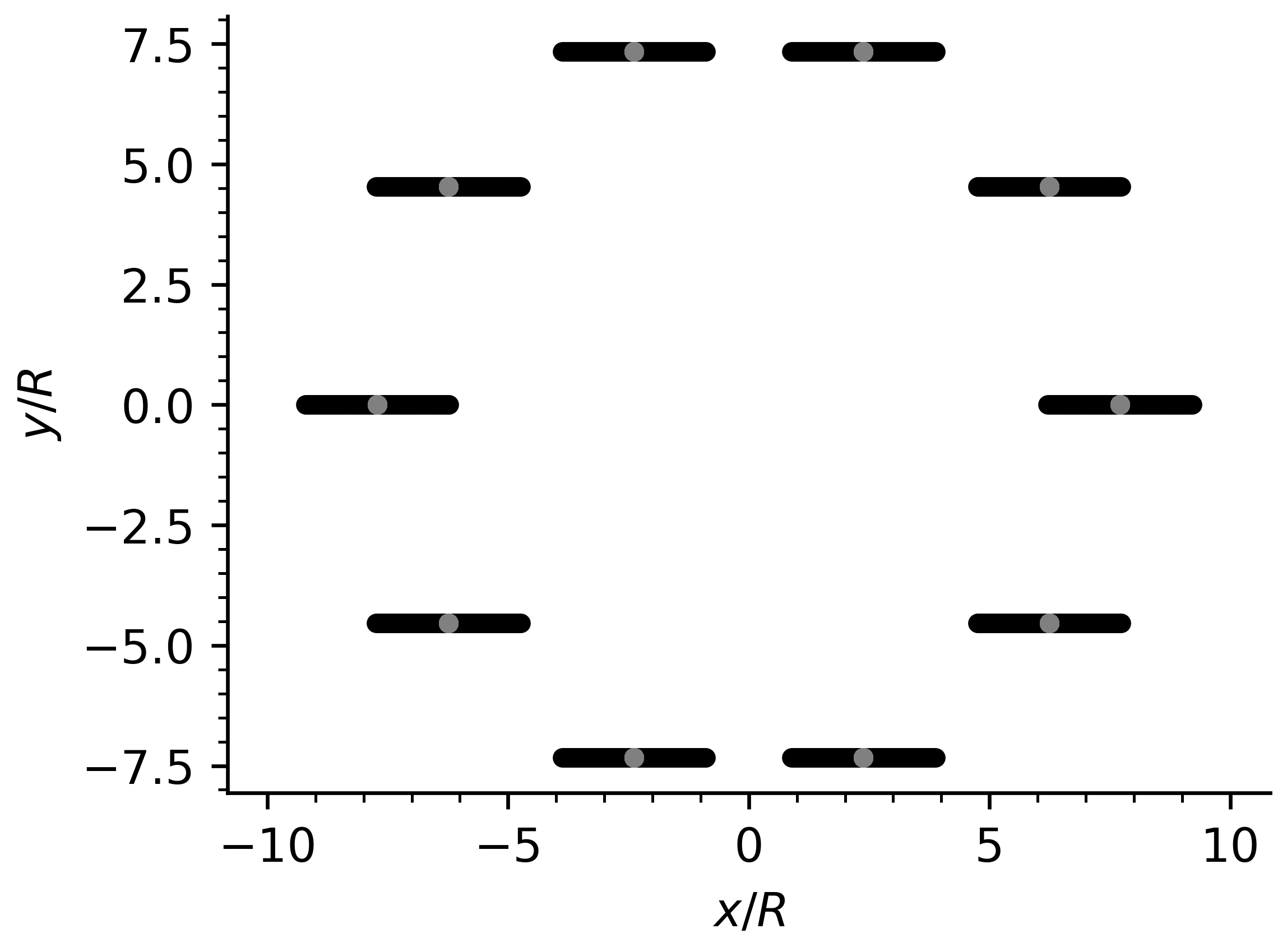}
\caption{$L_x=3.0R$}
\label{fig:lx15}
\end{subfigure}
\begin{subfigure}{0.49\linewidth}
\includegraphics[width=\textwidth]{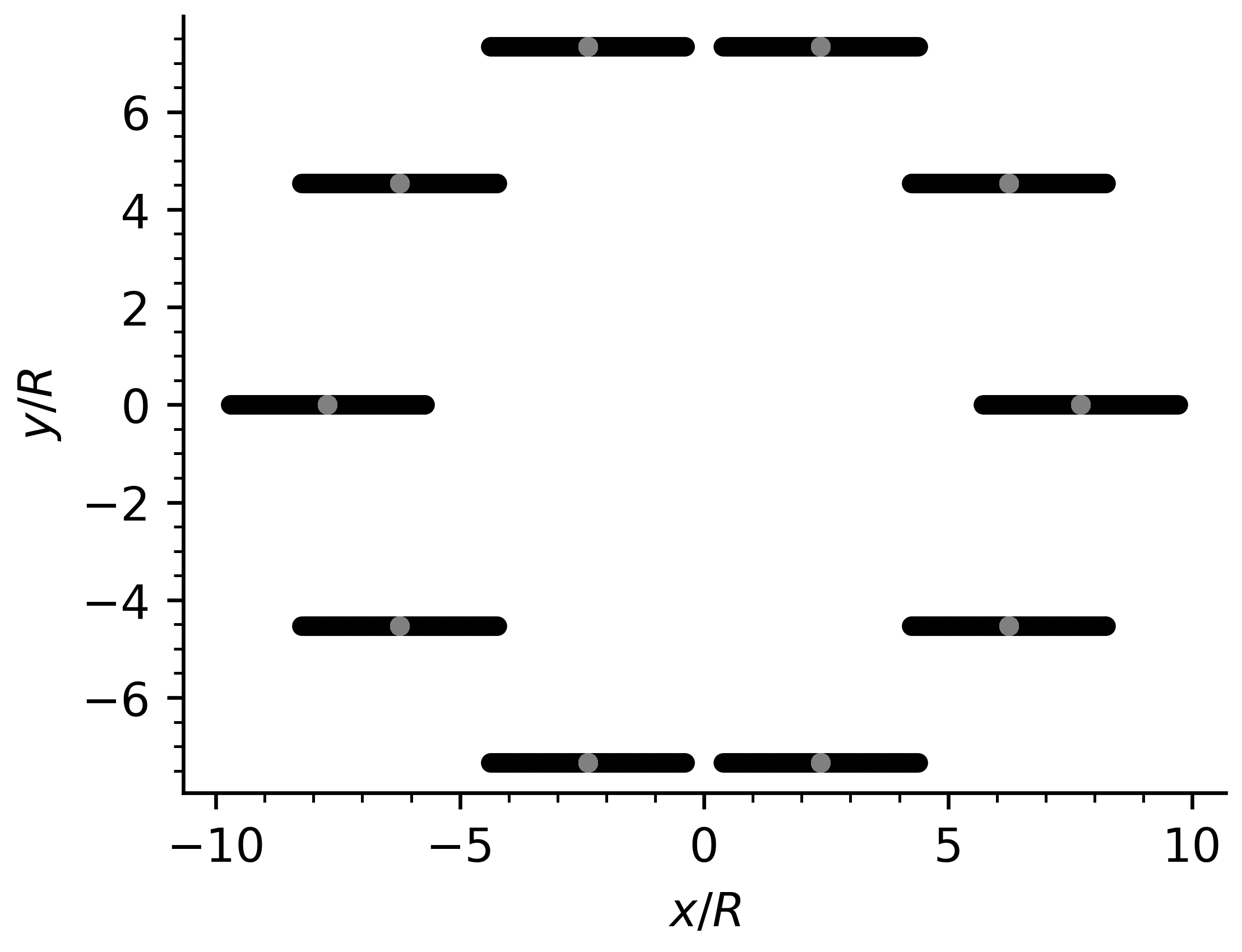}
\caption{$L_x=4.0R$}
\label{fig:lx20}
\end{subfigure}
\caption{Tilted nanotube projection on the $(x,y)$-plane for $A=10$: black dots denote projection of atoms positions of the tilted nanotube, grey dots denote projection of atoms positions of the straight nanotube }
\label{fig:lx}
\end{figure}

As we see from \eqref{eq0}, $x$-component of atom position depends linearly on its position along the string. For large enough amount of atoms, the string in $(x,y)$-plane looks similar to a segment of a solid line of length $L_x=L_z \sin \theta$ (Fig. \ref{fig:lx}). For this case, we can consider the $\chi_0$-function for a string as averaged over $x$-component of atoms positions $\chi_0$-function. That means, 
\begin{eqnarray}\label{eq21}
\sum_{k=1}^{N_z} e^{-i \mu_x\tilde{x}_k} \approx N_z \int_{-L_x/2}^{L_x/2} \frac{d \tilde{x}}{L_x} e^{-i \mu_x\tilde{x}}.
\end{eqnarray}

\noindent When we perform averaging \eqref{eq21}, the dependence of $\tilde{\chi}_0^{(t)}$ on $k$ vanishes. This leads function $\tilde{\chi}_0^{(t)}$ to be of the analogous dependence on $N_z$ and indices $j$, $k$ as in $\tilde{\chi}_0^{(s)}$. So, formulas \eqref{eq8}-\eqref{eq13_2} can be used for the case of a tilted nanotube as well after substitution $\tilde{\chi}_0^{(s)} \rightarrow \tilde{\chi}_0^{(t)}$.  

After integrating over $\tilde{x}$, $u_x$, $u_y$ in \eqref{eq20} and keeping only even on $\mu_x$, $\mu_y$ terms, we obtain 
\begin{eqnarray}\label{eq22}
\tilde{\chi}_0^{(t)} = \frac{A}{\pi L_x}\int_{\mathbb{R}^2}  d^2 \mu \ \cos(\mu_x x) \cos(\mu_y y) \frac{\sin\left(\mu_x \frac{L_x}{2} \right)}{\mu_x}  \frac{e^{- \ \frac{\mu^2_x \langle {u^2_x} \rangle}{2}} e^{- \ \frac{\mu^2_y \langle {u^2_x} \rangle}{2}}}{\mu^2+R^{-2}}.
\end{eqnarray}
Using identity $\cos(\mu_x x) \sin \left(\mu_x \frac{L_x}{2} \right)=\frac{1}{2} \{\sin\left[\mu_x \left(x+\frac{L_x}{2} \right) \right]-\sin\left[\mu_x \left(x-\frac{L_x}{2} \right) \right] \}$, we can rewrite \eqref{eq22} as
\begin{eqnarray}\label{eq23}
\tilde{\chi}_0^{(t)} = \frac{A}{2 L_x} \left[\tilde{I}(L_x)- \tilde{I}(-L_x) \right],
\end{eqnarray}
where
\begin{eqnarray}\label{eq24}
\tilde{I}(L_x) = \frac{1}{\pi} \int_{\mathbb{R}^2}  d^2 \mu \ \cos(\mu_y y) \frac{\sin \left[\mu_x \left(x+\frac{L_x}{2}\right) \right]}{\mu_x}  \frac{e^{- \ \frac{\mu^2_x \langle {u^2_x} \rangle}{2}} e^{- \ \frac{\mu^2_y \langle {u^2_x} \rangle}{2}}}{\mu^2+R^{-2}}.
\end{eqnarray}
Performing integration over $\mu_y$ leads to
\begin{eqnarray}\label{eq25}
\tilde{I}(L_x) = e^{ \frac{\langle {u^2_x} \rangle}{2R^2}} \int_{0}^{\infty}  d \mu_x \frac{\sin \left[\mu_x \left(x+\frac{L_x}{2}\right) \right]}{\mu_x \sqrt{\mu_x^2+R^{-2}}}  \left\{e^{-y \sqrt{\mu_x^2+R^{-2}}} \text{erfc}\left[\sqrt{\frac{\langle {u^2_x} \rangle}{2}} \sqrt{\mu_x^2+R^{-2}} -\frac{y}{\sqrt{2\langle {u^2_x} \rangle}} \right] +\right. \nonumber \\
\left. + e^{y \sqrt{\mu_x^2+R^{-2}}} \text{erfc}\left[\sqrt{\frac{\langle {u^2_x} \rangle}{2}} \sqrt{\mu_x^2+R^{-2}} +\frac{y}{\sqrt{2\langle {u^2_x} \rangle}} \right] \right\}.
\end{eqnarray}
where $\text{erfc}(z)=1-\text{erf}(z)$ is complementary error function. 

Integration over $\mu_x$ in \eqref{eq25} we perform numerically.

\section{Results and Discussion}
\begin{figure}[!t]
	\begin{subfigure}{0.32\linewidth}
	\includegraphics[width=\textwidth]{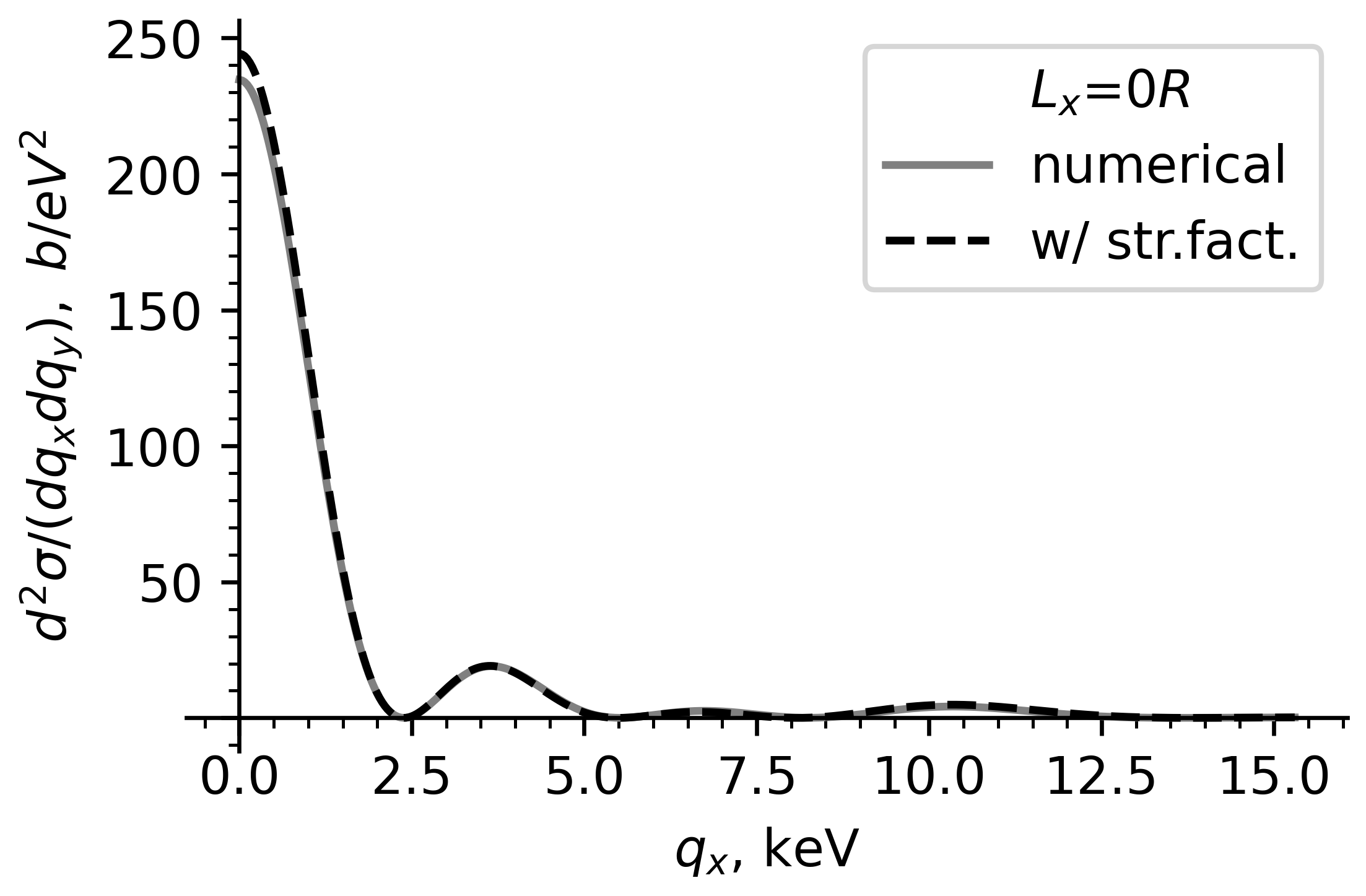}
	\caption{$q_y=0.0$ keV}
	\label{fig:cs_0_0}
	\end{subfigure}
	\begin{subfigure}{0.32\linewidth}
	\includegraphics[width=\textwidth]{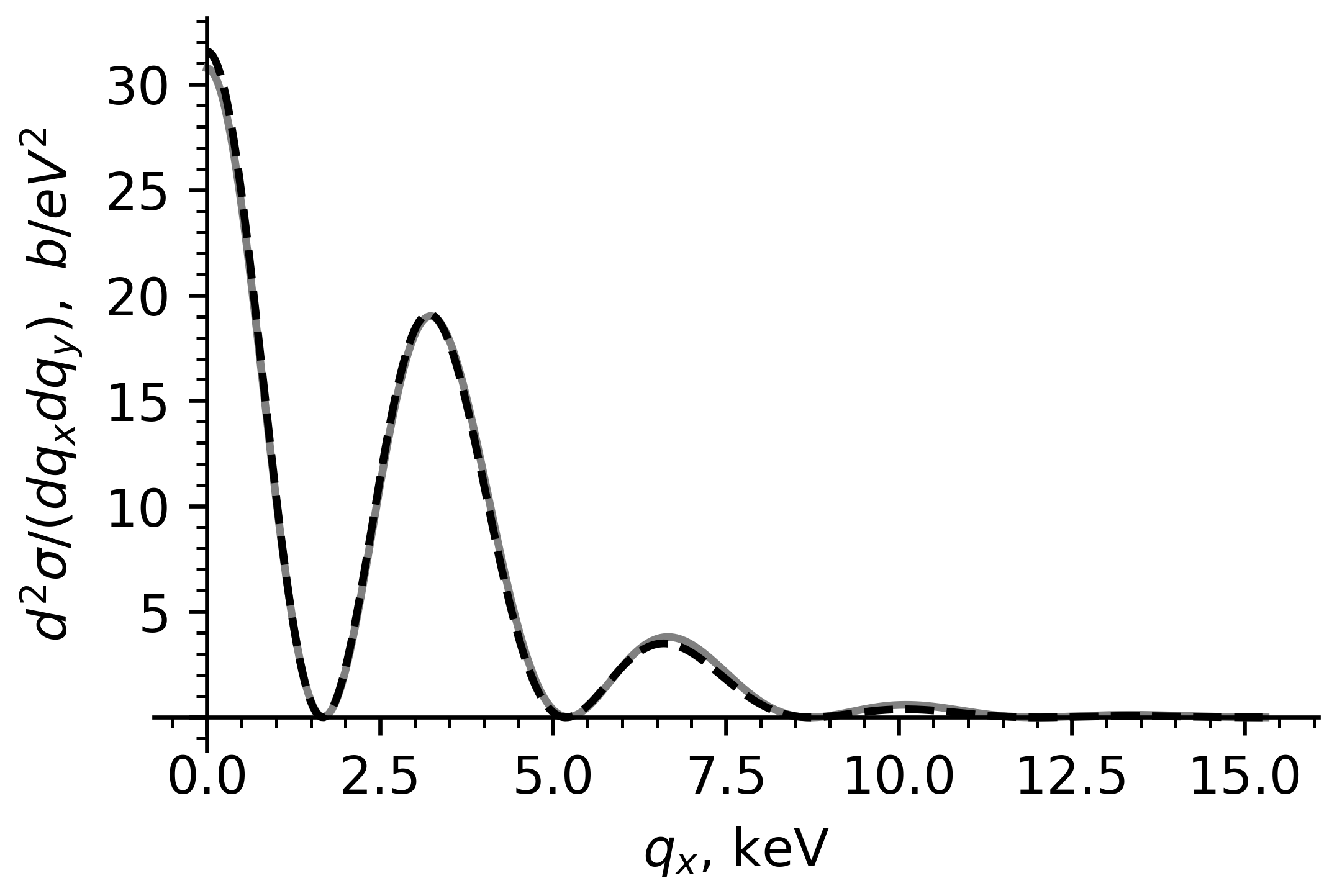}
	\caption{$q_y=1.7$ keV}
	\label{fig:cs_0_1}
	\end{subfigure}
	\begin{subfigure}{0.32\linewidth}
	\includegraphics[width=\textwidth]{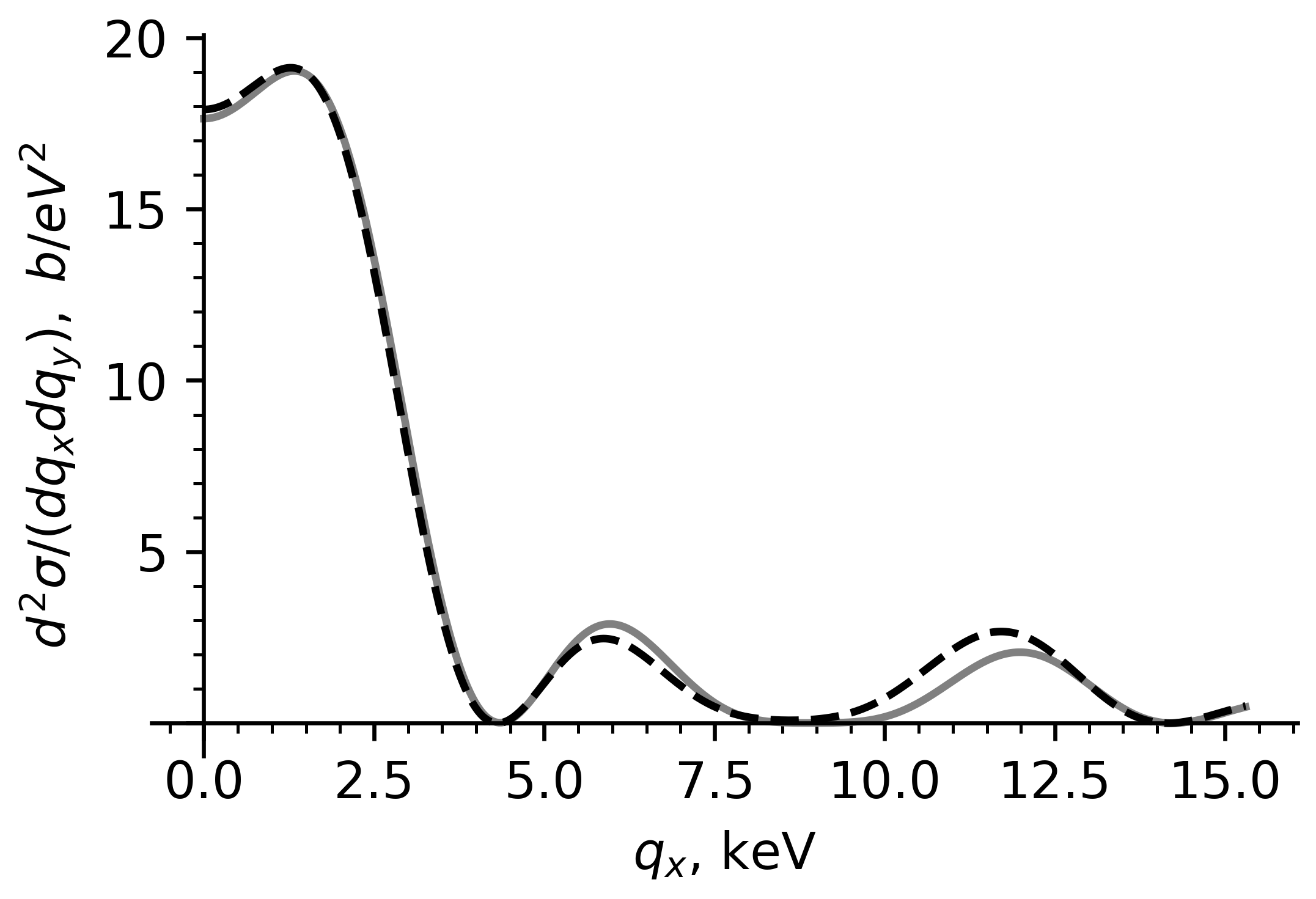}
	\caption{$q_y=3.4$ keV}
	\label{fig:cs_0_2}
	\end{subfigure}
\vfill
	\begin{subfigure}{0.32\linewidth}
	\includegraphics[width=\textwidth]{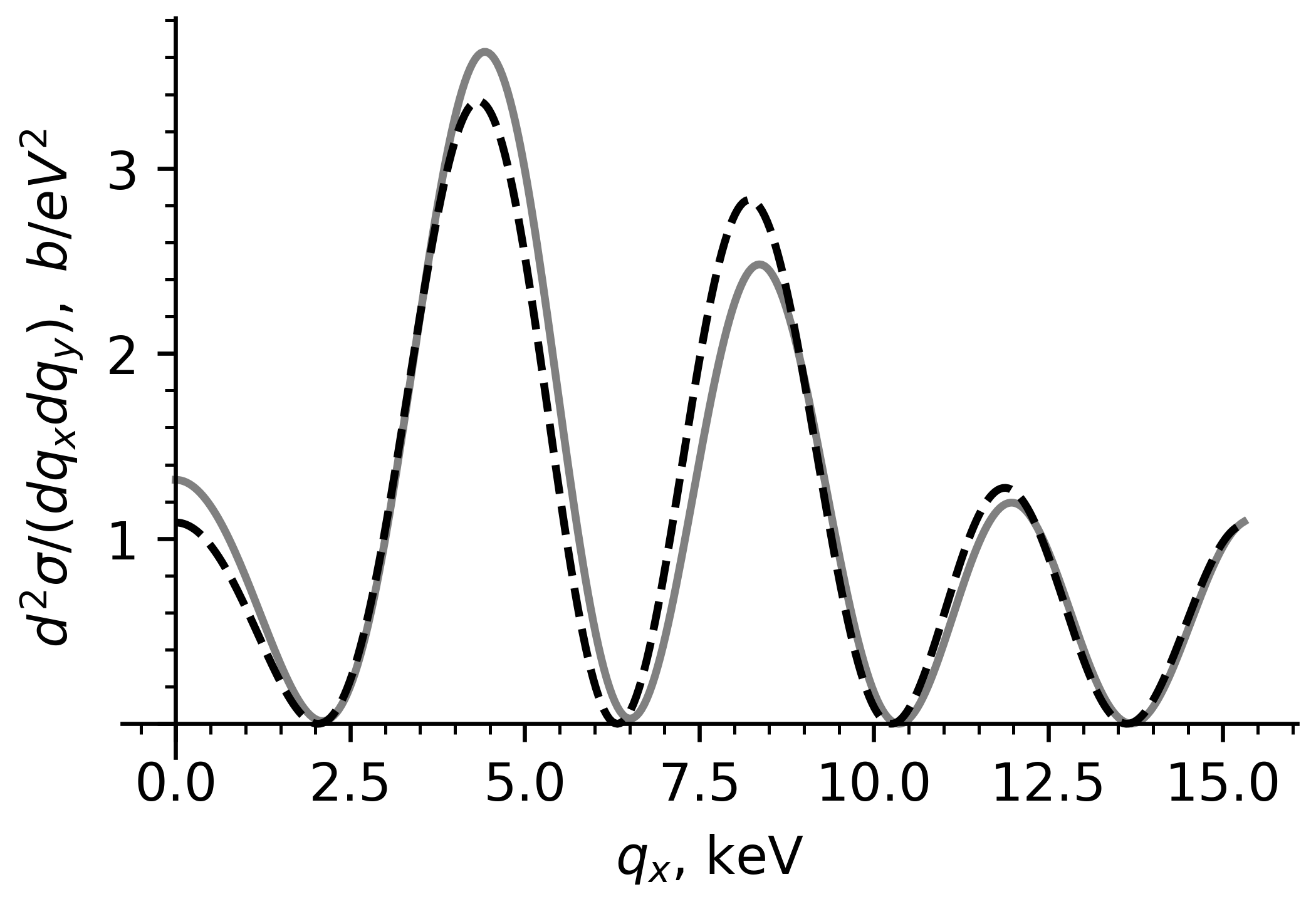}
	\caption{$q_y=5.1$ keV}
	\label{fig:cs_0_3}
	\end{subfigure}
	\begin{subfigure}{0.32\linewidth}
	\includegraphics[width=\textwidth]{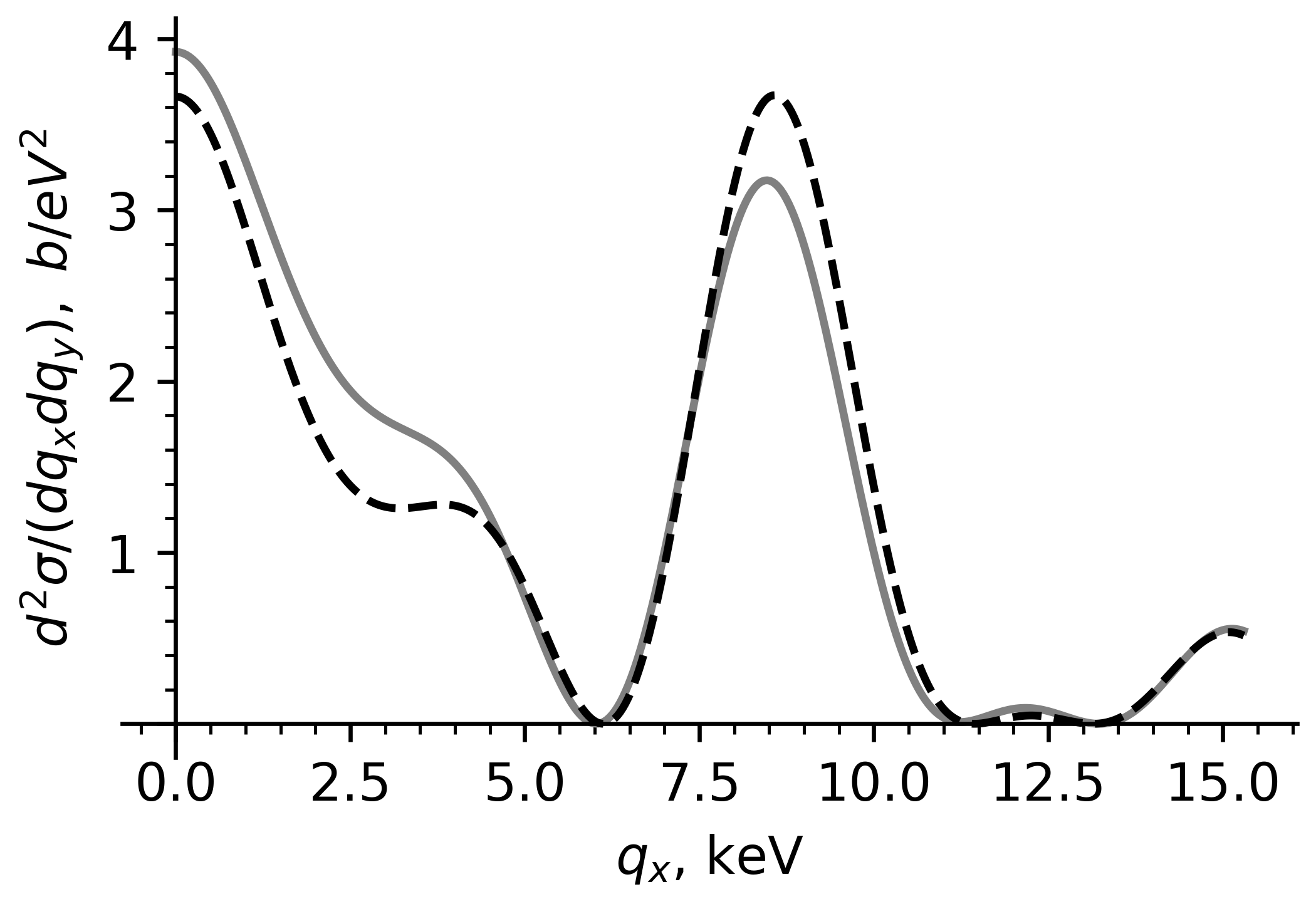}
	\caption{$q_y=6.8$ keV}
	\label{fig:cs_0_4}
	\end{subfigure}
	\begin{subfigure}{0.32\linewidth}
	\includegraphics[width=\textwidth]{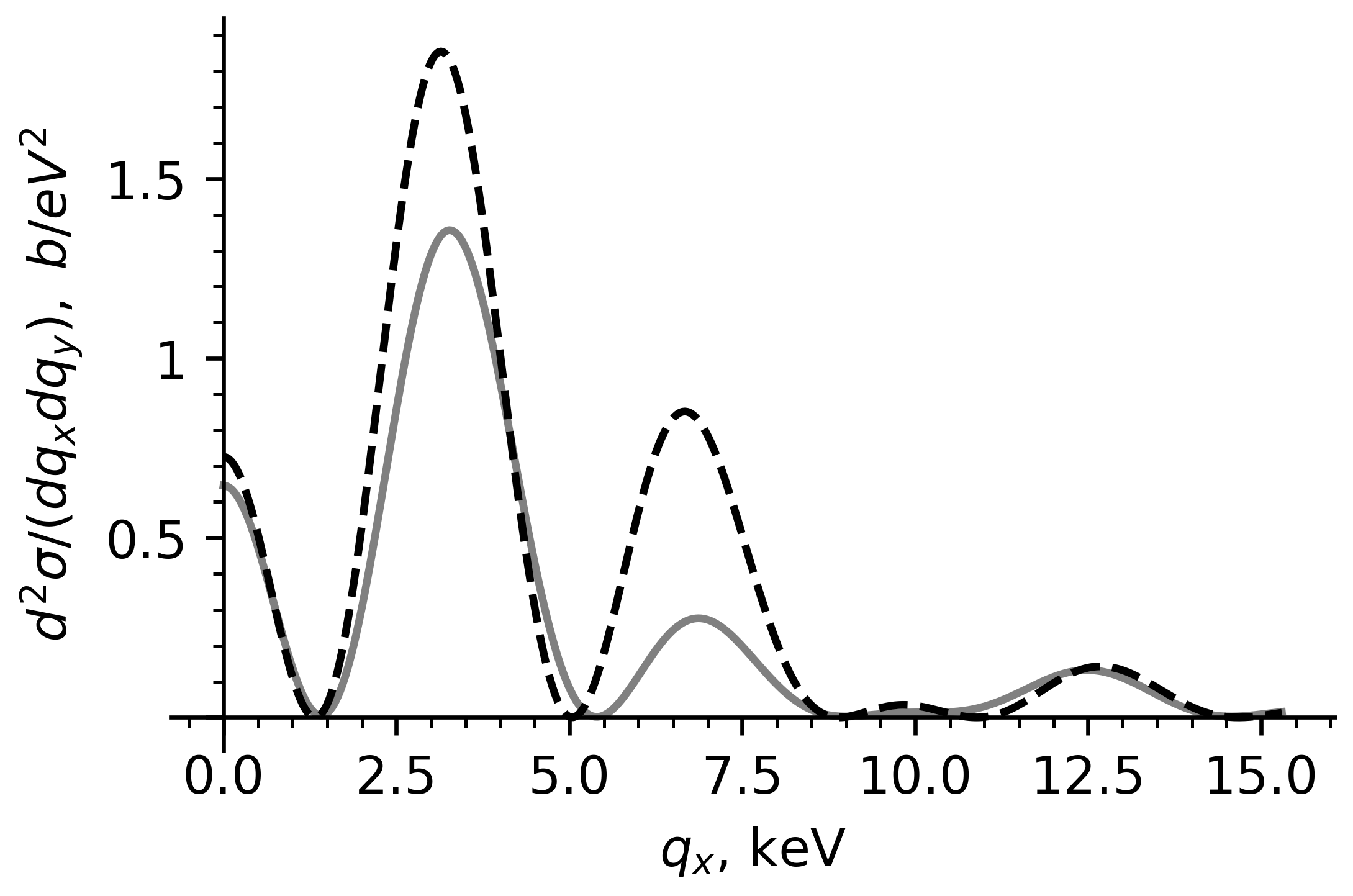}
	\caption{$q_y=8.5$ keV}
	\label{fig:cs_0_5}
	\end{subfigure}
\vfill
	\begin{subfigure}{0.32\linewidth}
	\includegraphics[width=\textwidth]{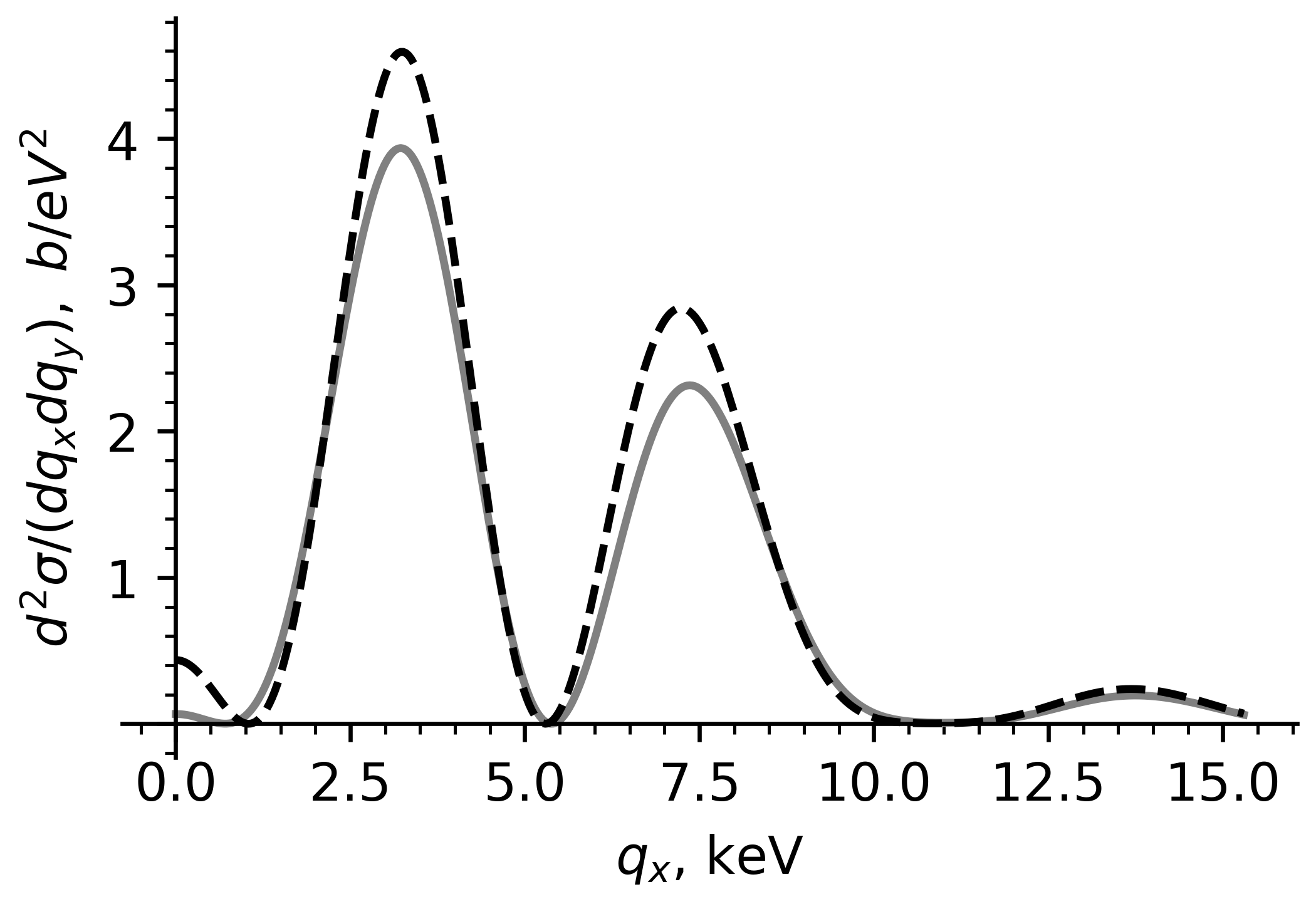}
	\caption{$q_y=10.2$ keV}
	\label{fig:cs_0_6}
	\end{subfigure}
	\begin{subfigure}{0.32\linewidth}
	\includegraphics[width=\textwidth]{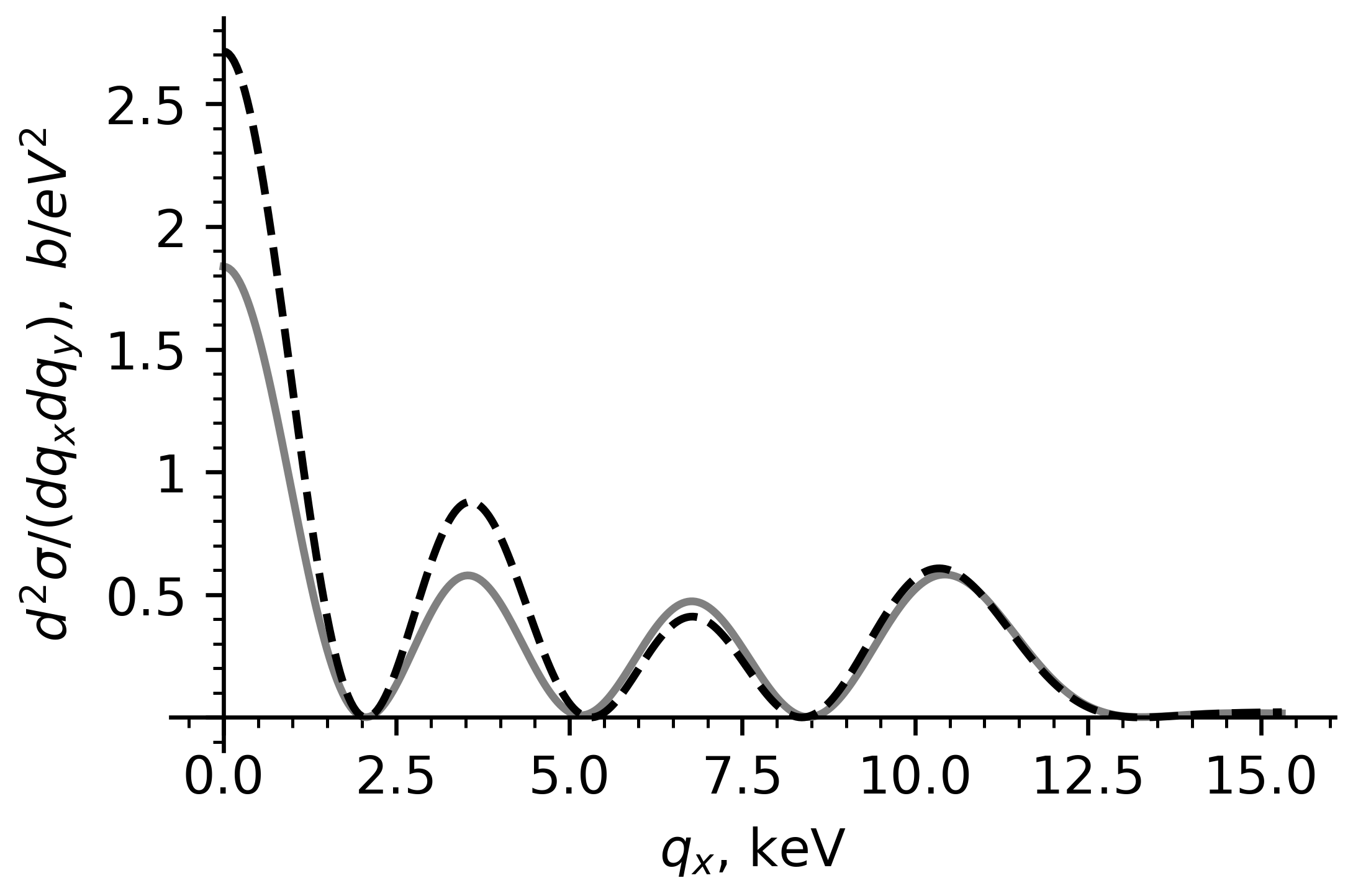}
	\caption{$q_y=11.9$ keV}
	\label{fig:cs_0_7}
	\end{subfigure}
	\begin{subfigure}{0.32\linewidth}
	\includegraphics[width=\textwidth]{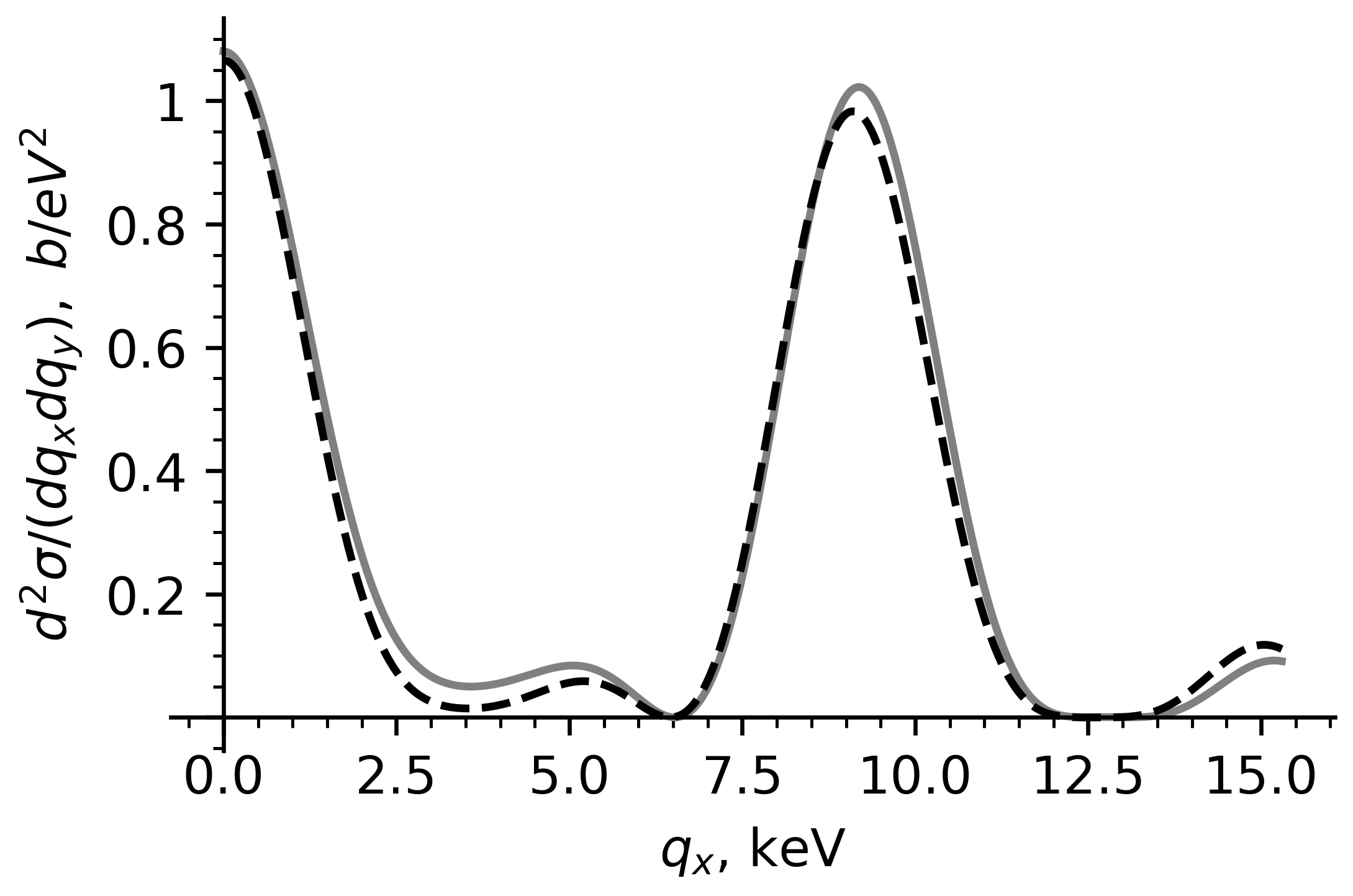}
	\caption{$q_y=13.6$ keV}
	\label{fig:cs_0_8}
	\end{subfigure}
\caption{Differential cross section of a fast charged particle scattering on a straight nanotube}
\label{fig_cs_l0}
\end{figure}

\begin{figure}[!h]
\begin{subfigure}{0.32\linewidth}
\includegraphics[width=\textwidth]{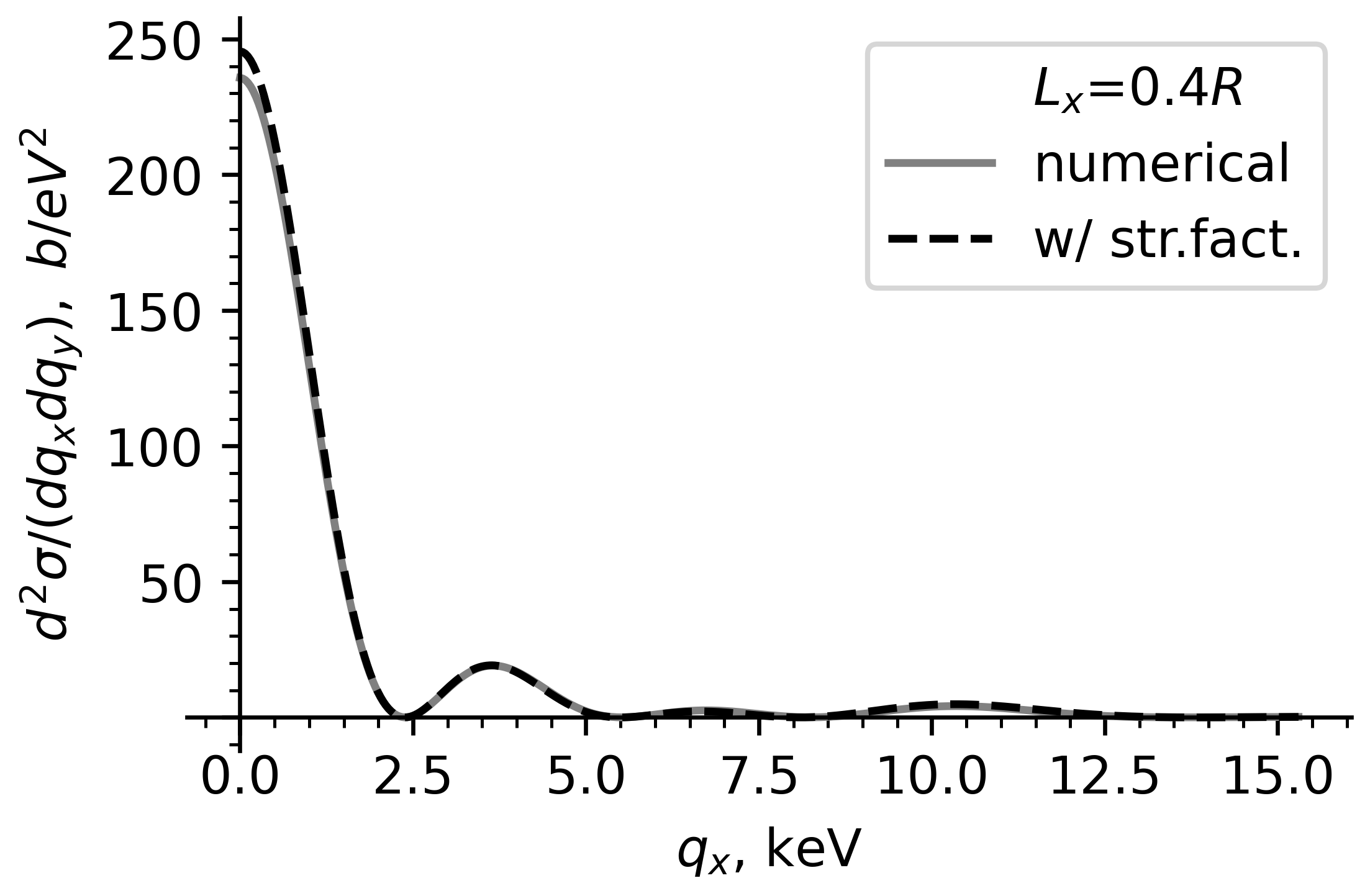}
\caption{$q_y=0.0$ keV}
\label{fig:cs_2_0}
\end{subfigure}
\begin{subfigure}{0.32\linewidth}
\includegraphics[width=\textwidth]{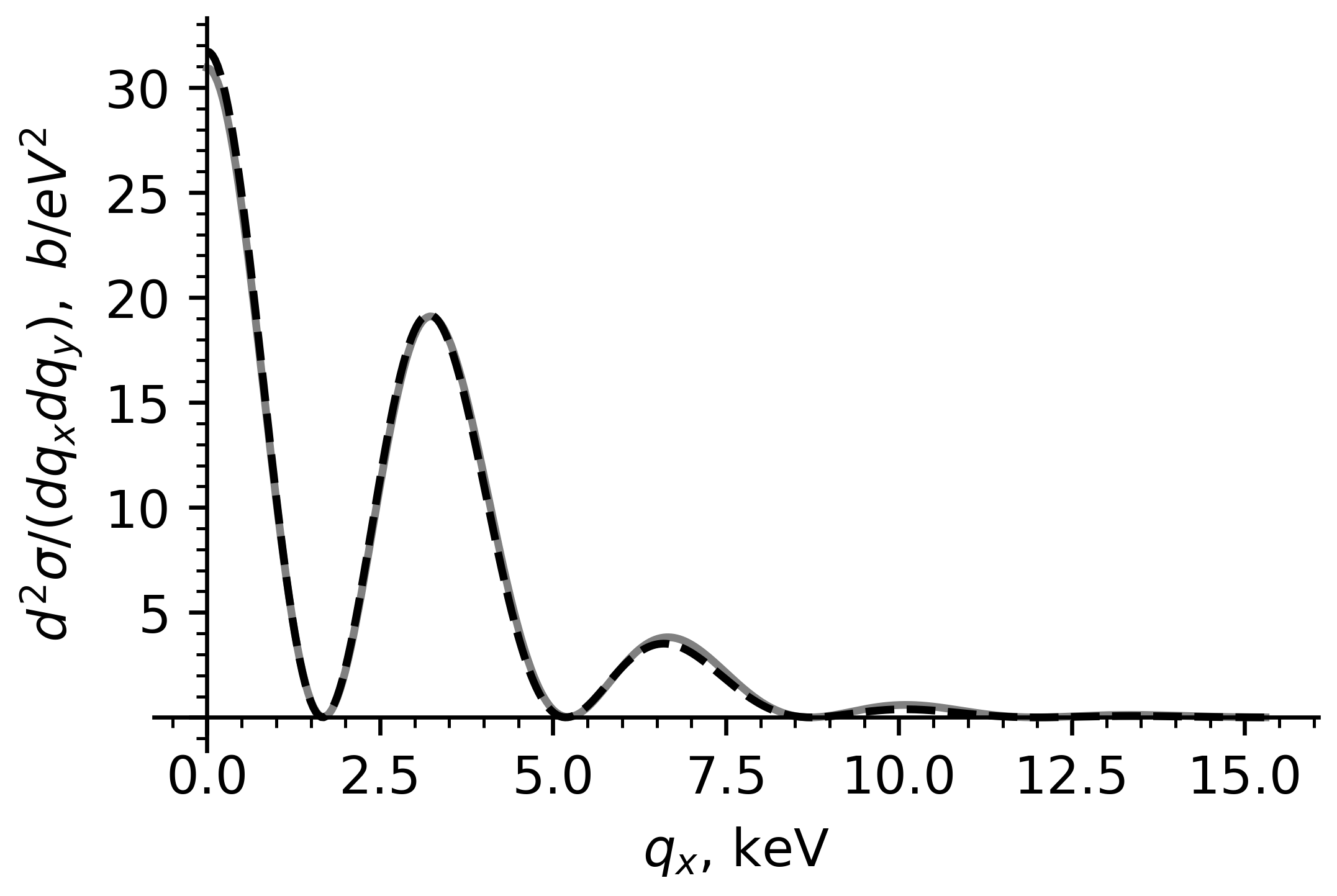}
\caption{$q_y=1.7$ keV}
\label{fig:cs_2_1}
\end{subfigure}
\begin{subfigure}{0.32\linewidth}
\includegraphics[width=\textwidth]{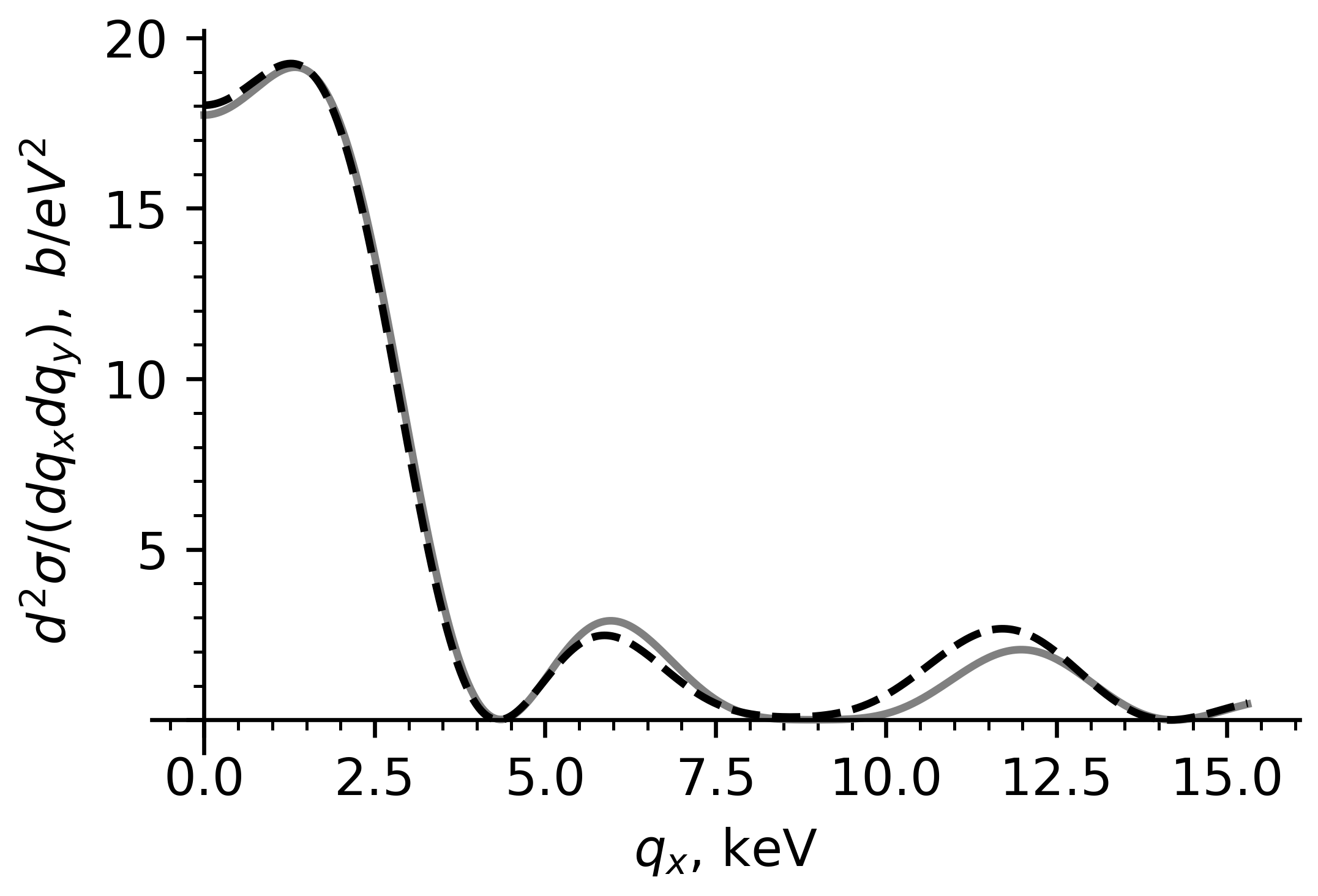}
\caption{$q_y=3.4$ keV}
\label{fig:cs_2_2}
\end{subfigure}
\vfill
\begin{subfigure}{0.32\linewidth}
\includegraphics[width=\textwidth]{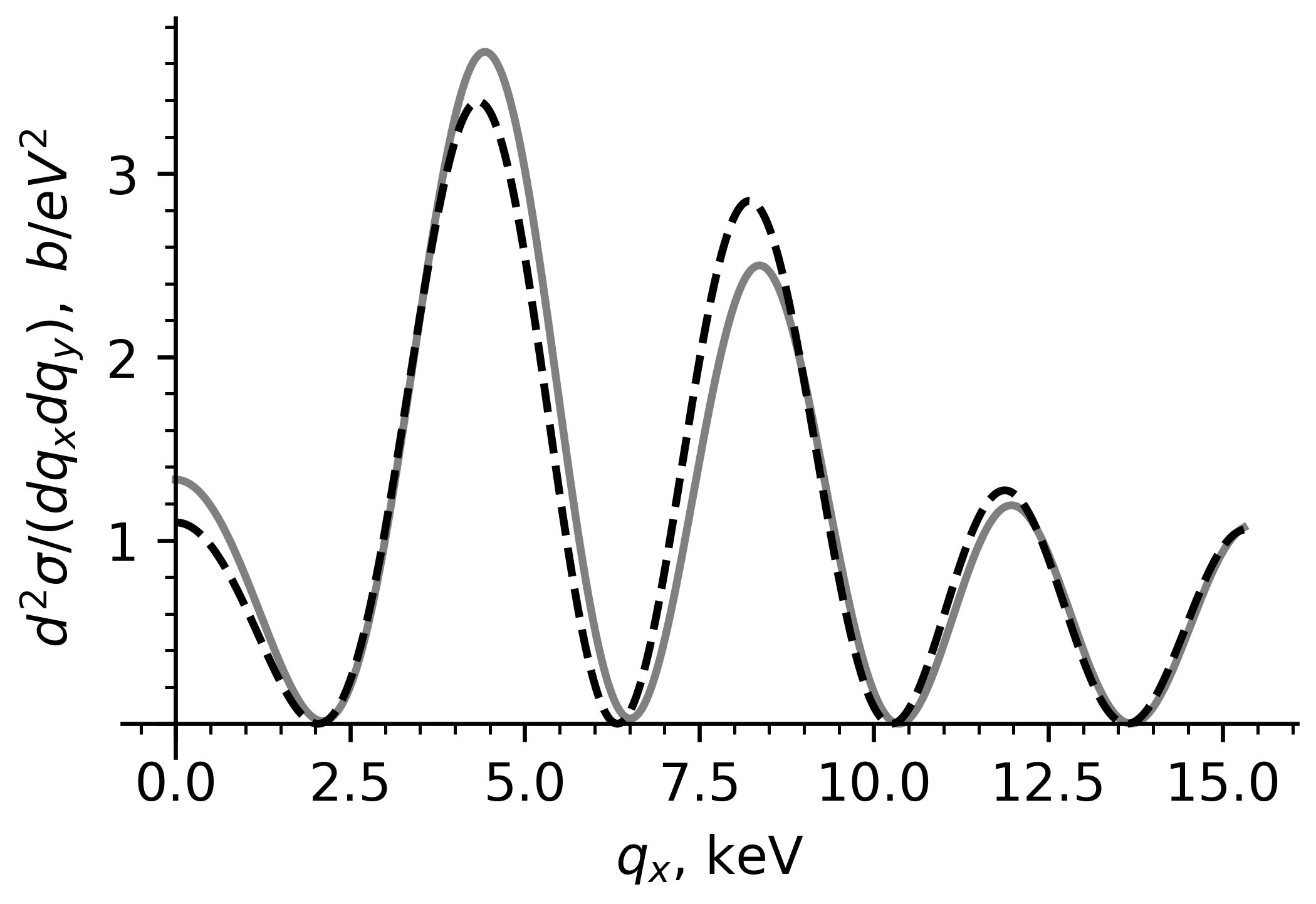}
\caption{$q_y=5.1$ keV}
\label{fig:cs_2_3}
\end{subfigure}
\begin{subfigure}{0.32\linewidth}
\includegraphics[width=\textwidth]{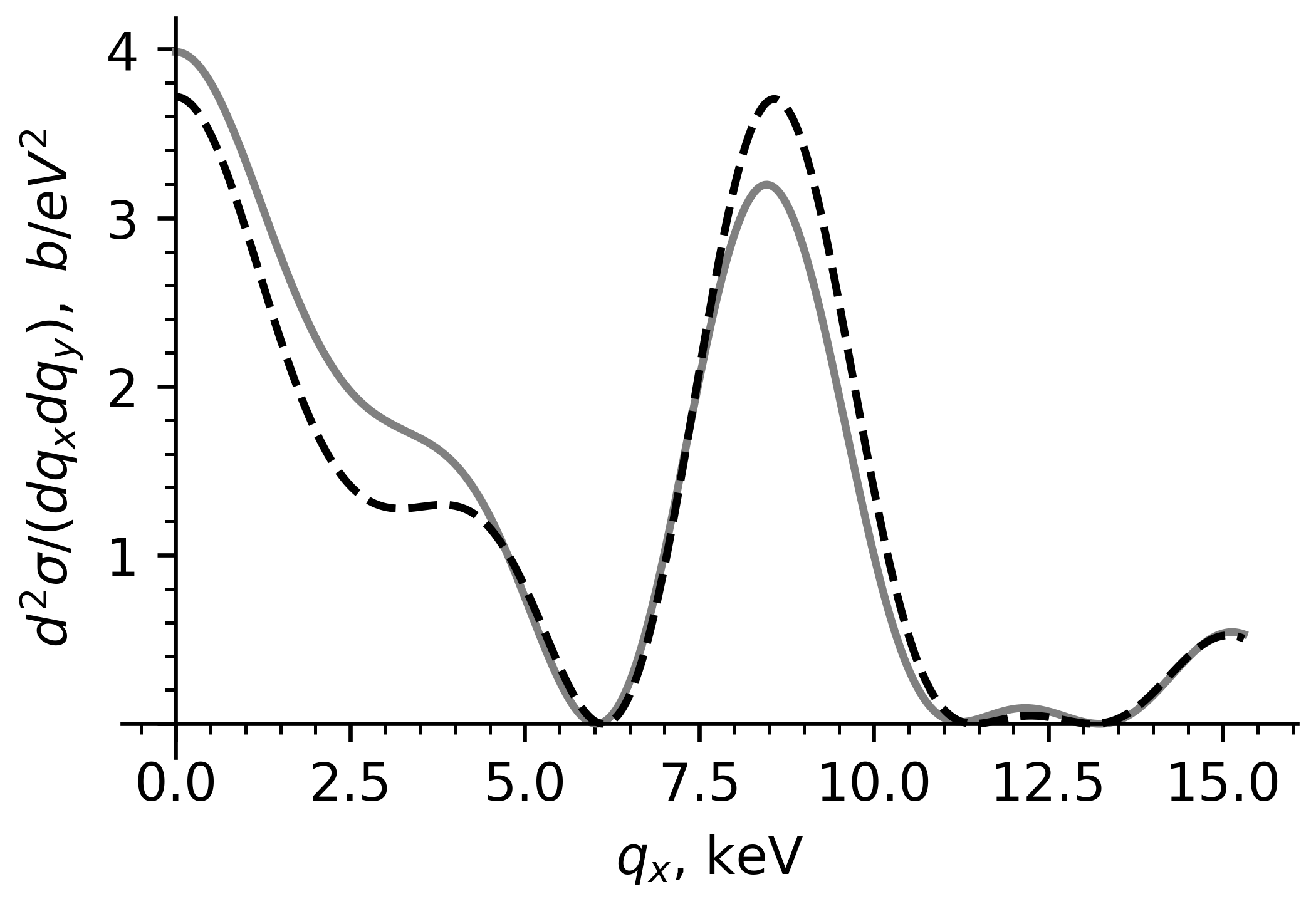}
\caption{$q_y=6.8$ keV}
\label{fig:cs_2_4}
\end{subfigure}
\begin{subfigure}{0.32\linewidth}
\includegraphics[width=\textwidth]{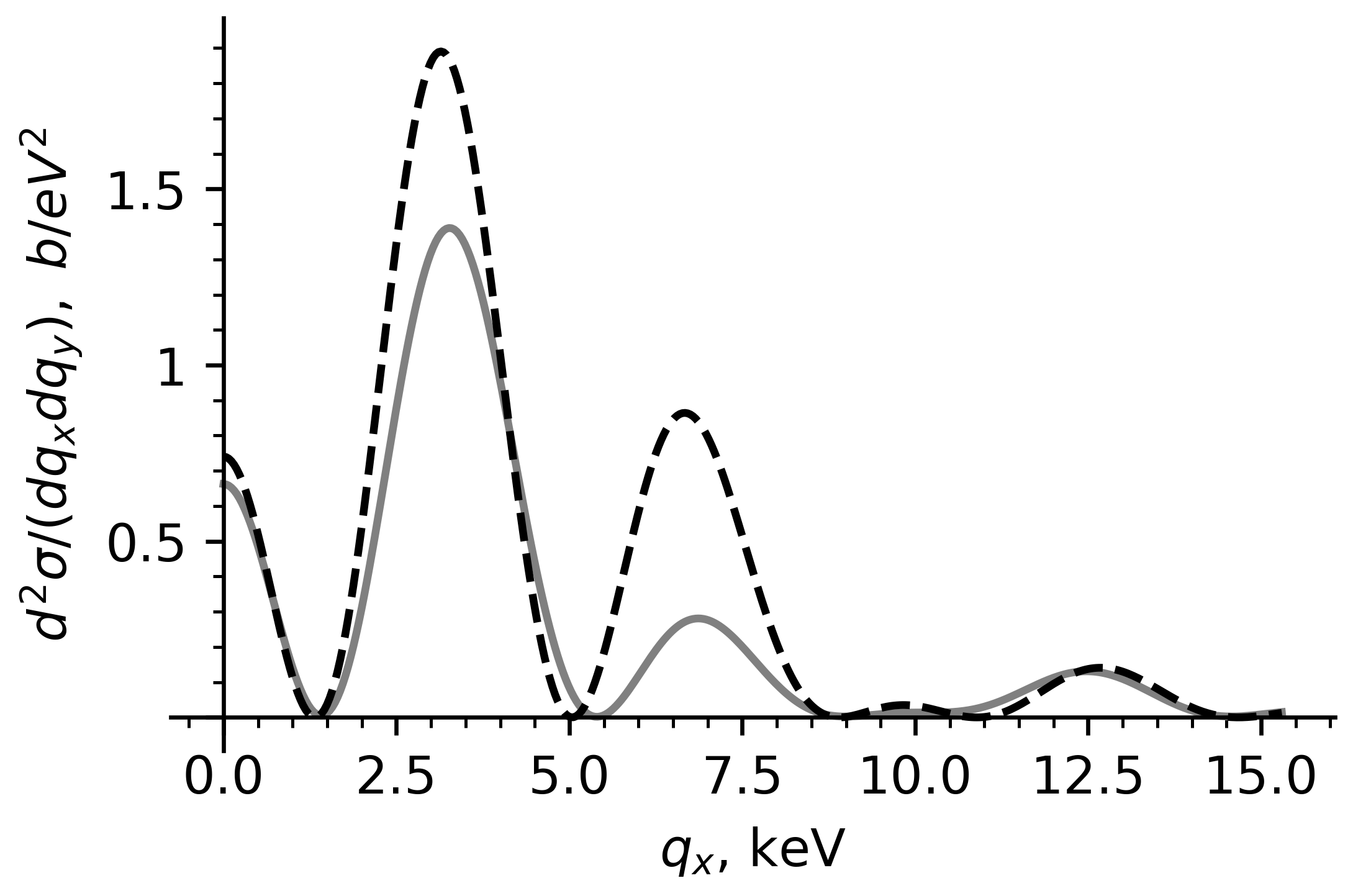}
\caption{$q_y=8.5$ keV}
\label{fig:cs_2_5}
\end{subfigure}
\vfill
\begin{subfigure}{0.32\linewidth}
\includegraphics[width=\textwidth]{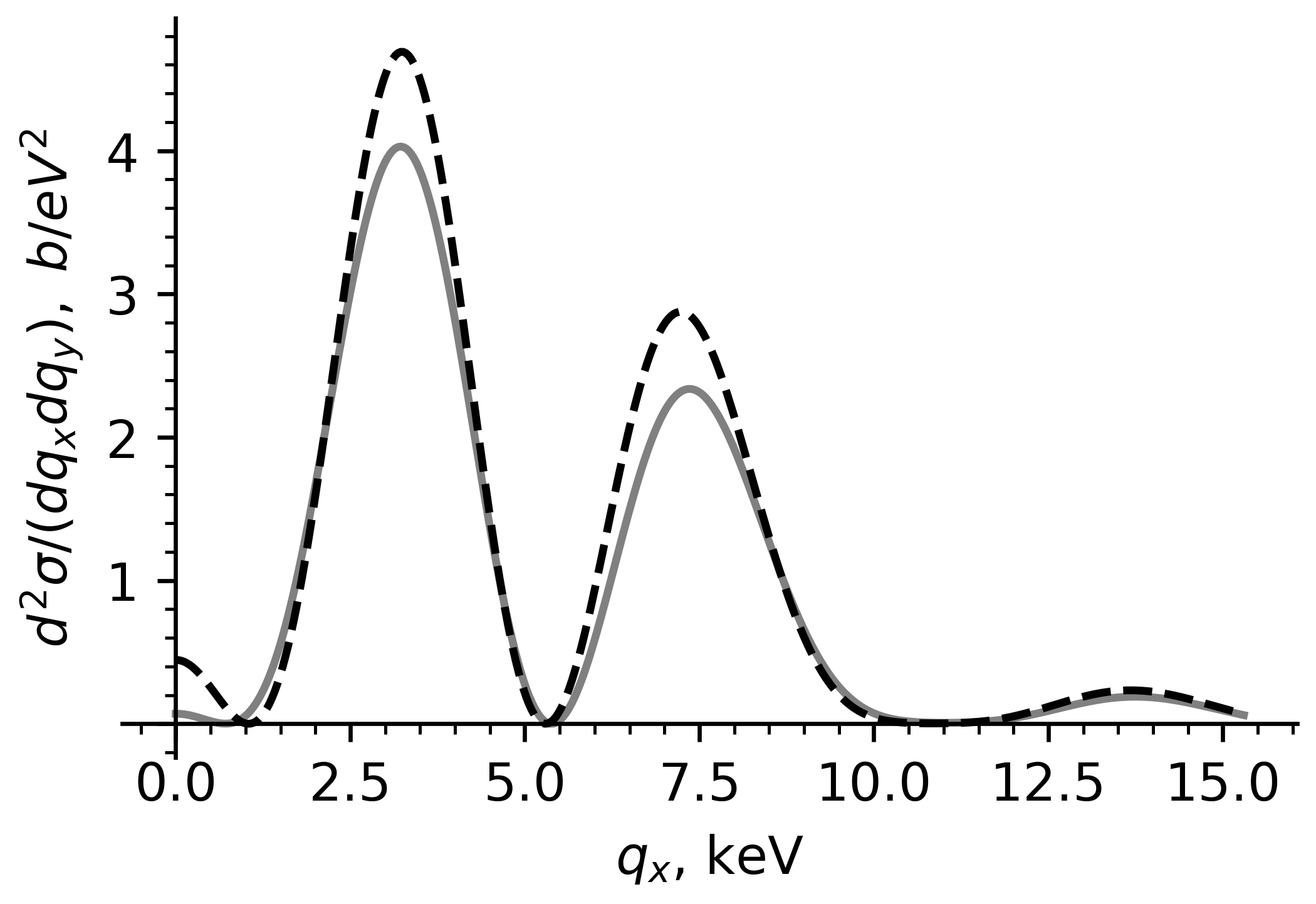}
\caption{$q_y=10.2$ keV}
\label{fig:cs_2_6}
\end{subfigure}
\begin{subfigure}{0.32\linewidth}
\includegraphics[width=\textwidth]{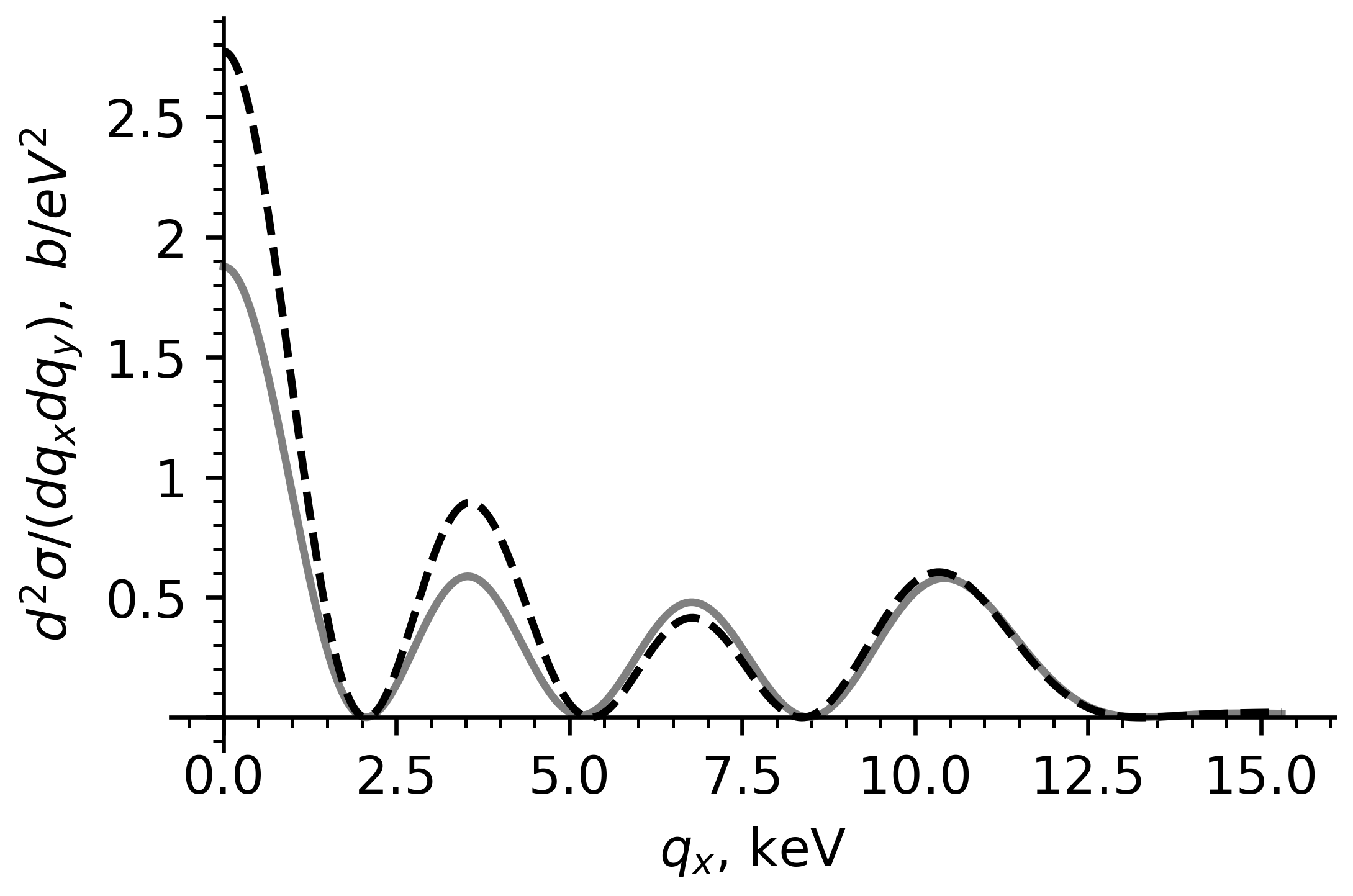}
\caption{$q_y=11.9$ keV}
\label{fig:cs_2_7}
\end{subfigure}
\begin{subfigure}{0.32\linewidth}
\includegraphics[width=\textwidth]{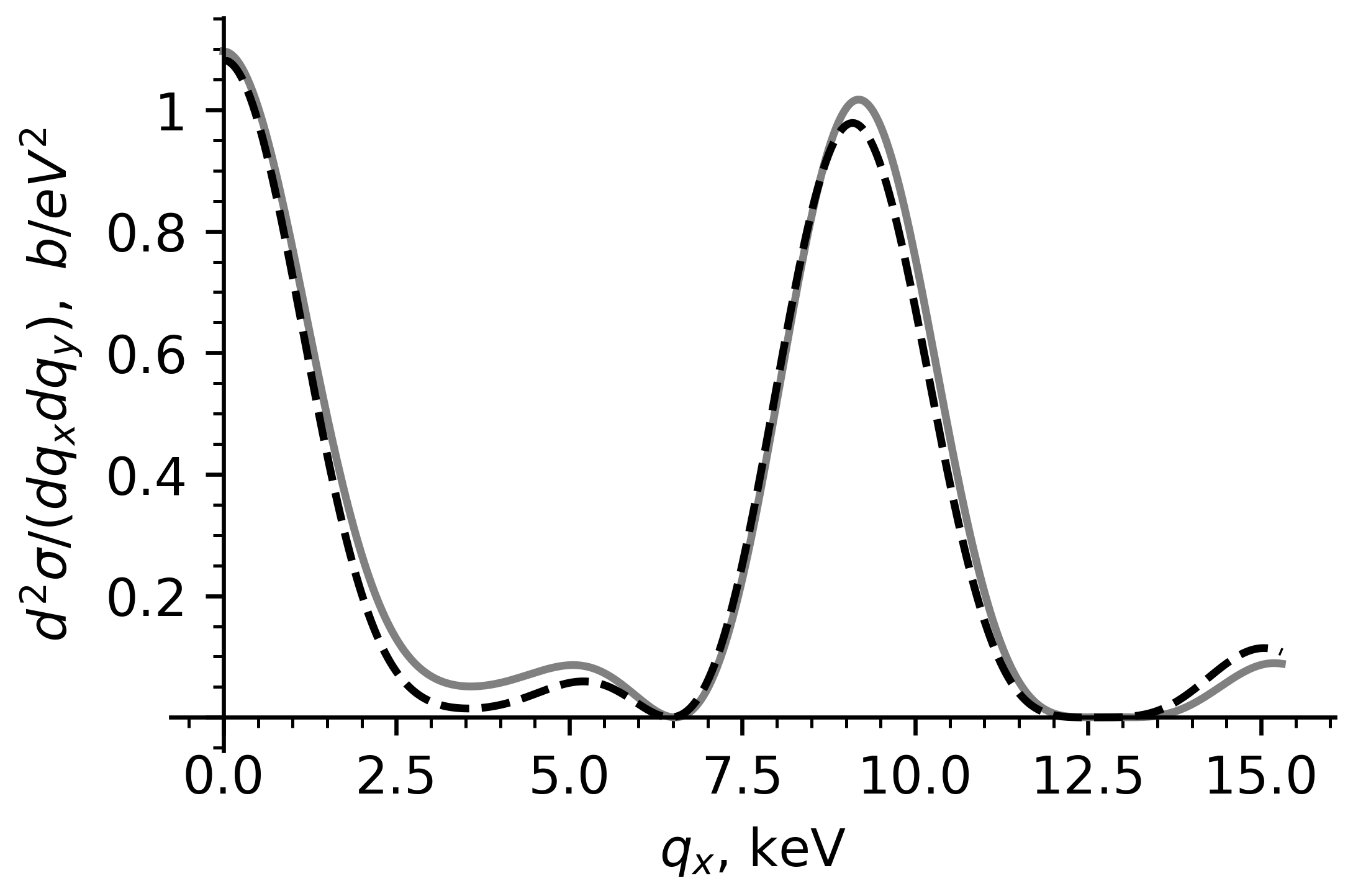}
\caption{$q_y=13.6$ keV}
\label{fig:cs_2_8}
\end{subfigure}
\caption{Differential cross section of a fast charged particle scattering on a tilted nanotube, $L_x=0.4R$}
\label{fig_cs_l2}
\end{figure}

\begin{figure}[!h]
\begin{subfigure}{0.32\linewidth}
\includegraphics[width=\textwidth]{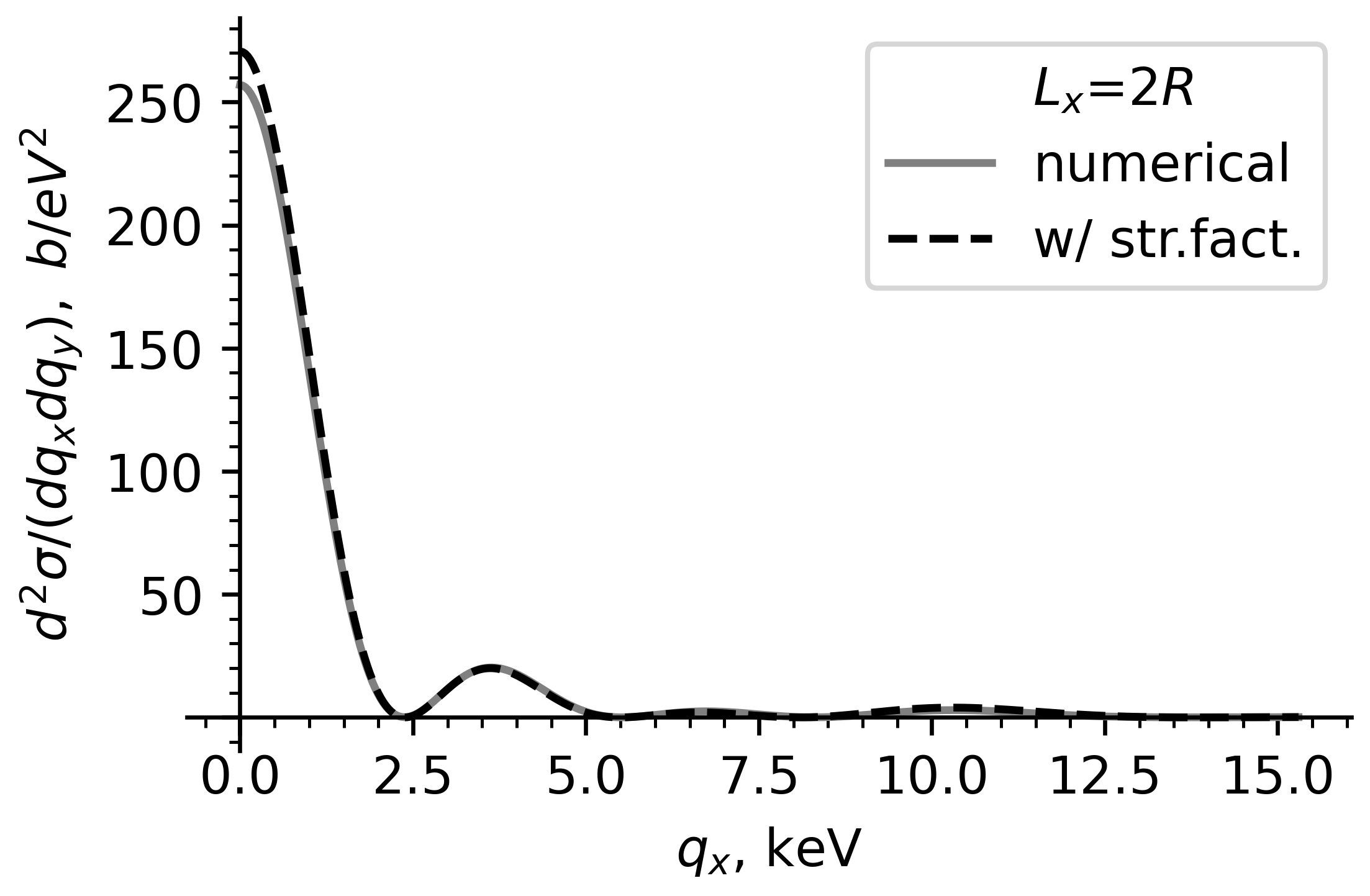}
\caption{$q_y=0.0$ keV}
\label{fig:cs_10_0}
\end{subfigure}
\begin{subfigure}{0.32\linewidth}
\includegraphics[width=\textwidth]{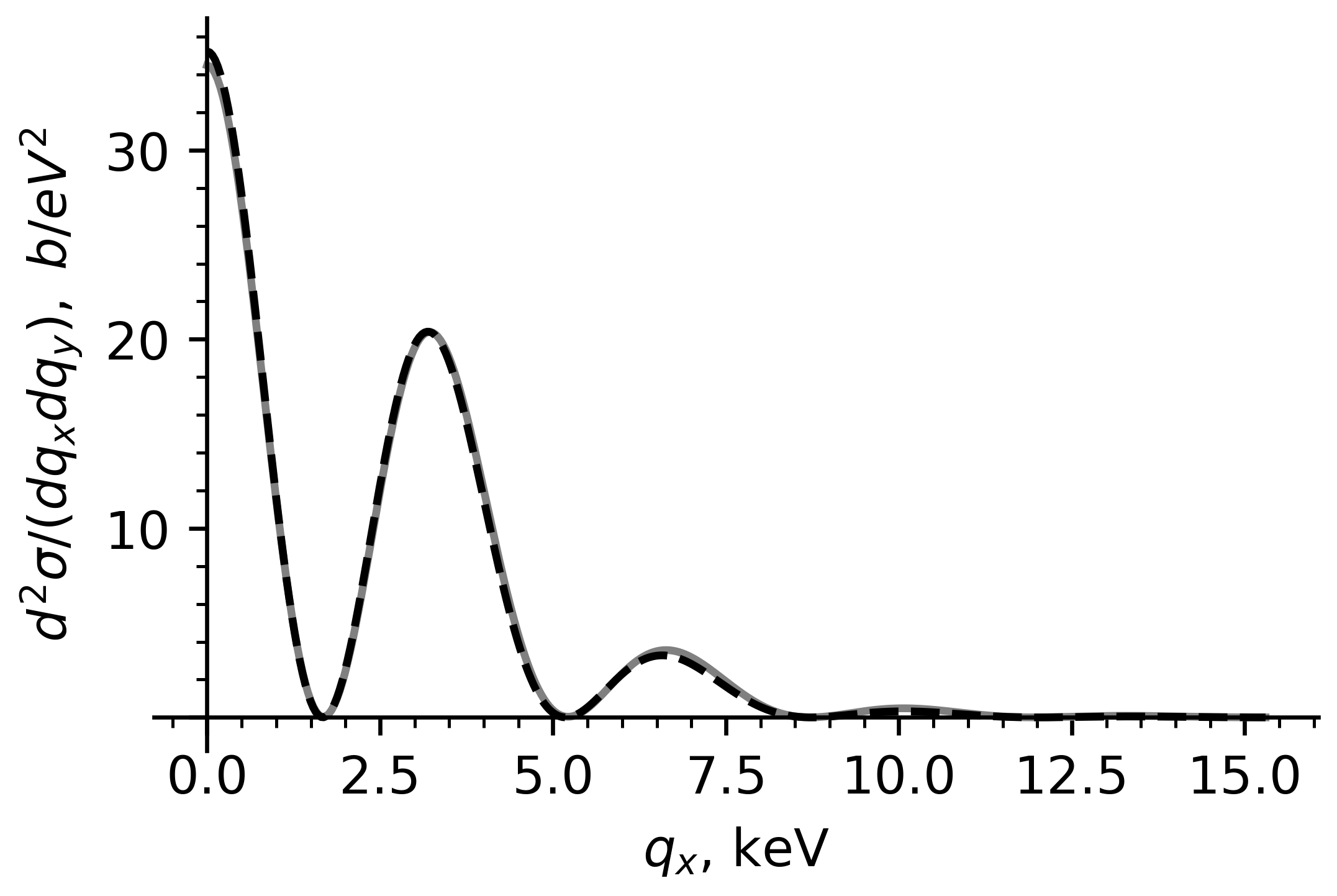}
\caption{$q_y=1.7$ keV}
\label{fig:cs_10_1}
\end{subfigure}
\begin{subfigure}{0.32\linewidth}
\includegraphics[width=\textwidth]{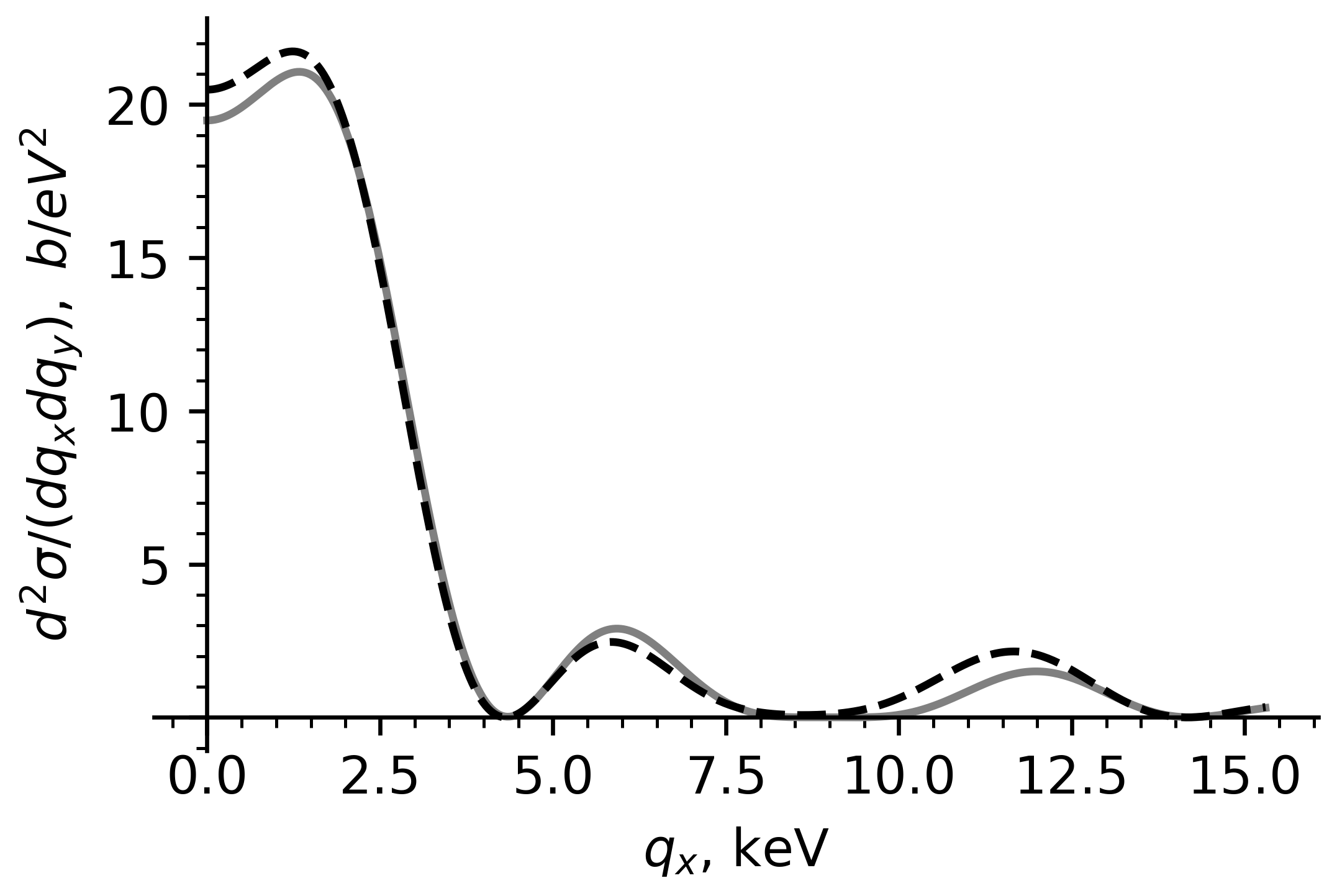}
\caption{$q_y=3.4$ keV}
\label{fig:cs_10_2}
\end{subfigure}
\vfill
\begin{subfigure}{0.32\linewidth}
\includegraphics[width=\textwidth]{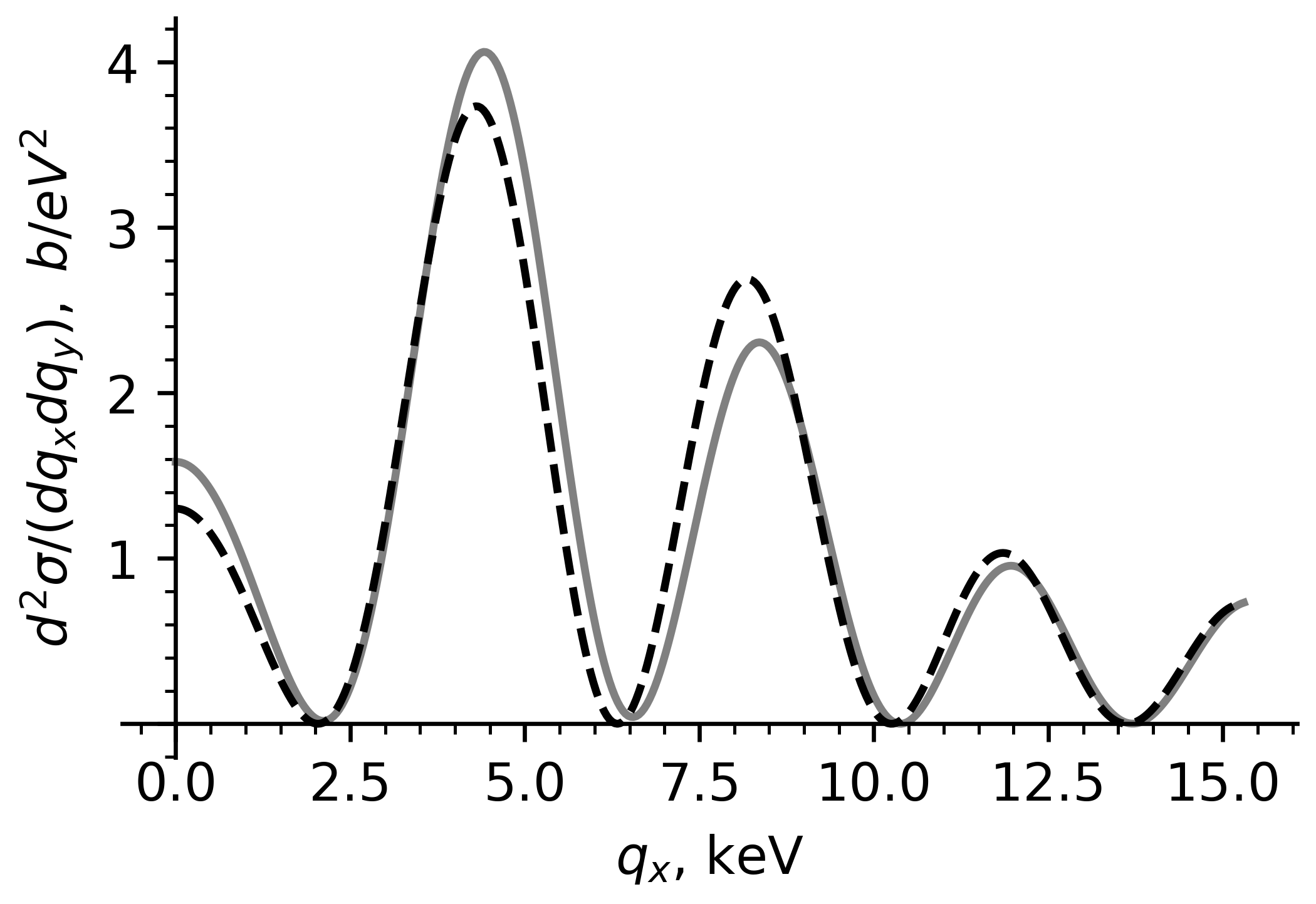}
\caption{$q_y=5.1$ keV}
\label{fig:cs_10_3}
\end{subfigure}
\begin{subfigure}{0.32\linewidth}
\includegraphics[width=\textwidth]{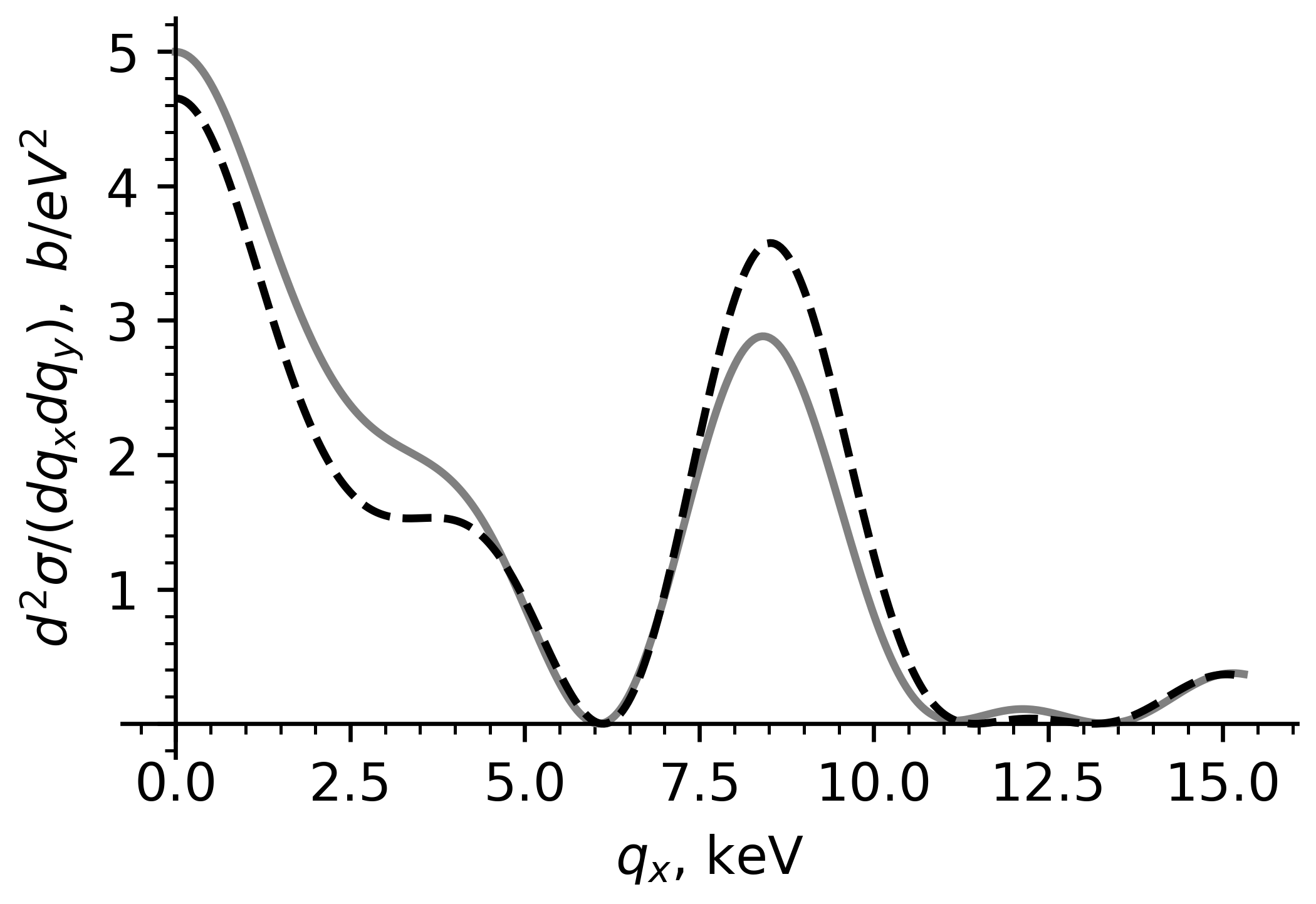}
\caption{$q_y=6.8$ keV}
\label{fig:cs_10_4}
\end{subfigure}
\begin{subfigure}{0.32\linewidth}
\includegraphics[width=\textwidth]{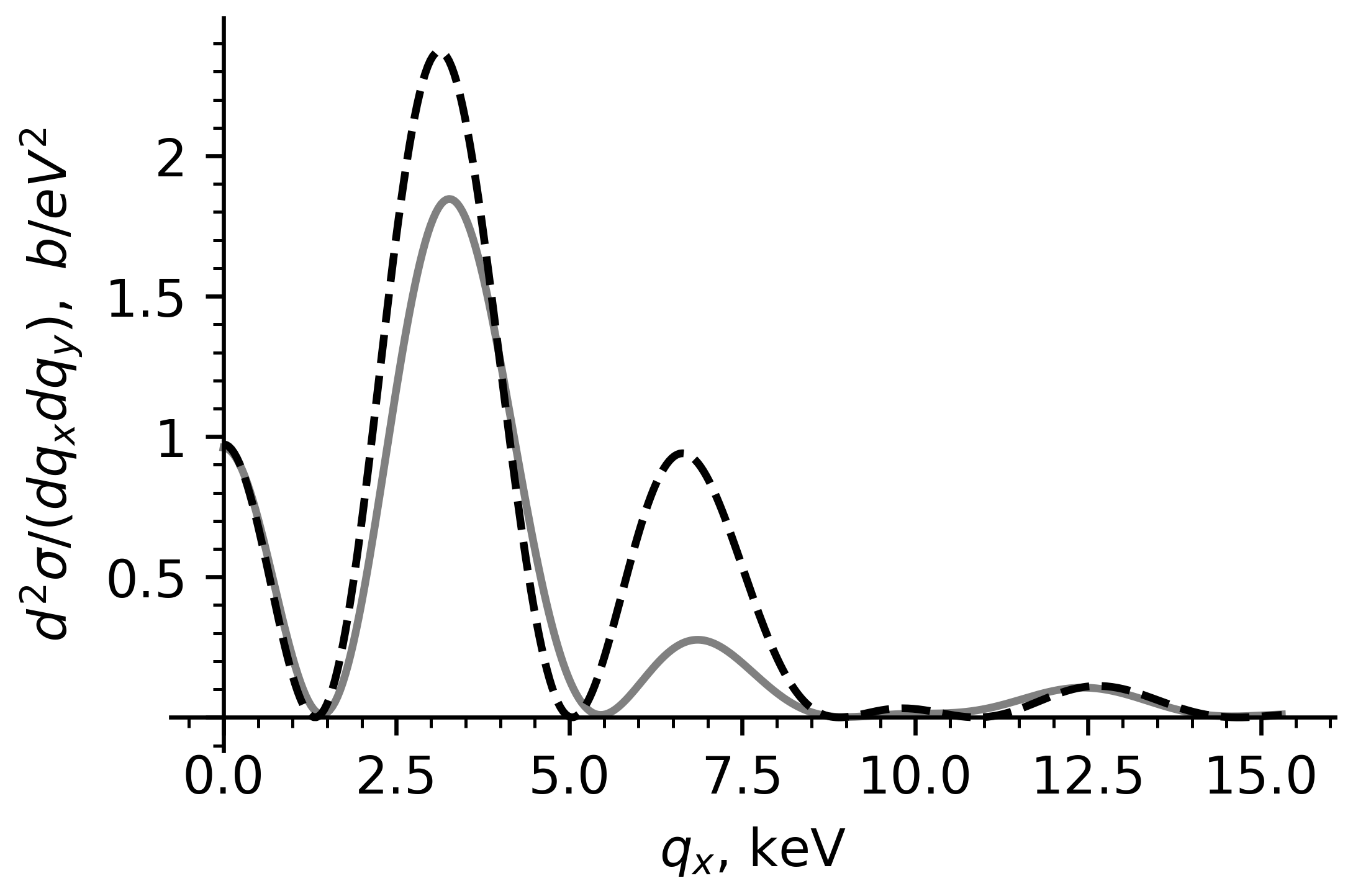}
\caption{$q_y=8.5$ keV}
\label{fig:cs_10_5}
\end{subfigure}
\vfill
\begin{subfigure}{0.32\linewidth}
\includegraphics[width=\textwidth]{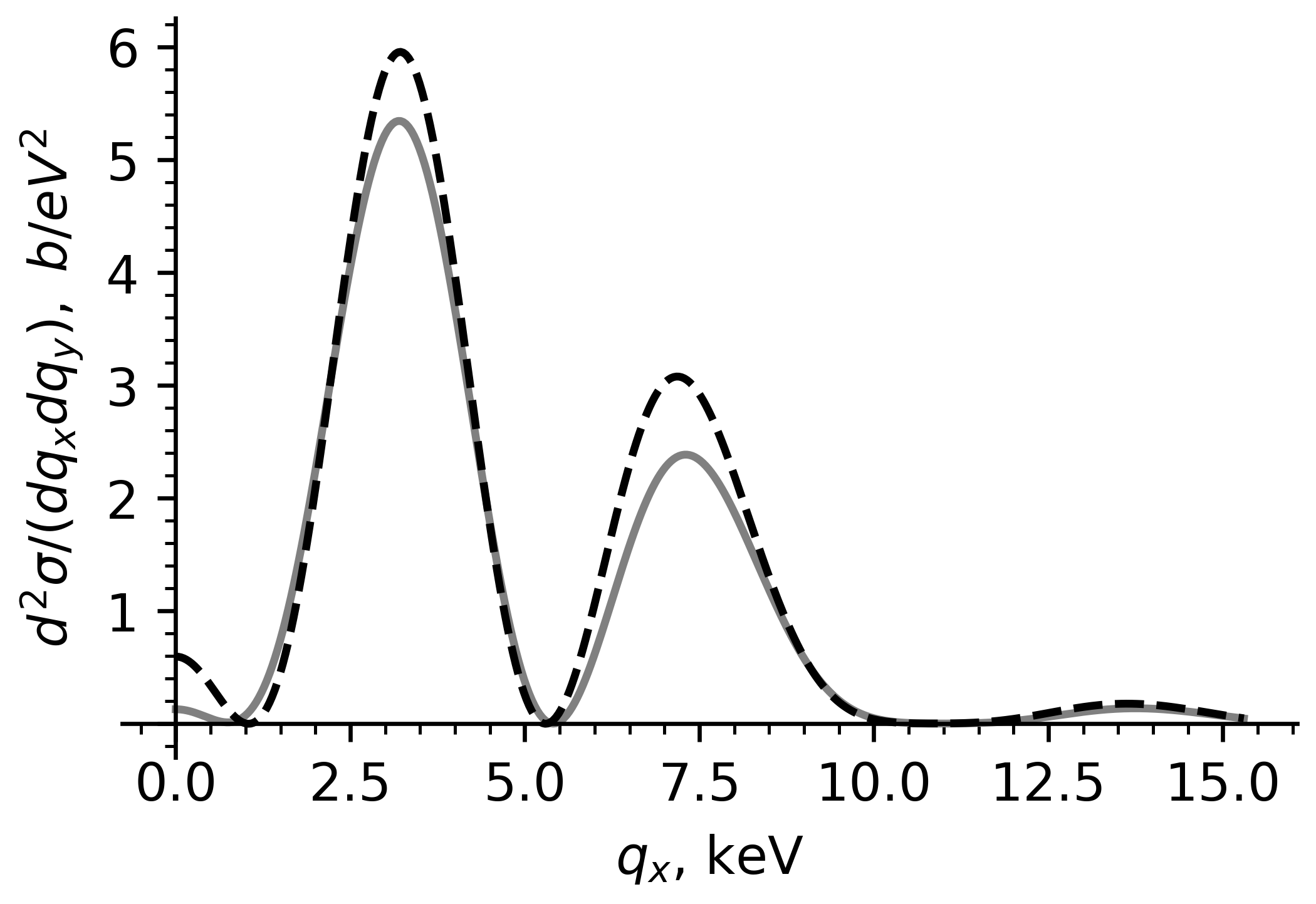}
\caption{$q_y=10.2$ keV}
\label{fig:cs_10_6}
\end{subfigure}
\begin{subfigure}{0.32\linewidth}
\includegraphics[width=\textwidth]{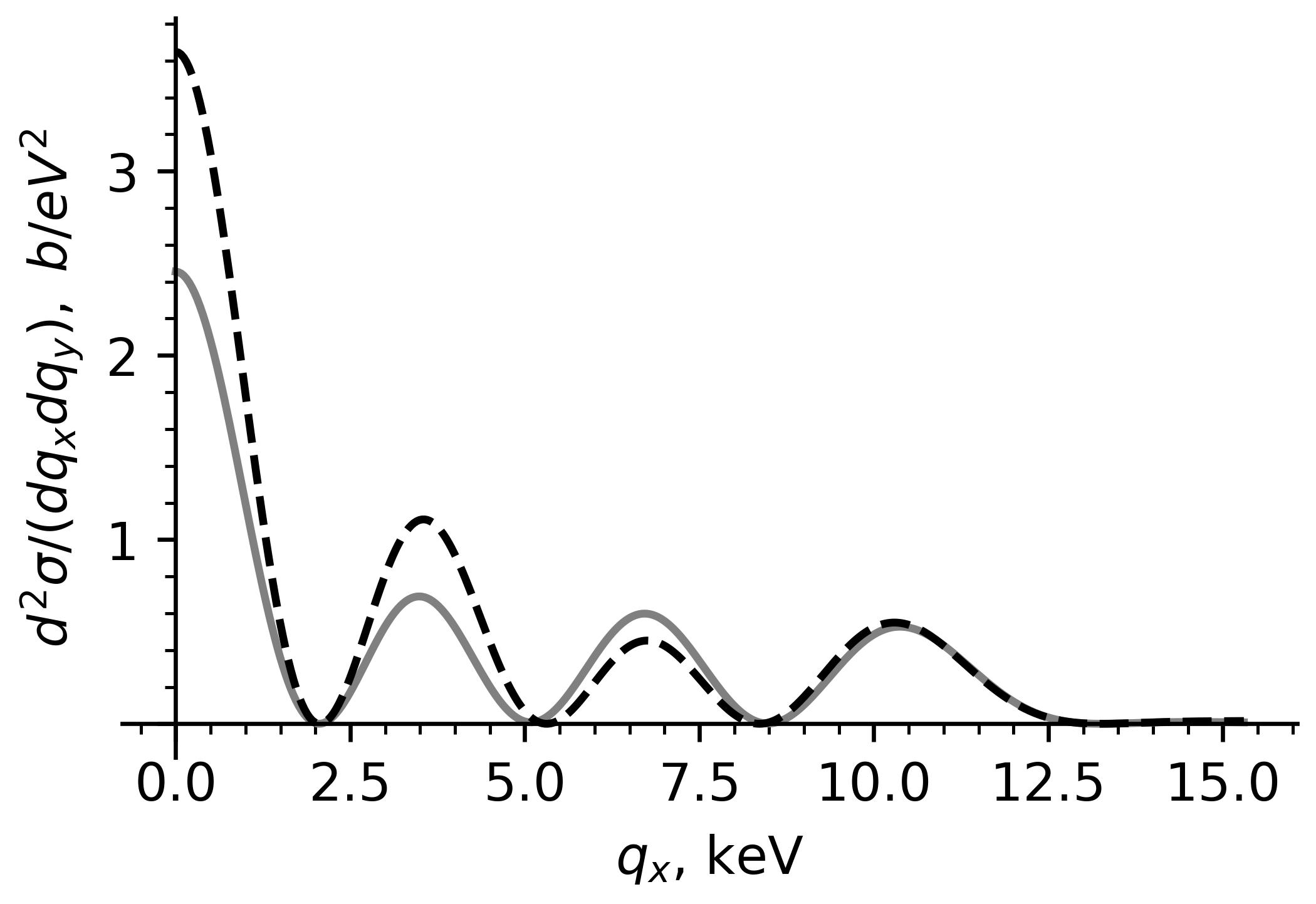}
\caption{$q_y=11.9$ keV}
\label{fig:cs_10_7}
\end{subfigure}
\begin{subfigure}{0.32\linewidth}
\includegraphics[width=\textwidth]{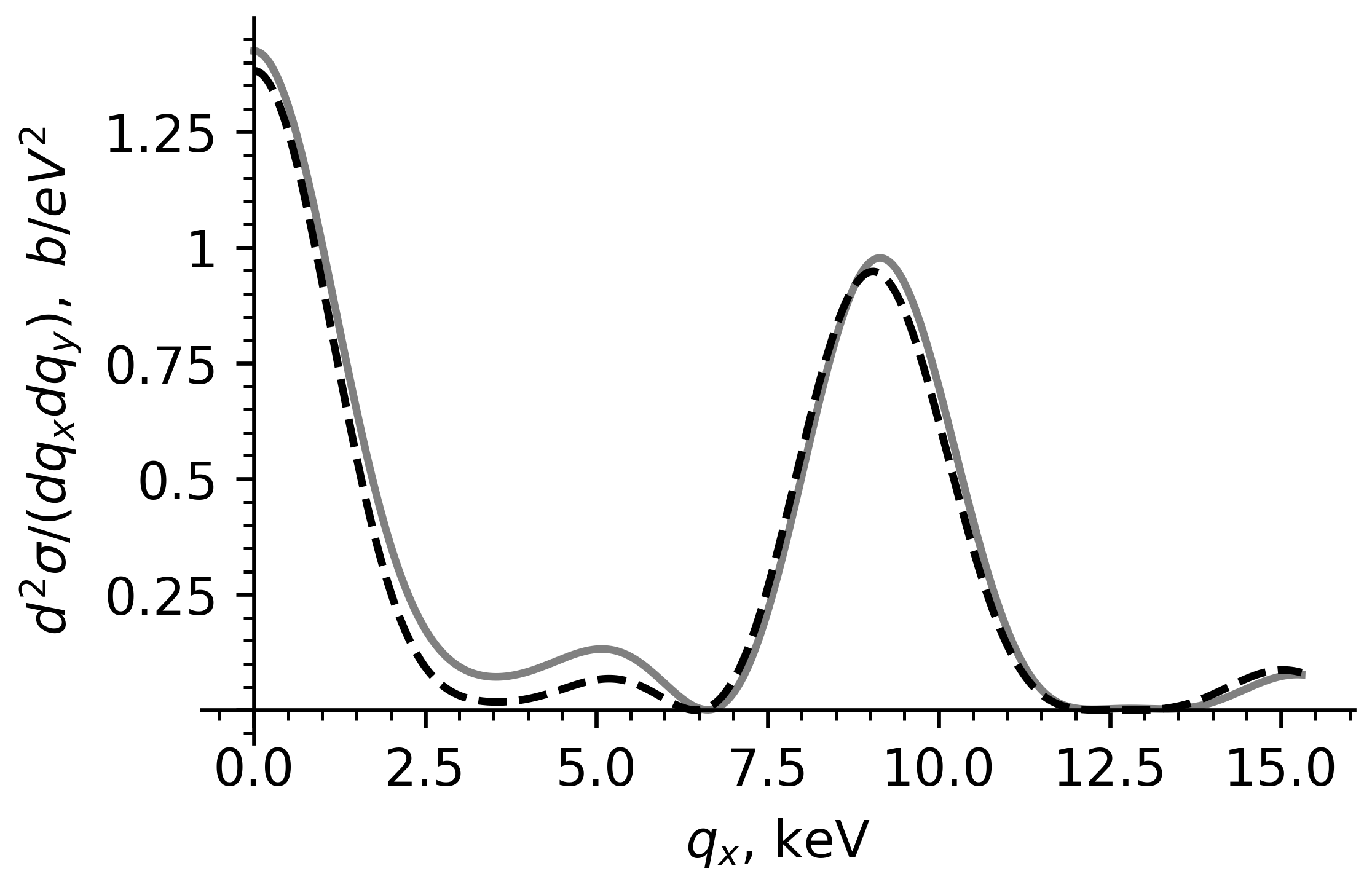}
\caption{$q_y=13.6$ keV}
\label{fig:cs_10_8}
\end{subfigure}
\caption{Differential cross section of a fast charged particle scattering on a tilted nanotube, $L_x=2R$}
\label{fig_cs_l10}
\end{figure}

\begin{figure}[!h]
\begin{subfigure}{0.32\linewidth}
\includegraphics[width=\textwidth]{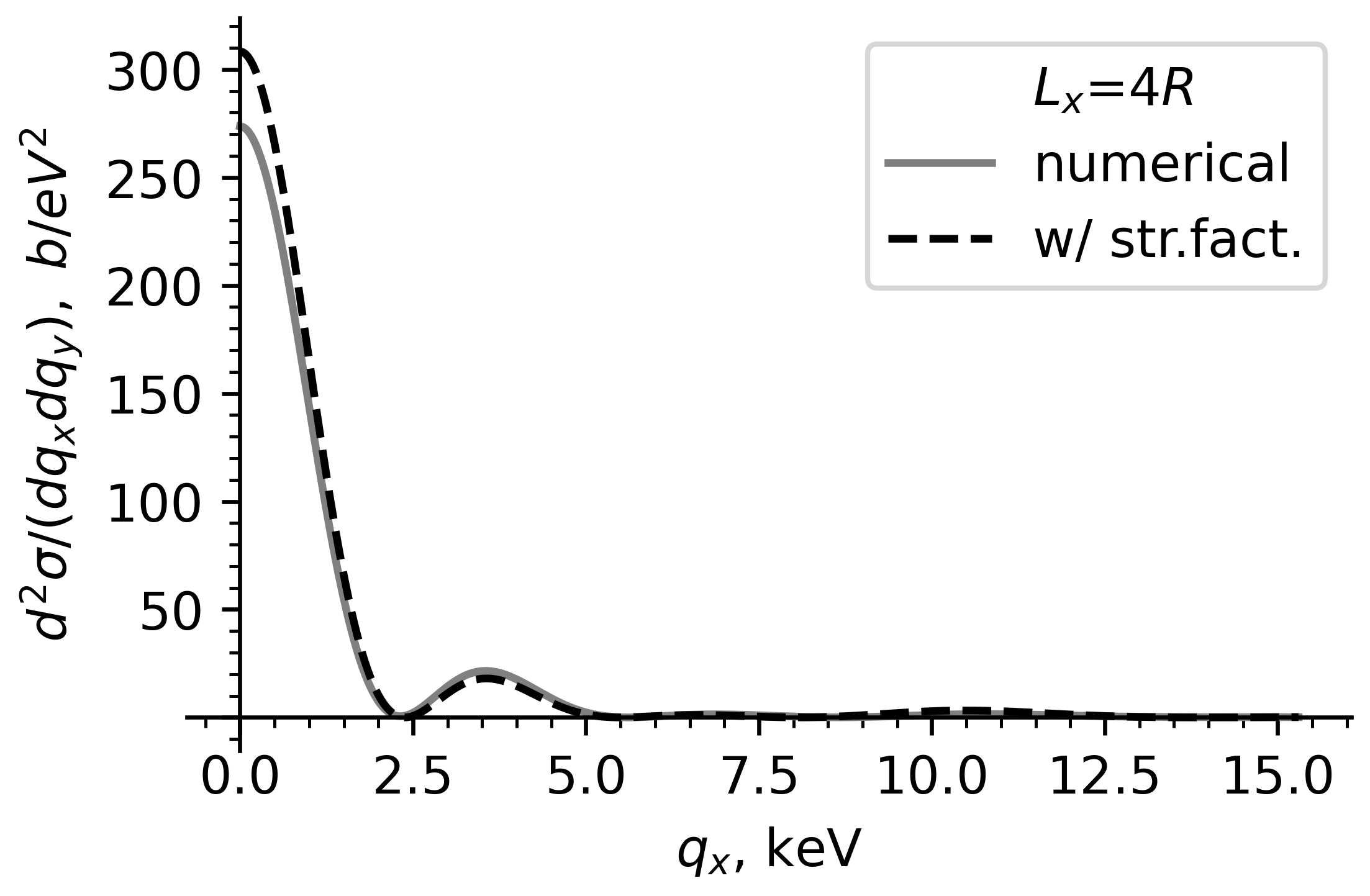}
\caption{$q_y=0.0$ keV}
\label{fig:cs_20_0}
\end{subfigure}
\begin{subfigure}{0.32\linewidth}
\includegraphics[width=\textwidth]{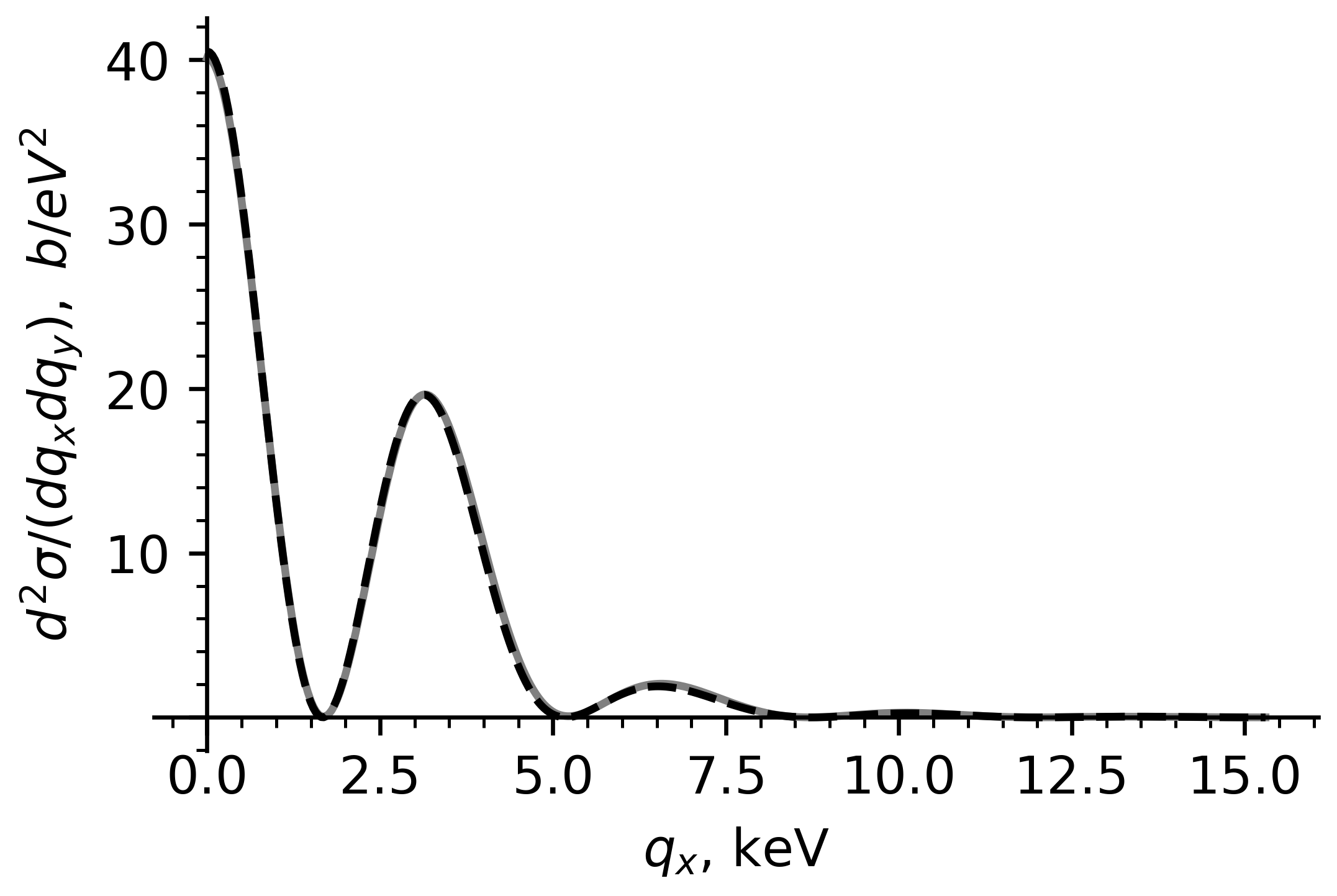}
\caption{$q_y=1.7$ keV}
\label{fig:cs_20_1}
\end{subfigure}
\begin{subfigure}{0.32\linewidth}
\includegraphics[width=\textwidth]{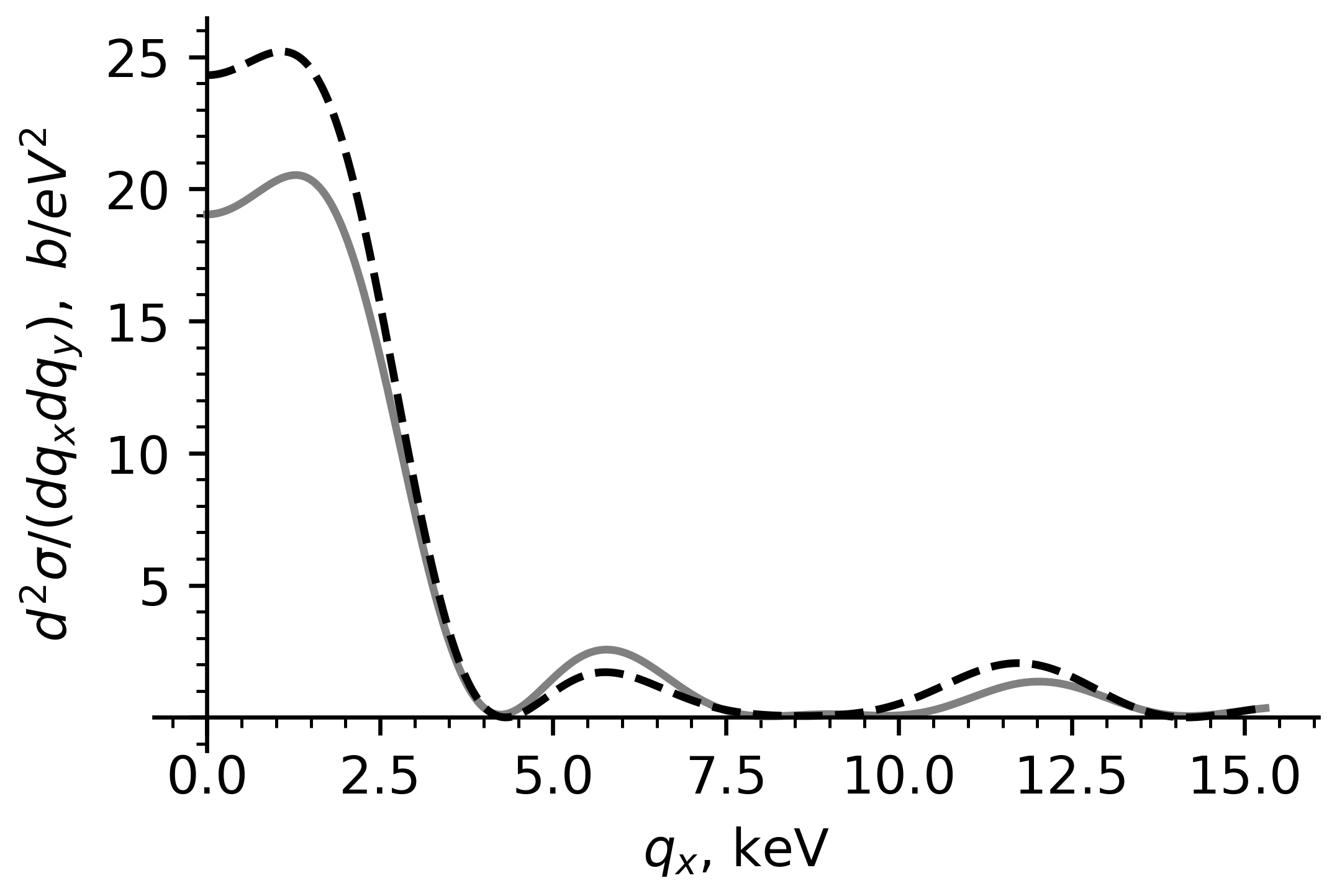}
\caption{$q_y=3.4$ keV}
\label{fig:cs_20_2}
\end{subfigure}
\vfill
\begin{subfigure}{0.32\linewidth}
\includegraphics[width=\textwidth]{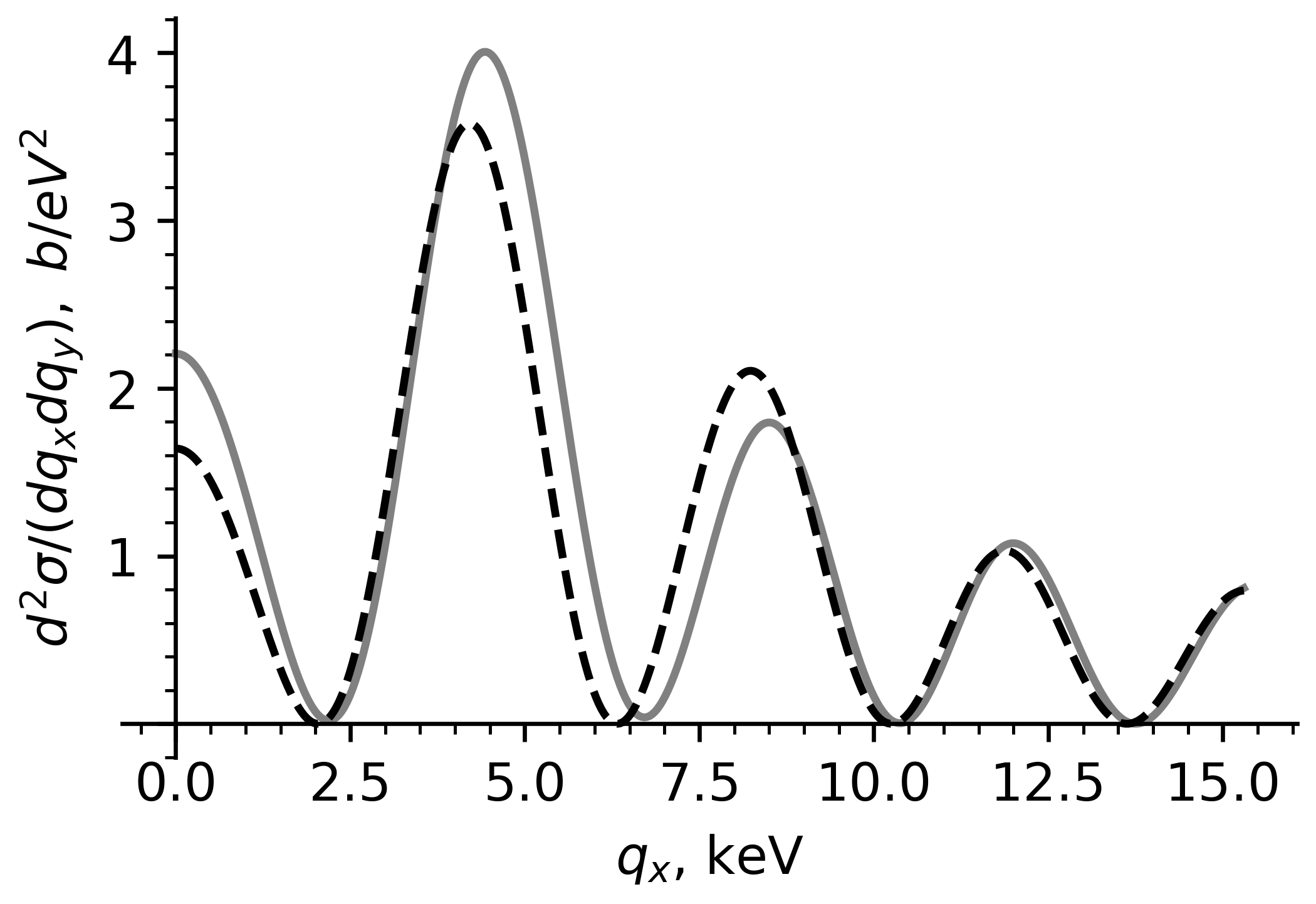}
\caption{$q_y=5.1$ keV}
\label{fig:cs_20_3}
\end{subfigure}
\begin{subfigure}{0.32\linewidth}
\includegraphics[width=\textwidth]{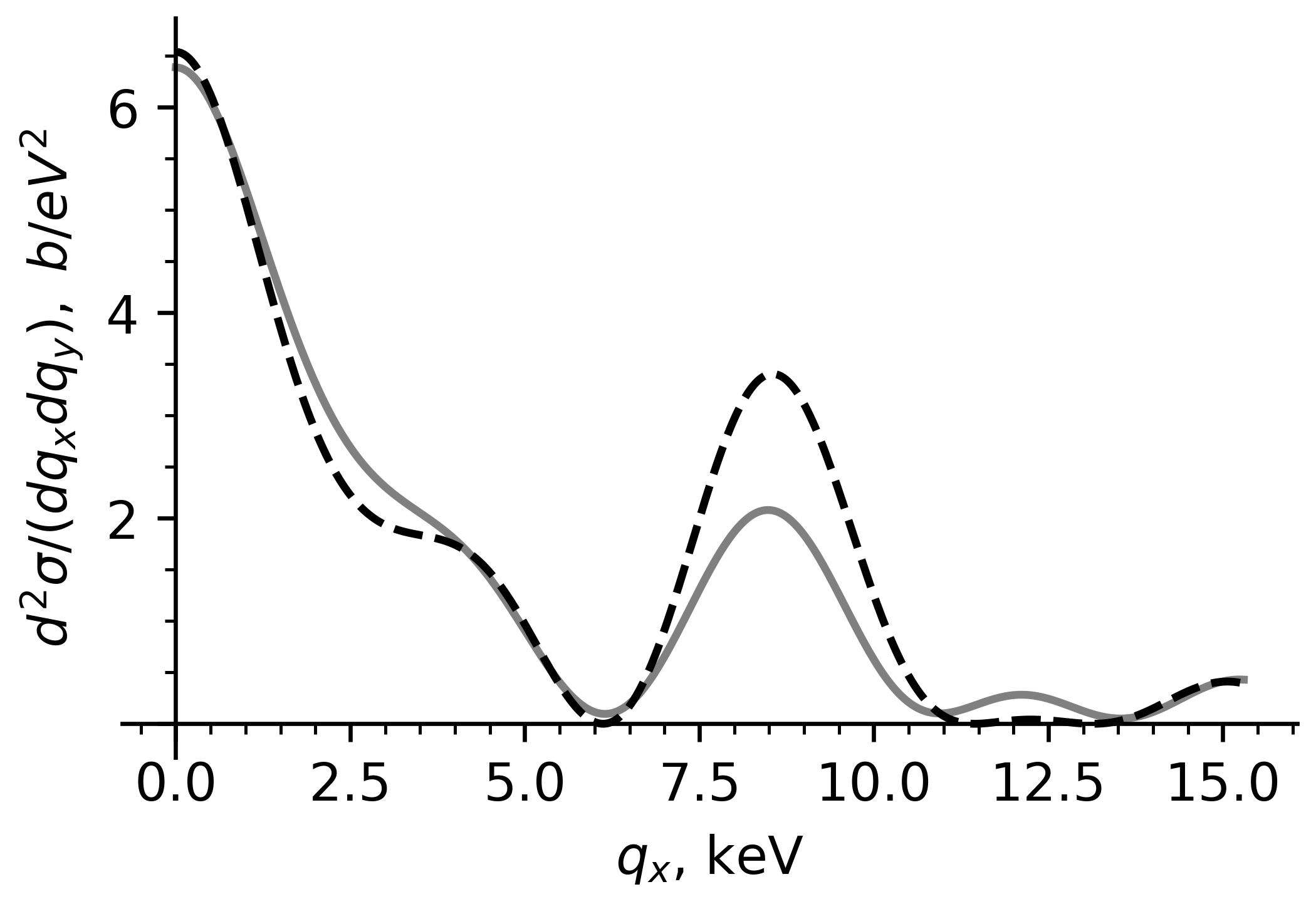}
\caption{$q_y=6.8$ keV}
\label{fig:cs_20_4}
\end{subfigure}
\begin{subfigure}{0.32\linewidth}
\includegraphics[width=\textwidth]{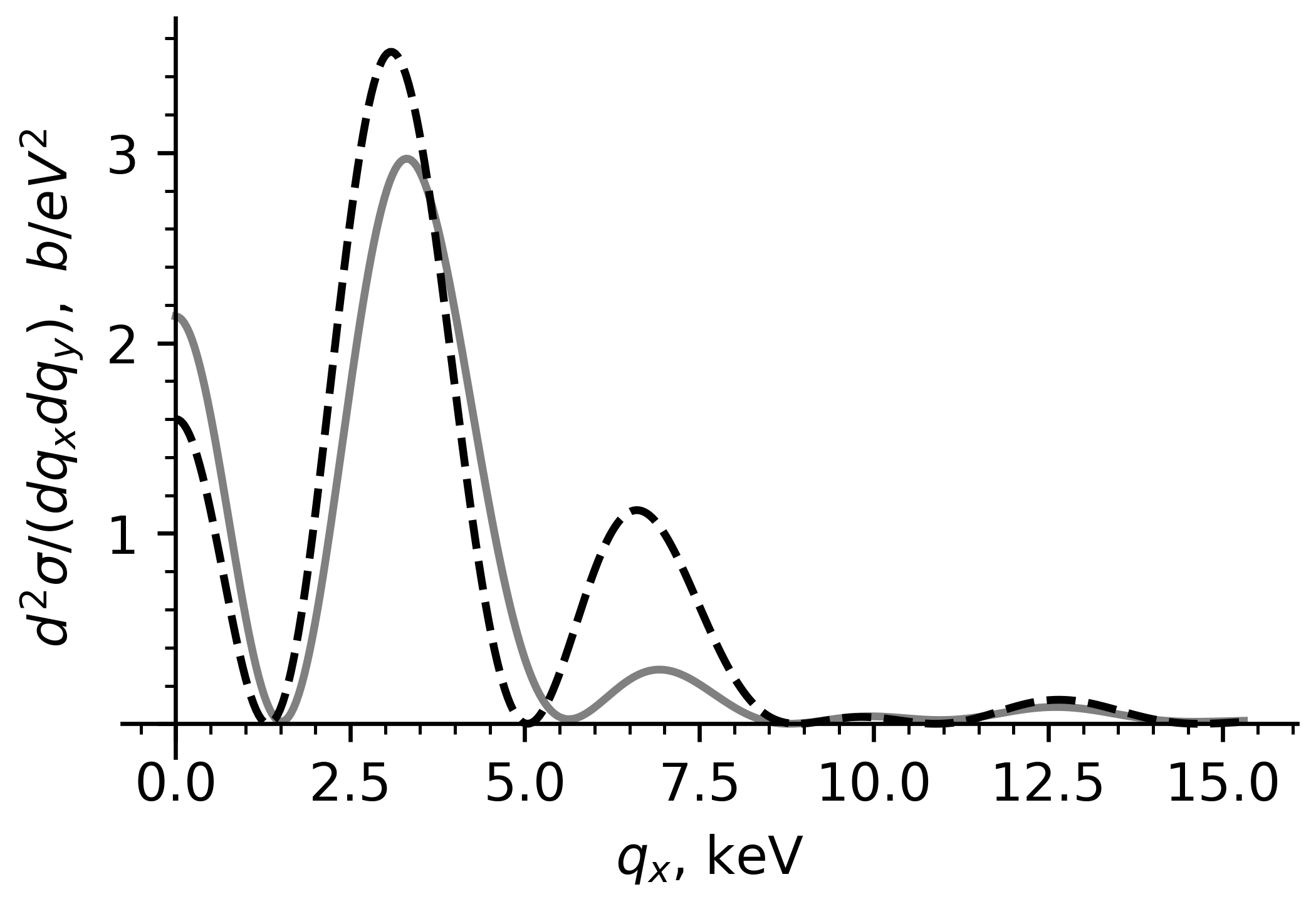}
\caption{$q_y=8.5$ keV}
\label{fig:cs_20_5}
\end{subfigure}
\vfill
\begin{subfigure}{0.32\linewidth}
\includegraphics[width=\textwidth]{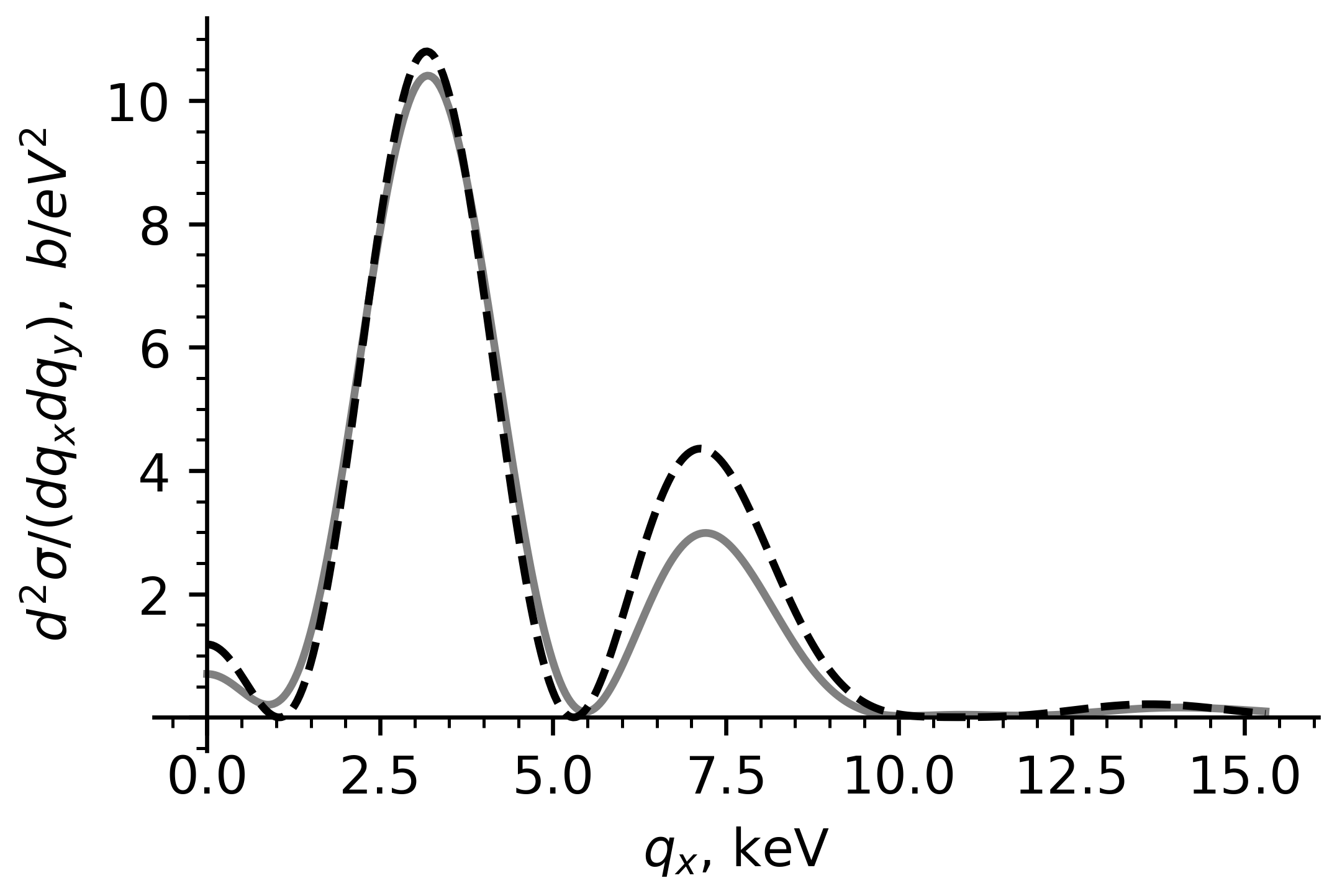}
\caption{$q_y=10.2$ keV}
\label{fig:cs_20_6}
\end{subfigure}
\begin{subfigure}{0.32\linewidth}
\includegraphics[width=\textwidth]{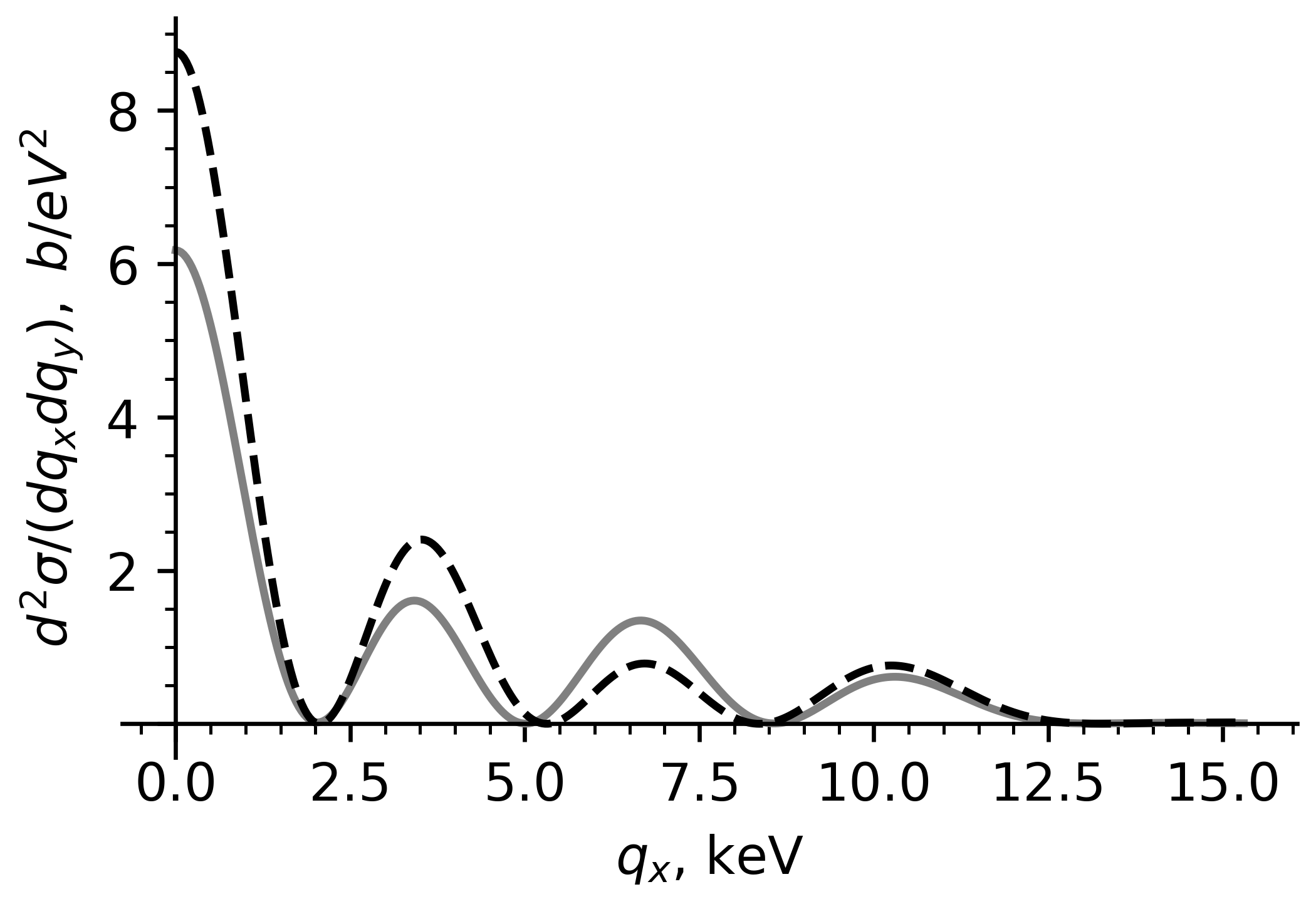}
\caption{$q_y=11.9$ keV}
\label{fig:cs_20_7}
\end{subfigure}
\begin{subfigure}{0.32\linewidth}
\includegraphics[width=\textwidth]{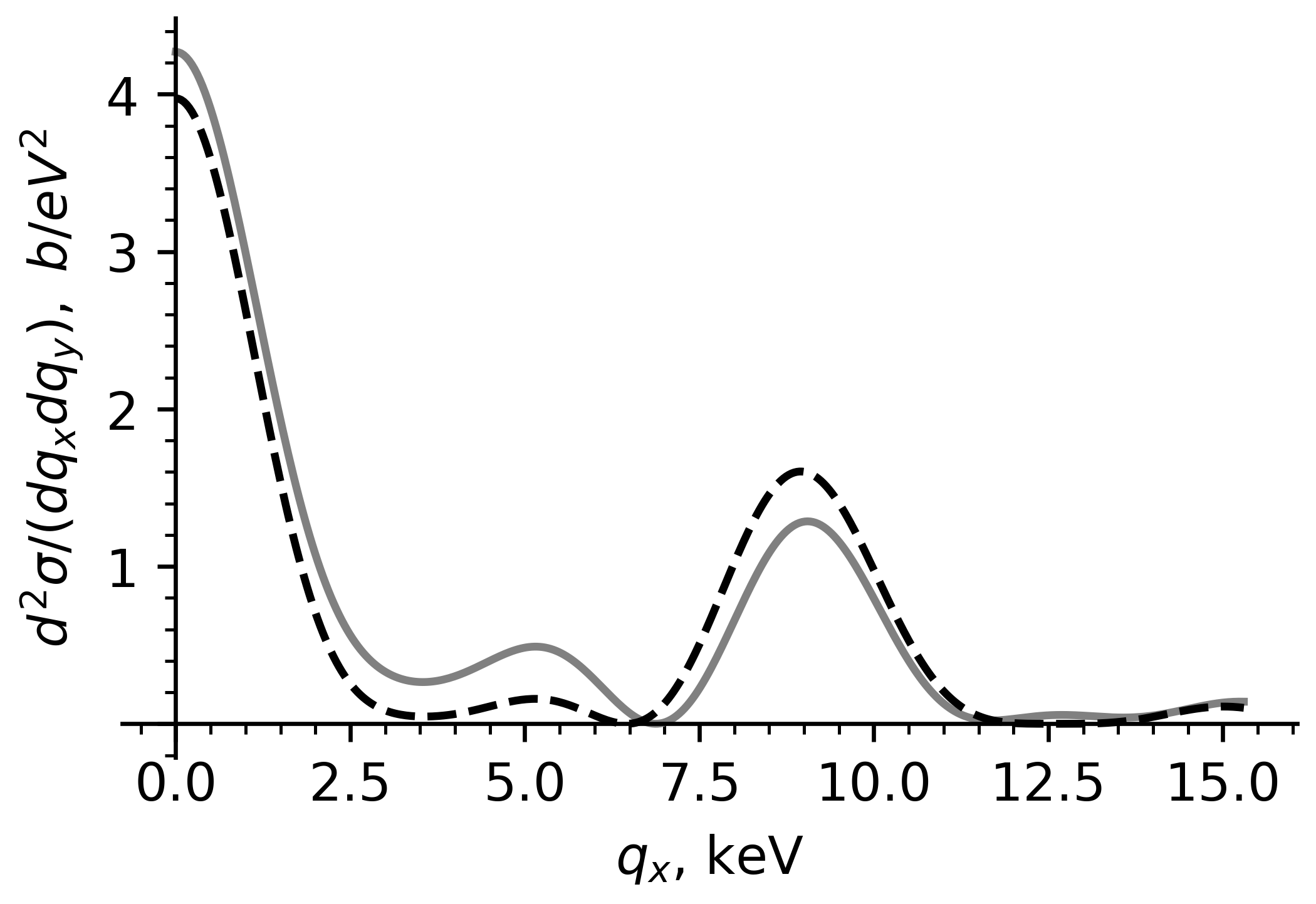}
\caption{$q_y=13.6$ keV}
\label{fig:cs_20_8}
\end{subfigure}
\caption{Differential cross section of a fast charged particle scattering on a tilted nanotube, $L_x=4R$}
\label{fig_cs_l20}
\end{figure}

In present paper, we calculated the differential scattering cross sections for straight and tilted $(5,0)$ zigzag nanotubes ($N_s=10$) numerically using the formula \eqref{eq4} with the scattering amplitude \eqref{eq10} and using the expression of the differential scattering cross section \eqref{eq13_1} defined through the differential cross section \eqref{eq13_2} and the structure factor $D_{N_s}$. For tilted nanotubes these formulas are used with substitution $\tilde{\chi}_0^{(s)} \rightarrow \tilde{\chi}_0^{(t)}$. The results of calculations using these formulas are presented on Figs. \ref{fig_cs_l0}-\ref{fig_cs_l20}: "numerical" corresponds to calculations using formulas \eqref{eq4}, \eqref{eq10}; "w/ str. fact." -- formula \eqref{eq13_1} involving the structure factor $D_{N_s}$. Different figures correspond to nanotubes tilted under different angles with respect to $z$-axis in $(z,x)$-plane. Fig. \ref{fig:lx} shows projections of atoms positions in nanotube to the $(x,y)$-plane. The angle between $z$-axis and strings of the nanotube is related to $L_x$, $L_z$ as
\begin{eqnarray}\label{eq26}
\theta=\text{arcsin}\left(\frac{L_x}{L_z}\right).
\end{eqnarray}

We performed all calculations for $A=10$ which is equivalent to $N_z=114$ and $L_z \approx 24$ nm. Maximum angle $\theta$ considered during this study was $\theta(L_x=4R)\approx 4.275 \cdot 10^{-3}$ radian.  

We see from Figs. \ref{fig_cs_l0}-\ref{fig_cs_l20} that for some regions of transferred momenta the cross sections obtained with both methods agree quite well. Speaking about all presented regions of transferred momenta, comparison indicates that formula \eqref{eq13_1} with structure factor carries a pattern of the differential cross section but does not represent the differential cross section precisely. The probable reason for this  is relatively small spacing between atomic strings, leading to overlap of regions of their influence. But calculations with formula \eqref{eq13_1} represent general features of the differential scattering cross section for this case. 

\begin{figure}[!h]
\begin{subfigure}{0.32\linewidth}
\includegraphics[width=\textwidth]{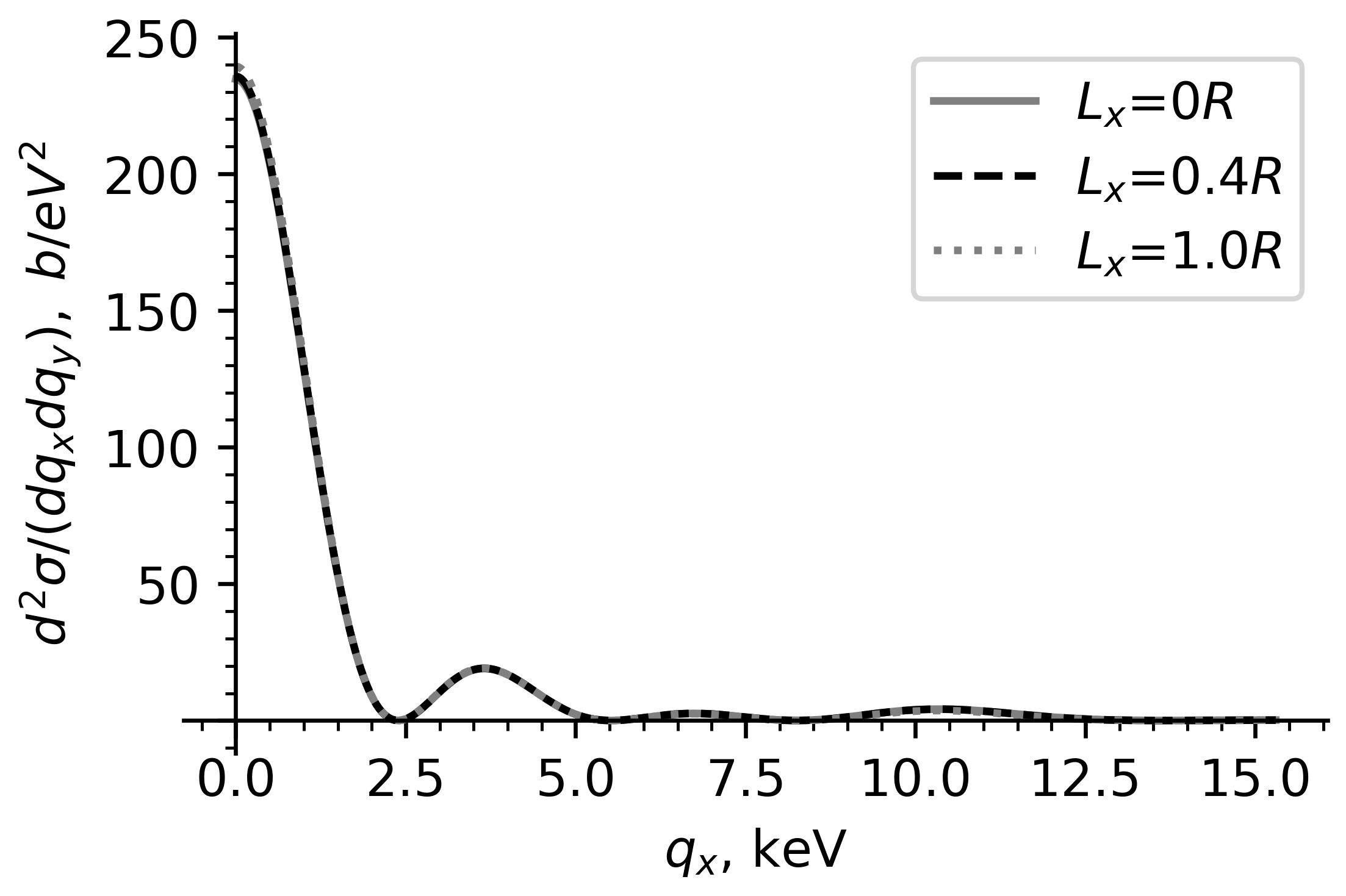}
\caption{$q_y=0.0$ keV}
\label{fig:cs_c1_0}
\end{subfigure}
\begin{subfigure}{0.32\linewidth}
\includegraphics[width=\textwidth]{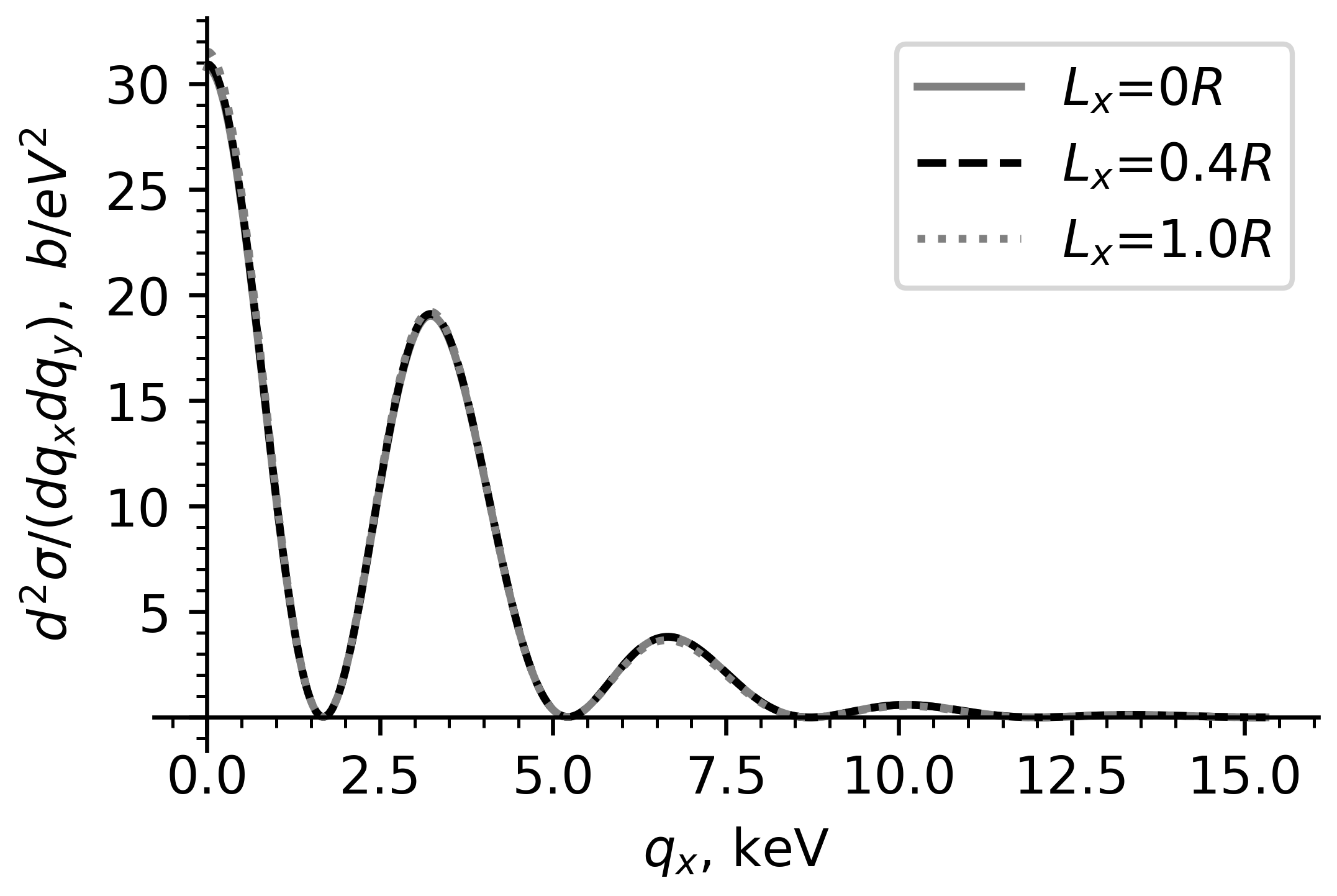}
\caption{$q_y=1.7$ keV}
\label{fig:cs_c1_1}
\end{subfigure}
\begin{subfigure}{0.32\linewidth}
\includegraphics[width=\textwidth]{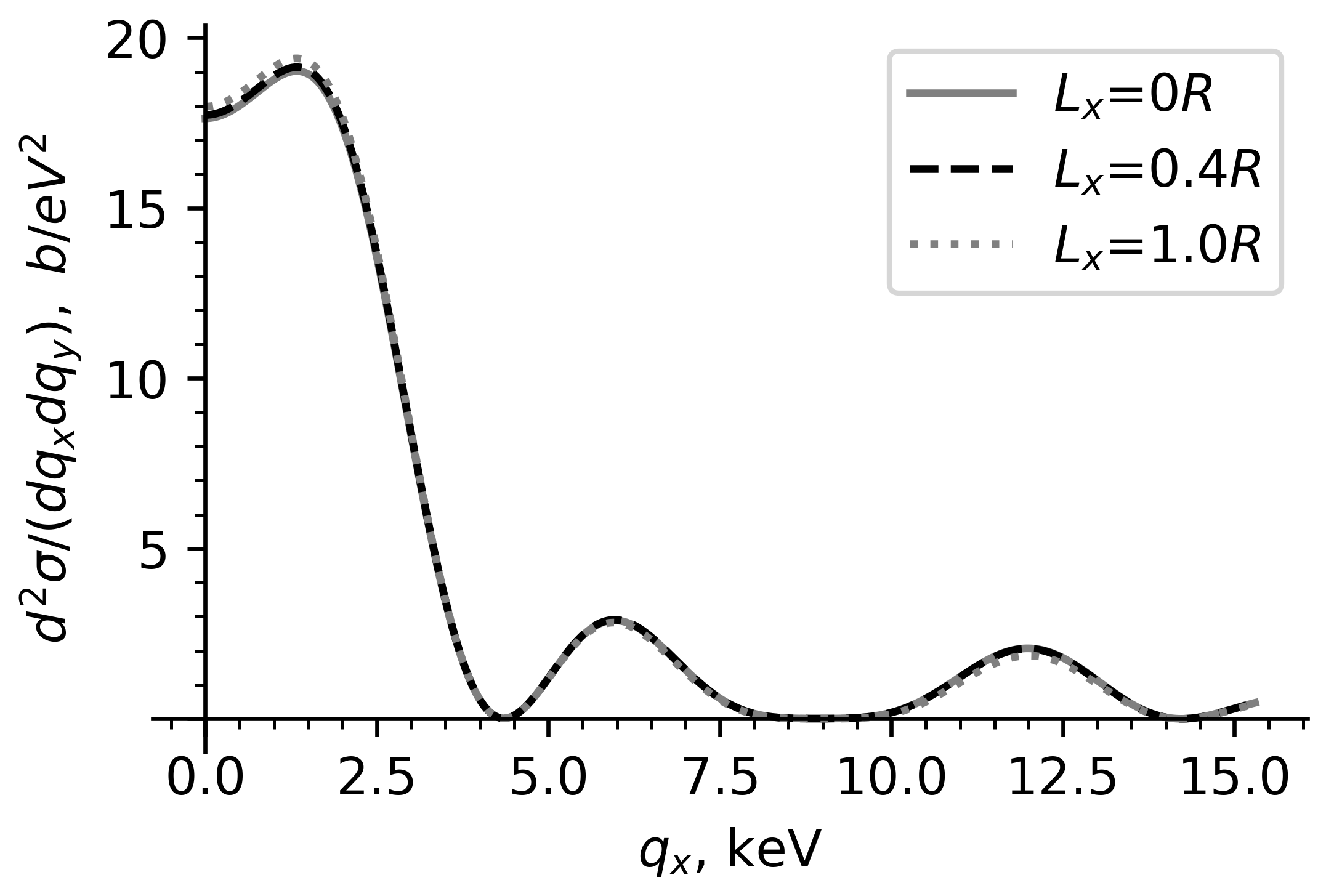}
\caption{$q_y=3.4$ keV}
\label{fig:cs_c1_2}
\end{subfigure}
\vfill
\begin{subfigure}{0.32\linewidth}
\includegraphics[width=\textwidth]{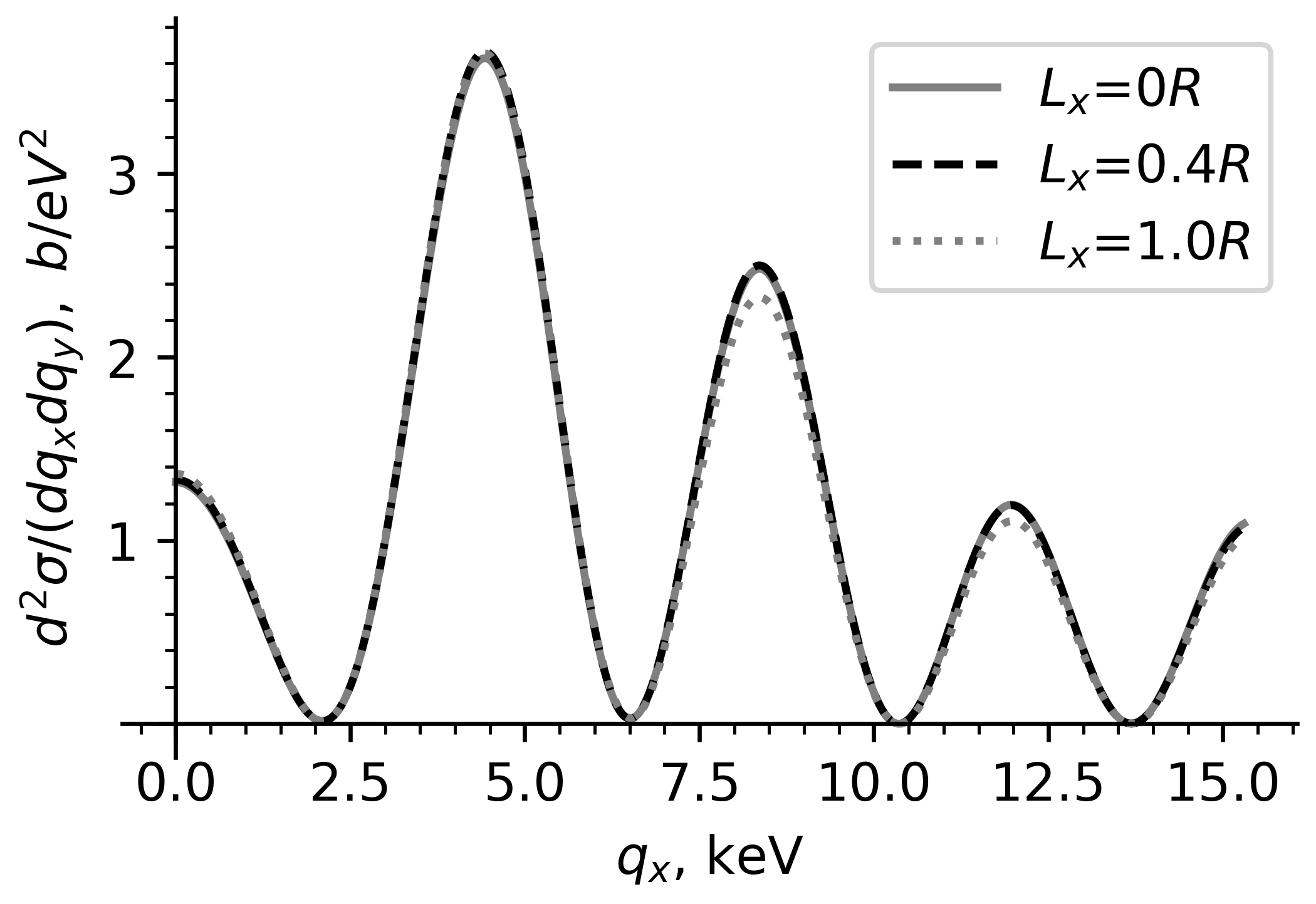}
\caption{$q_y=5.1$ keV}
\label{fig:cs_c1_3}
\end{subfigure}
\begin{subfigure}{0.32\linewidth}
\includegraphics[width=\textwidth]{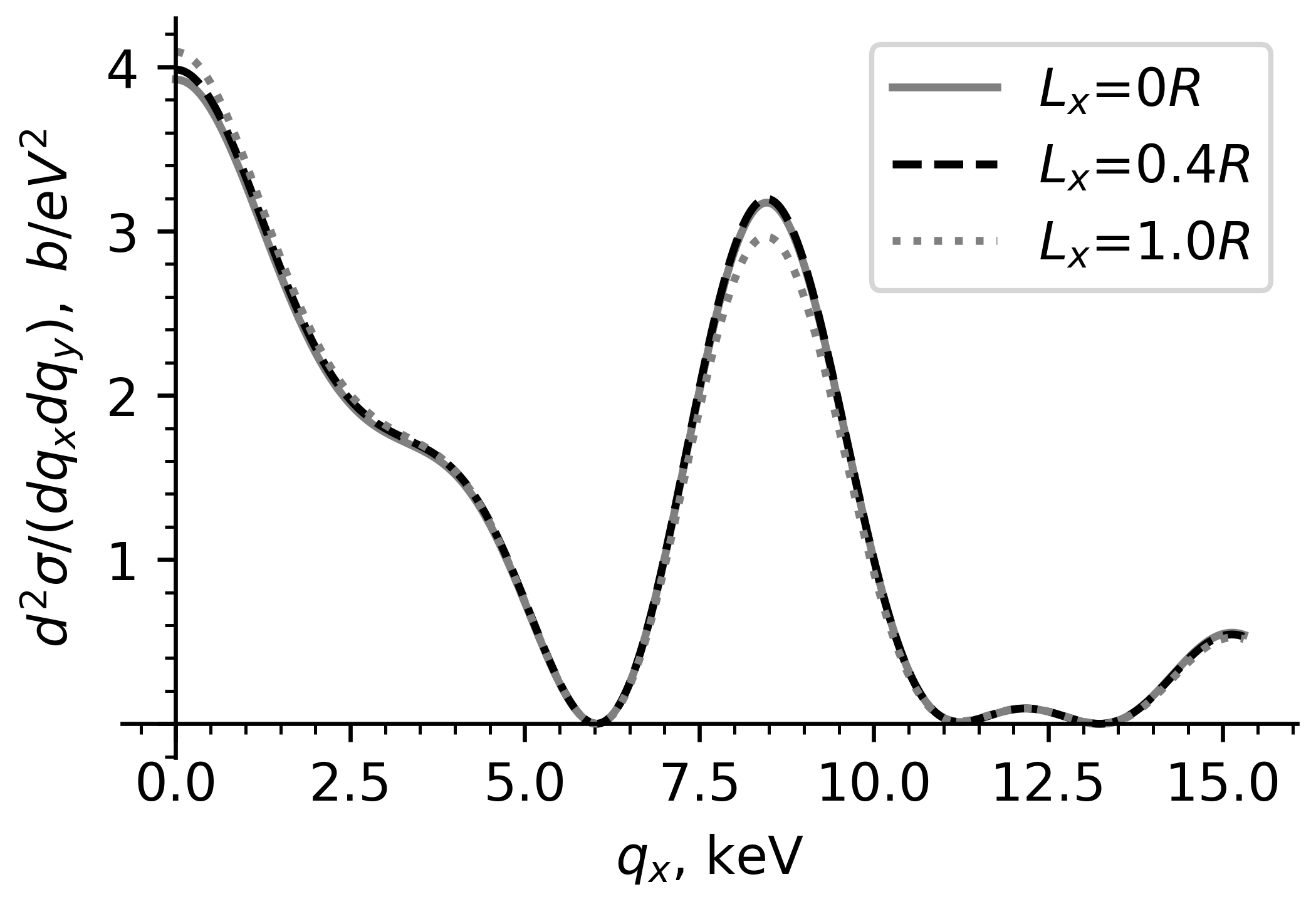}
\caption{$q_y=6.8$ keV}
\label{fig:cs_c1_4}
\end{subfigure}
\begin{subfigure}{0.32\linewidth}
\includegraphics[width=\textwidth]{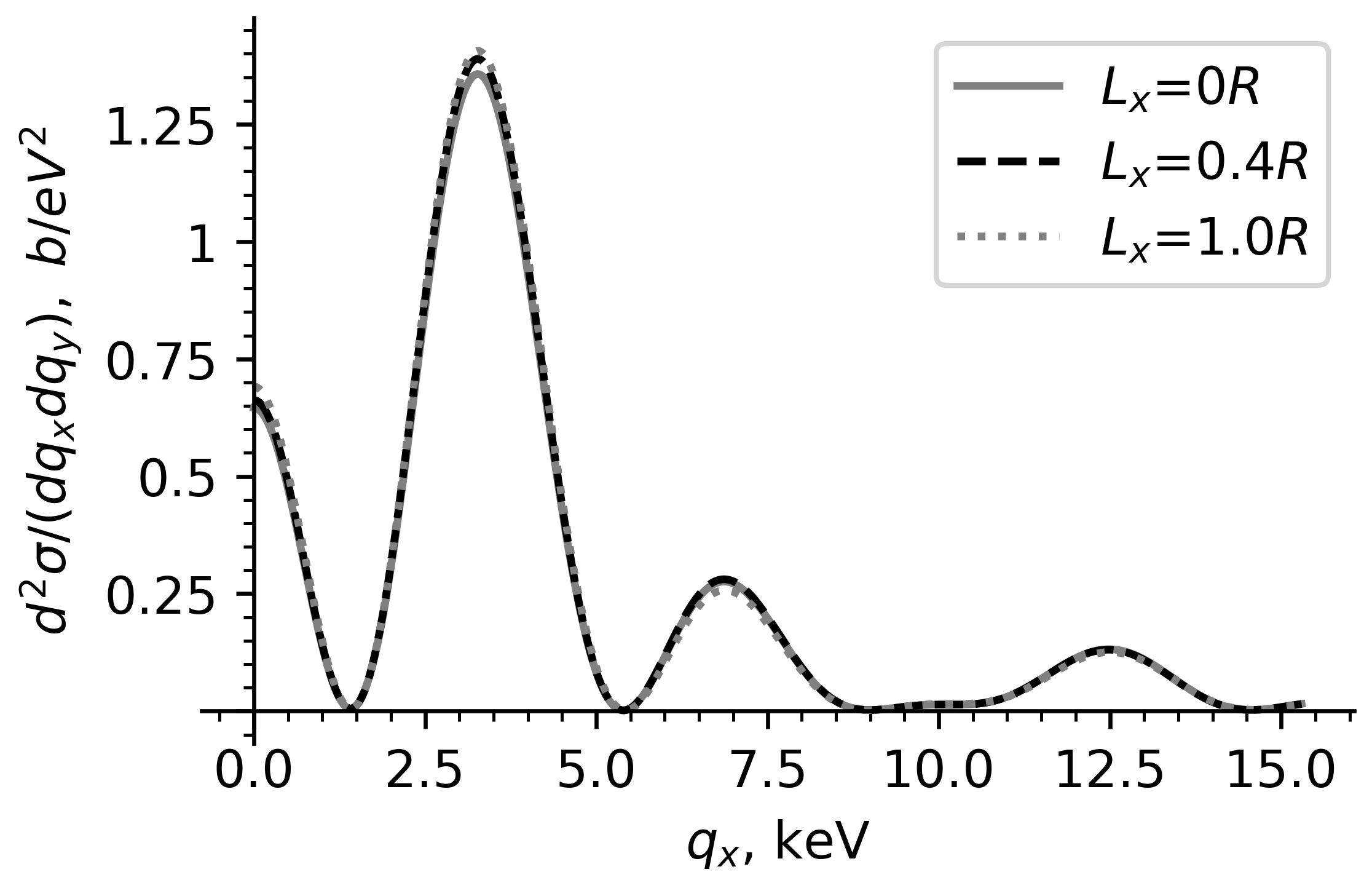}
\caption{$q_y=8.5$ keV}
\label{fig:cs_c1_5}
\end{subfigure}
\vfill
\begin{subfigure}{0.32\linewidth}
\includegraphics[width=\textwidth]{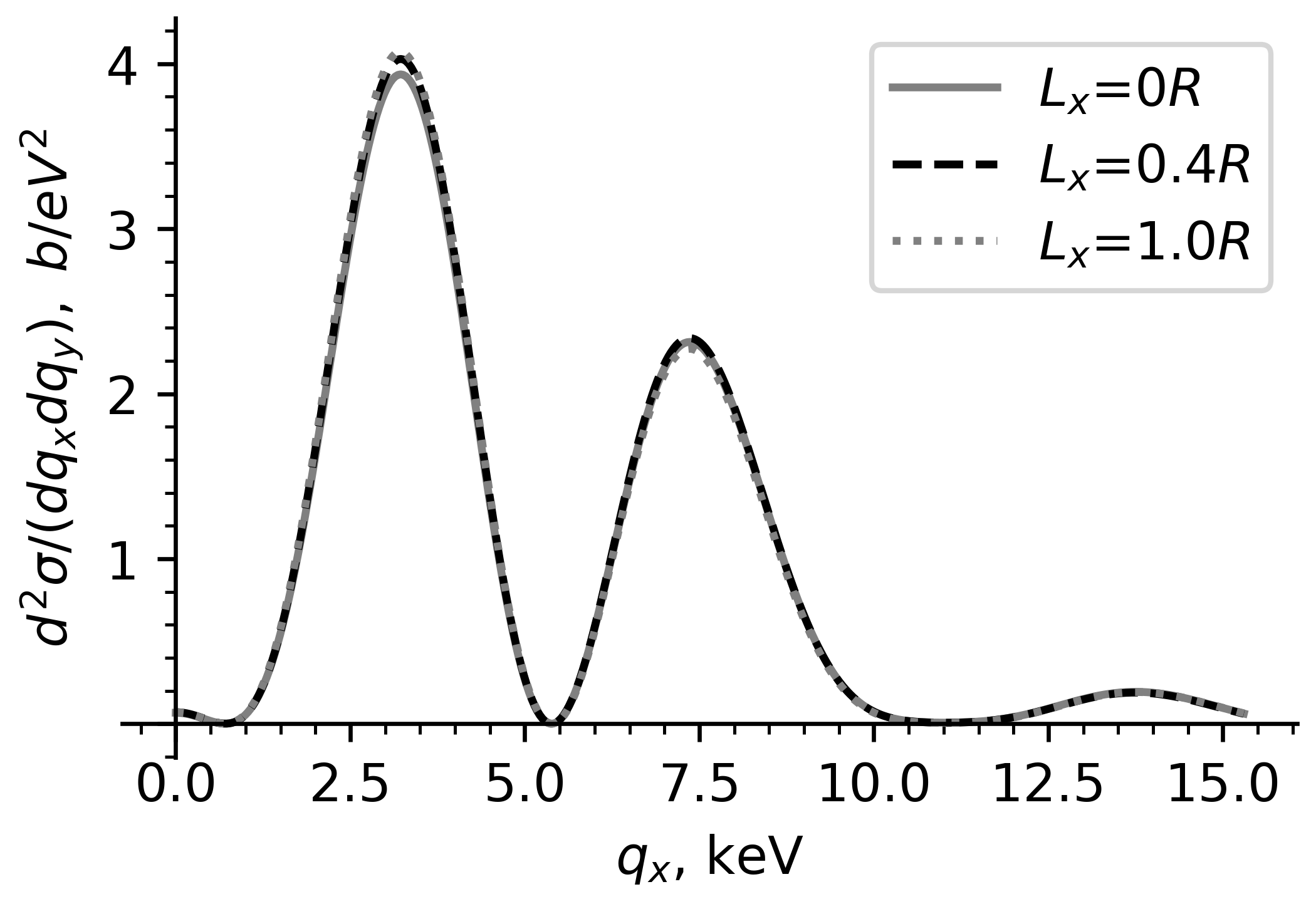}
\caption{$q_y=10.2$ keV}
\label{fig:cs_c1_6}
\end{subfigure}
\begin{subfigure}{0.32\linewidth}
\includegraphics[width=\textwidth]{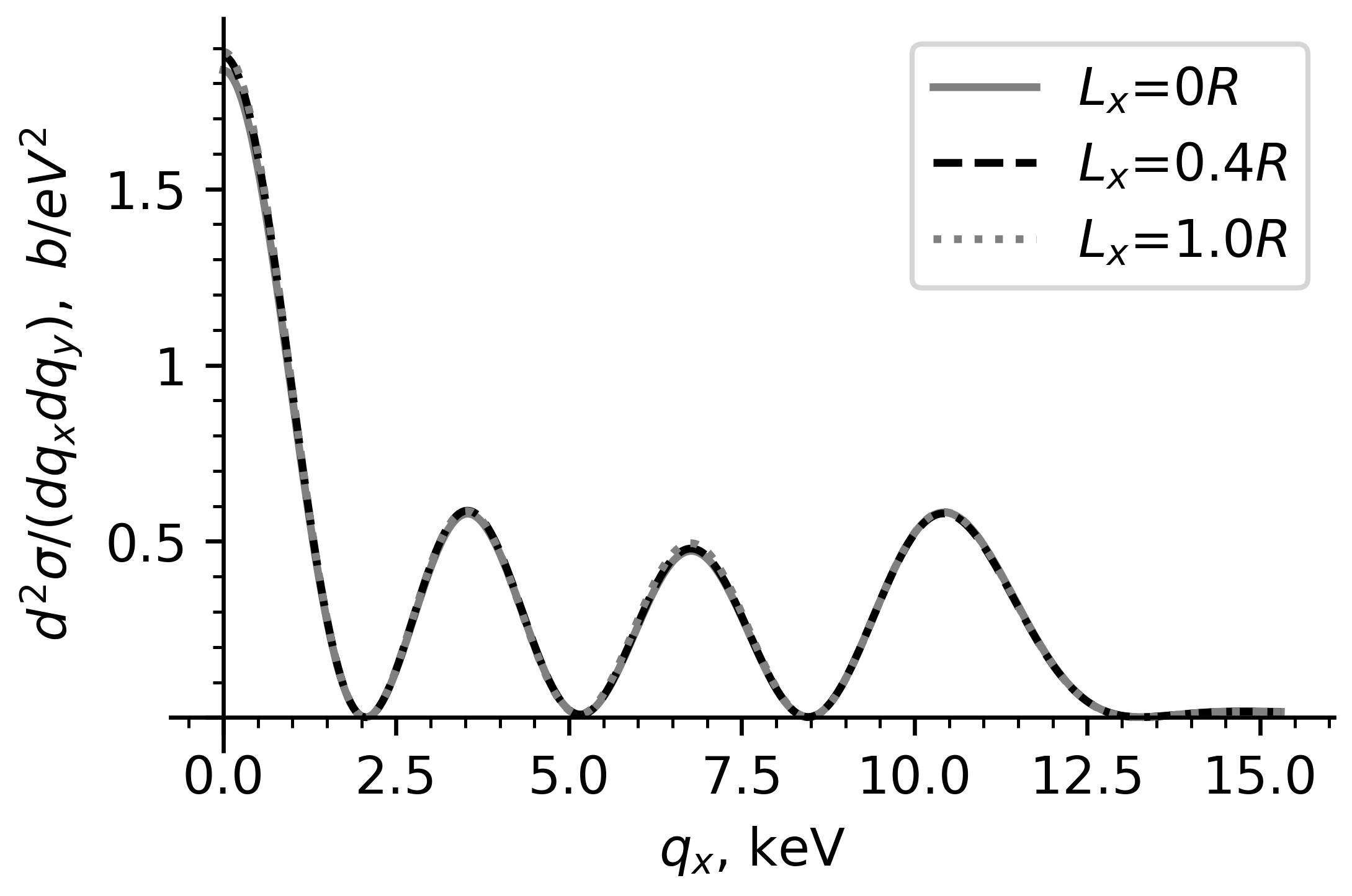}
\caption{$q_y=11.9$ keV}
\label{fig:cs_c1_7}
\end{subfigure}
\begin{subfigure}{0.32\linewidth}
\includegraphics[width=\textwidth]{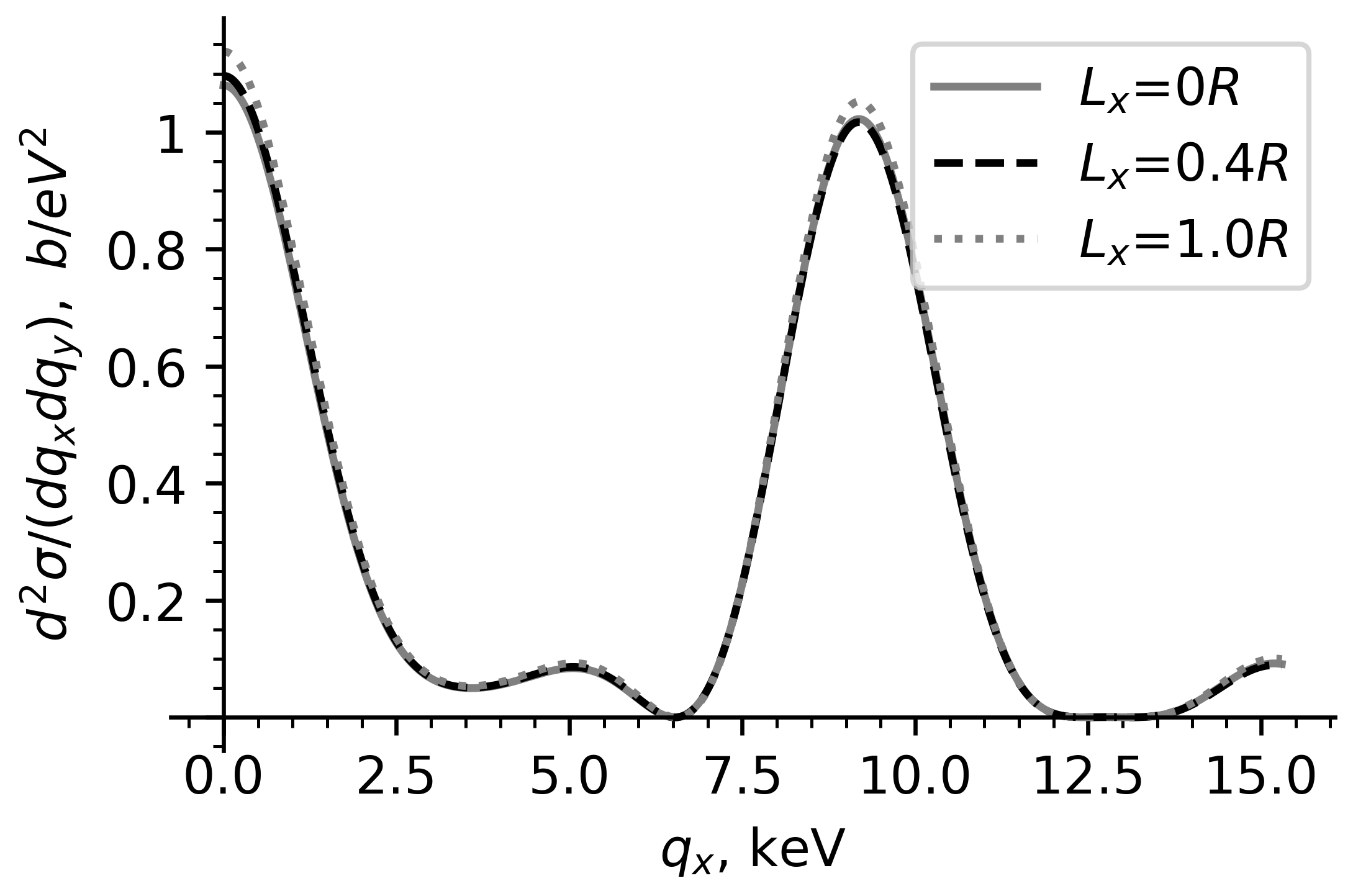}
\caption{$q_y=13.6$ keV}
\label{fig:cs_c1_8}
\end{subfigure}
\caption{Comparison of "numerical" differential cross sections of a fast charged particle scattering on a straight and tilted nanotubes with $L_x=\{0.4R, 1.0R\}$, $L_x=0R$ corresponds to the straight nanotube}
\label{fig_cs_c1}
\end{figure}

\begin{figure}[!h]
\begin{subfigure}{0.32\linewidth}
\includegraphics[width=\textwidth]{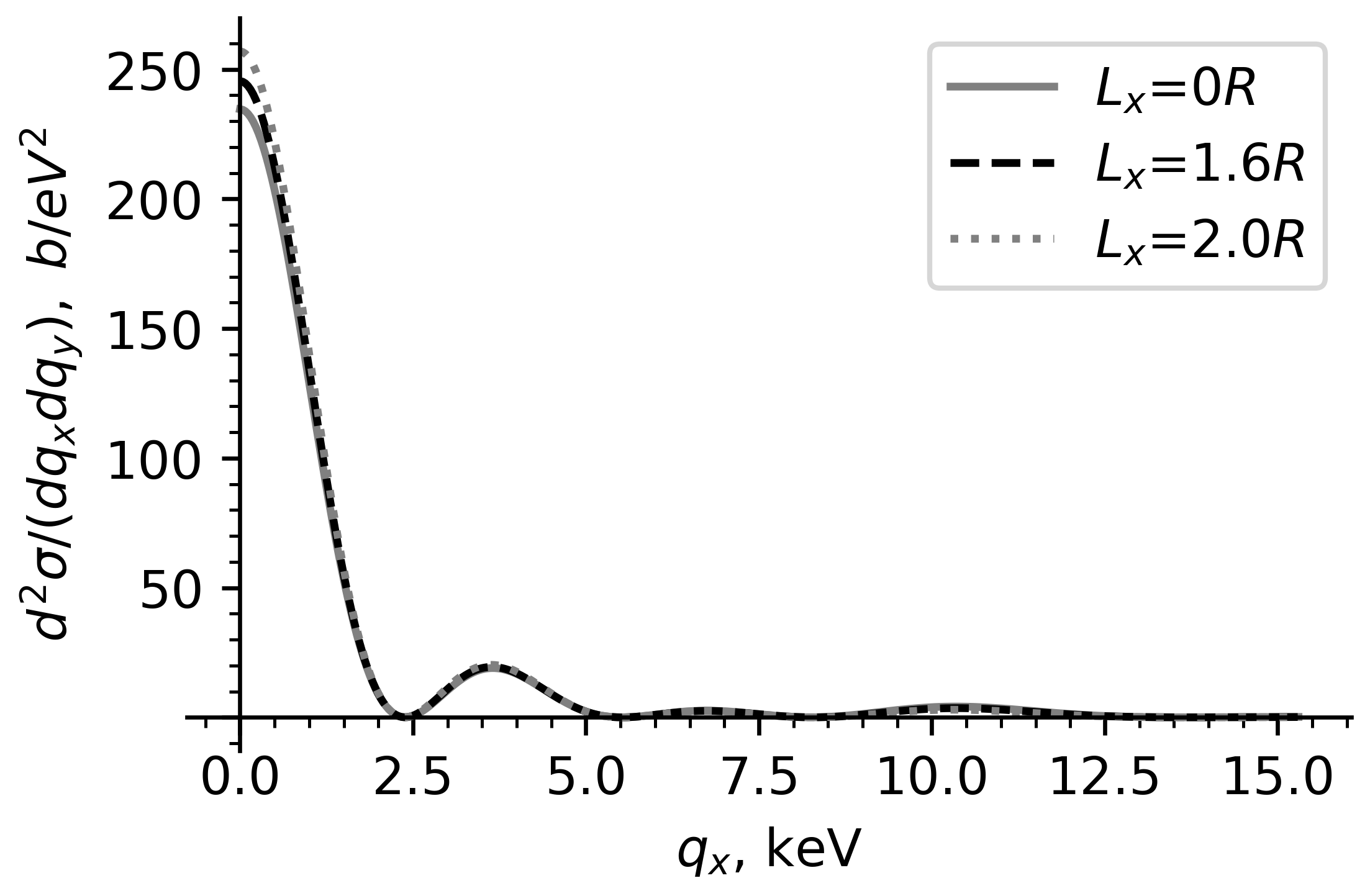}
\caption{$q_y=0.0$ keV}
\label{fig:cs_c2_0}
\end{subfigure}
\begin{subfigure}{0.32\linewidth}
\includegraphics[width=\textwidth]{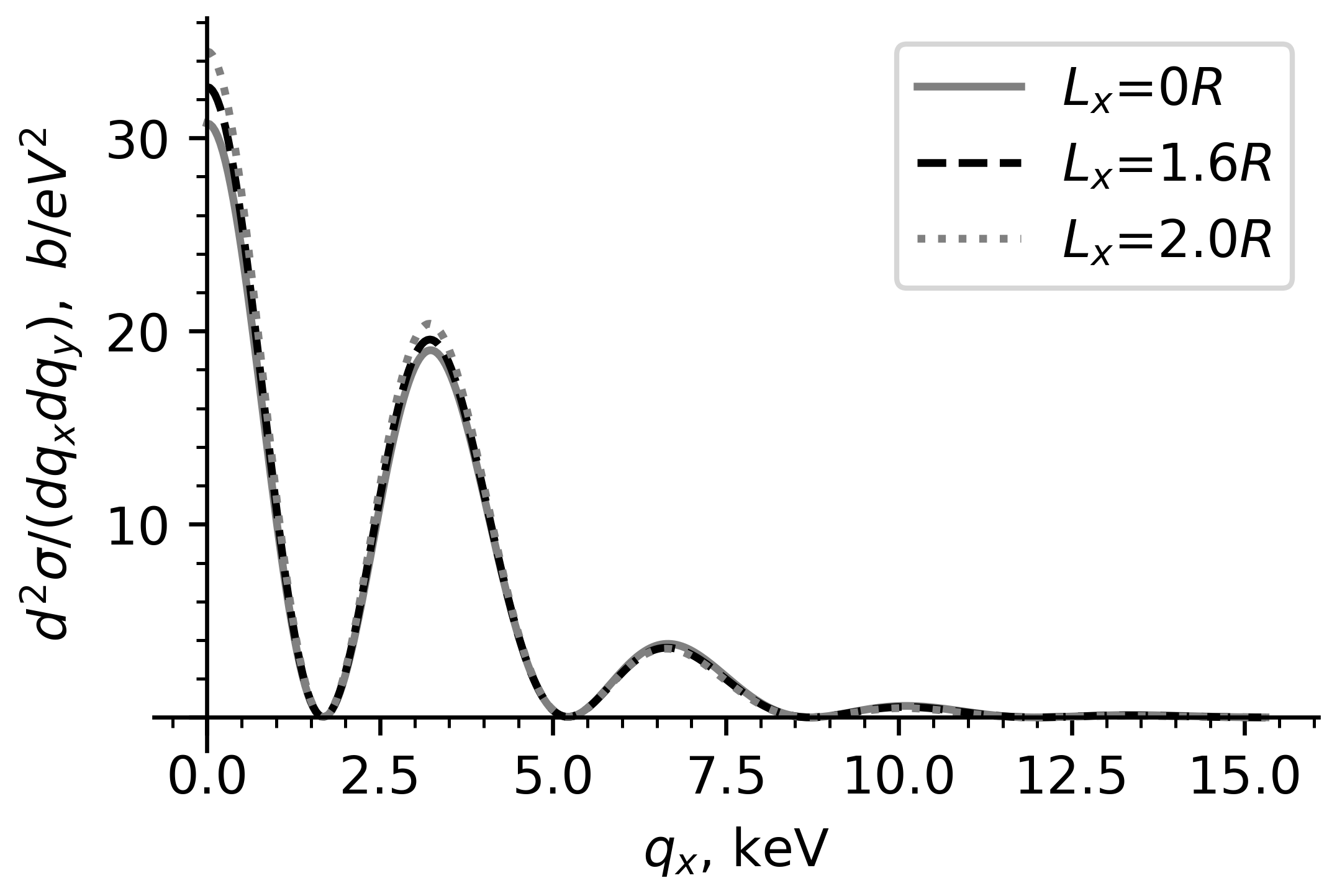}
\caption{$q_y=1.7$ keV}
\label{fig:cs_c2_1}
\end{subfigure}
\begin{subfigure}{0.32\linewidth}
\includegraphics[width=\textwidth]{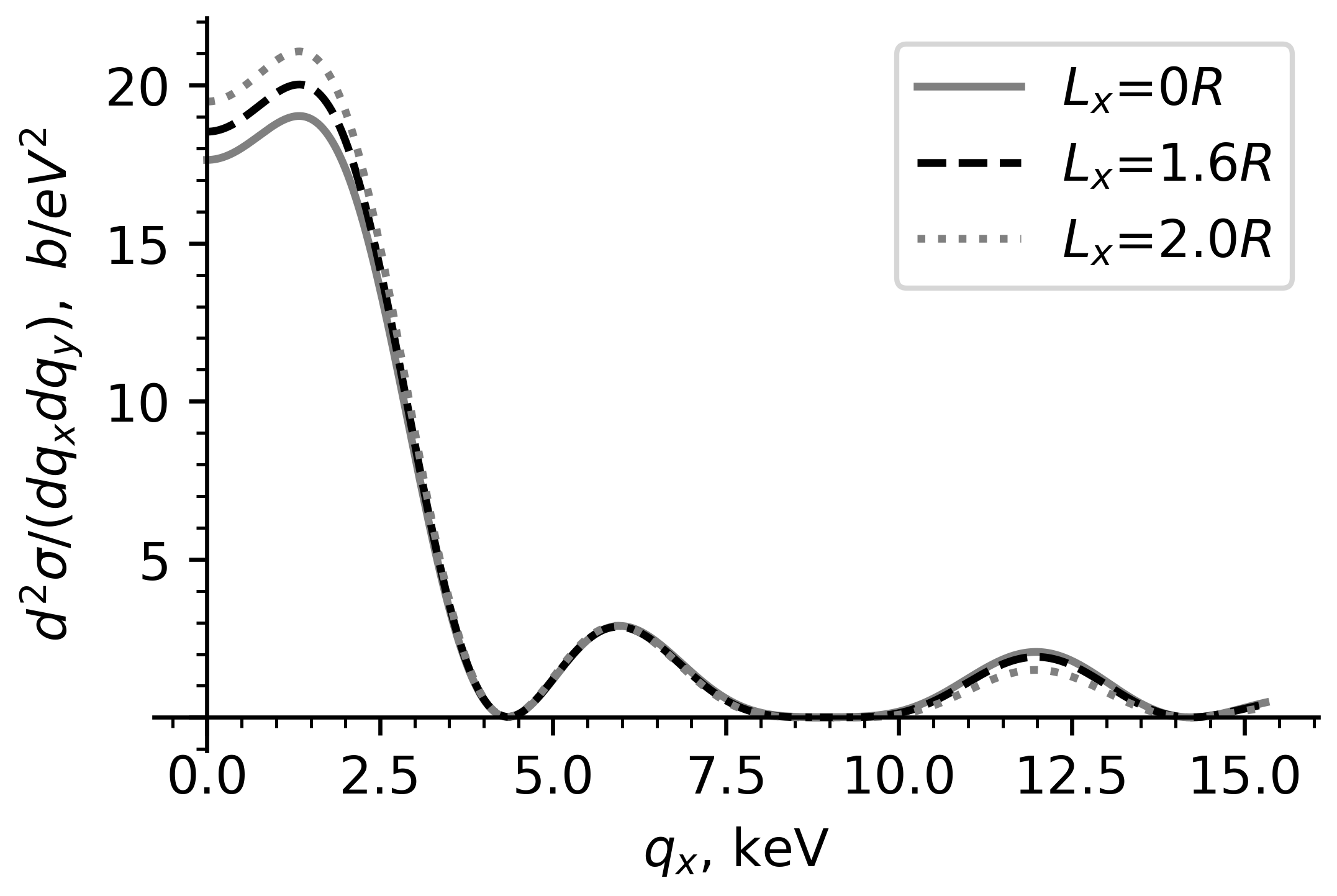}
\caption{$q_y=3.4$ keV}
\label{fig:cs_c2_2}
\end{subfigure}
\vfill
\begin{subfigure}{0.32\linewidth}
\includegraphics[width=\textwidth]{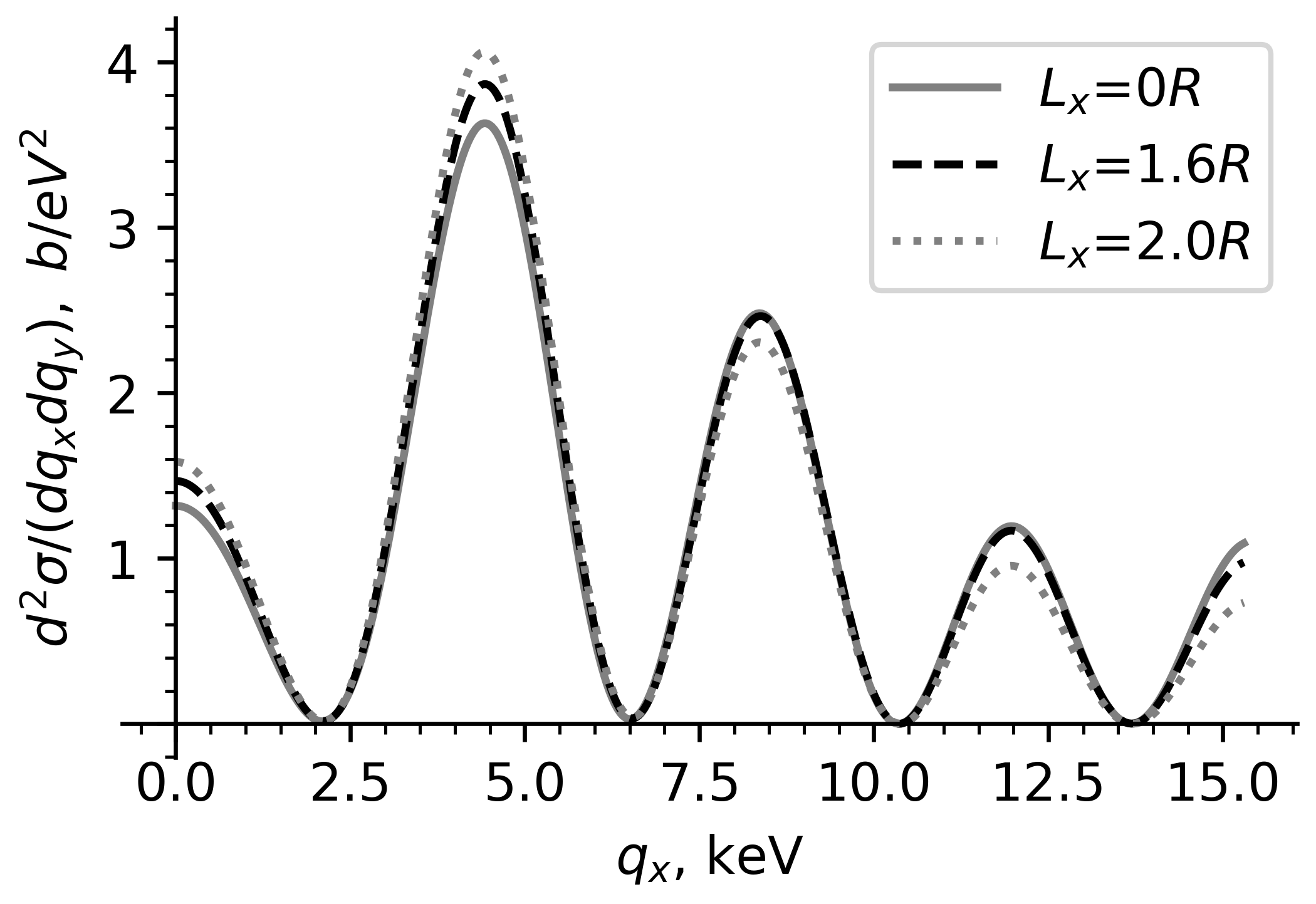}
\caption{$q_y=5.1$ keV}
\label{fig:cs_c2_3}
\end{subfigure}
\begin{subfigure}{0.32\linewidth}
\includegraphics[width=\textwidth]{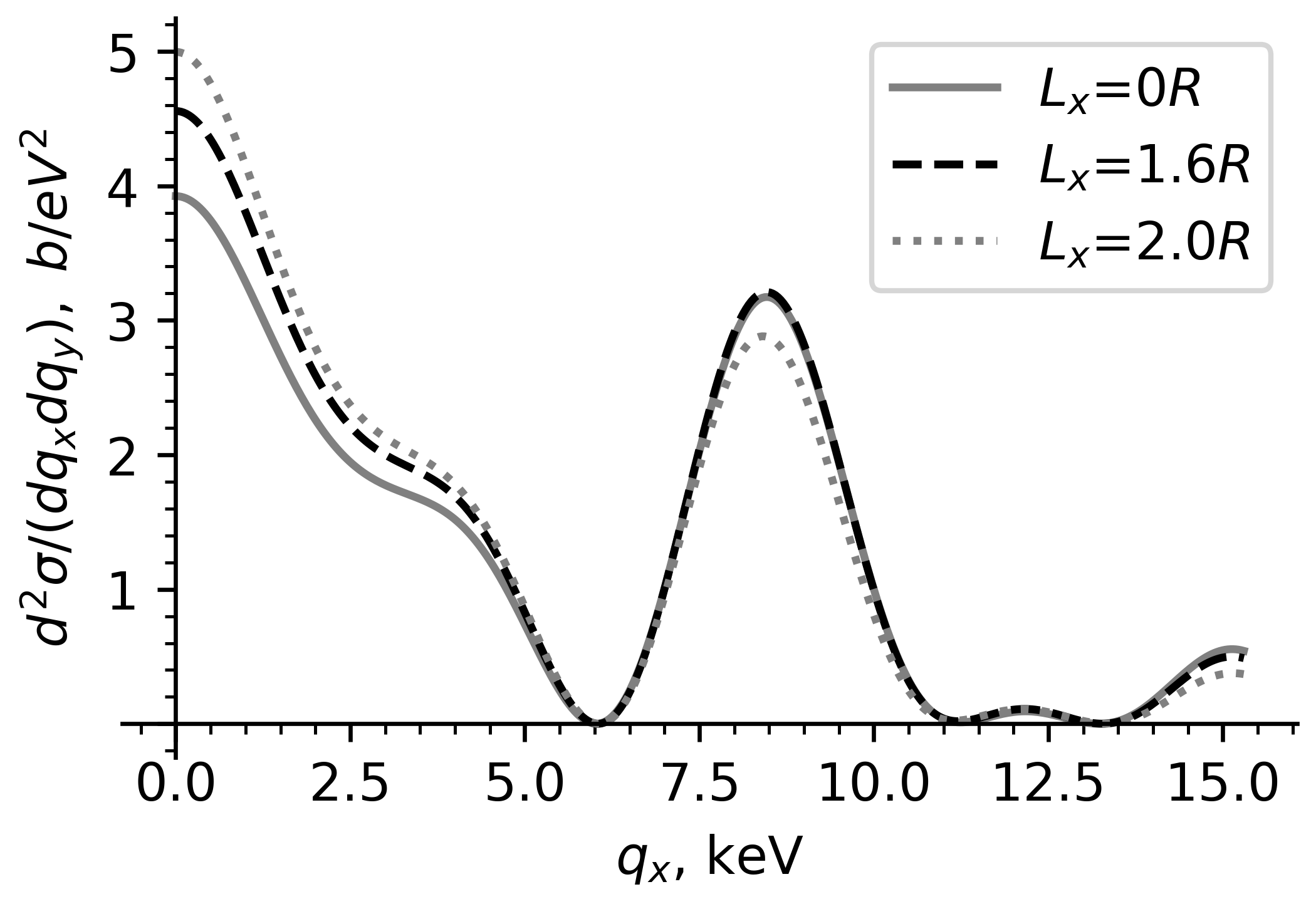}
\caption{$q_y=6.8$ keV}
\label{fig:cs_c2_4}
\end{subfigure}
\begin{subfigure}{0.32\linewidth}
\includegraphics[width=\textwidth]{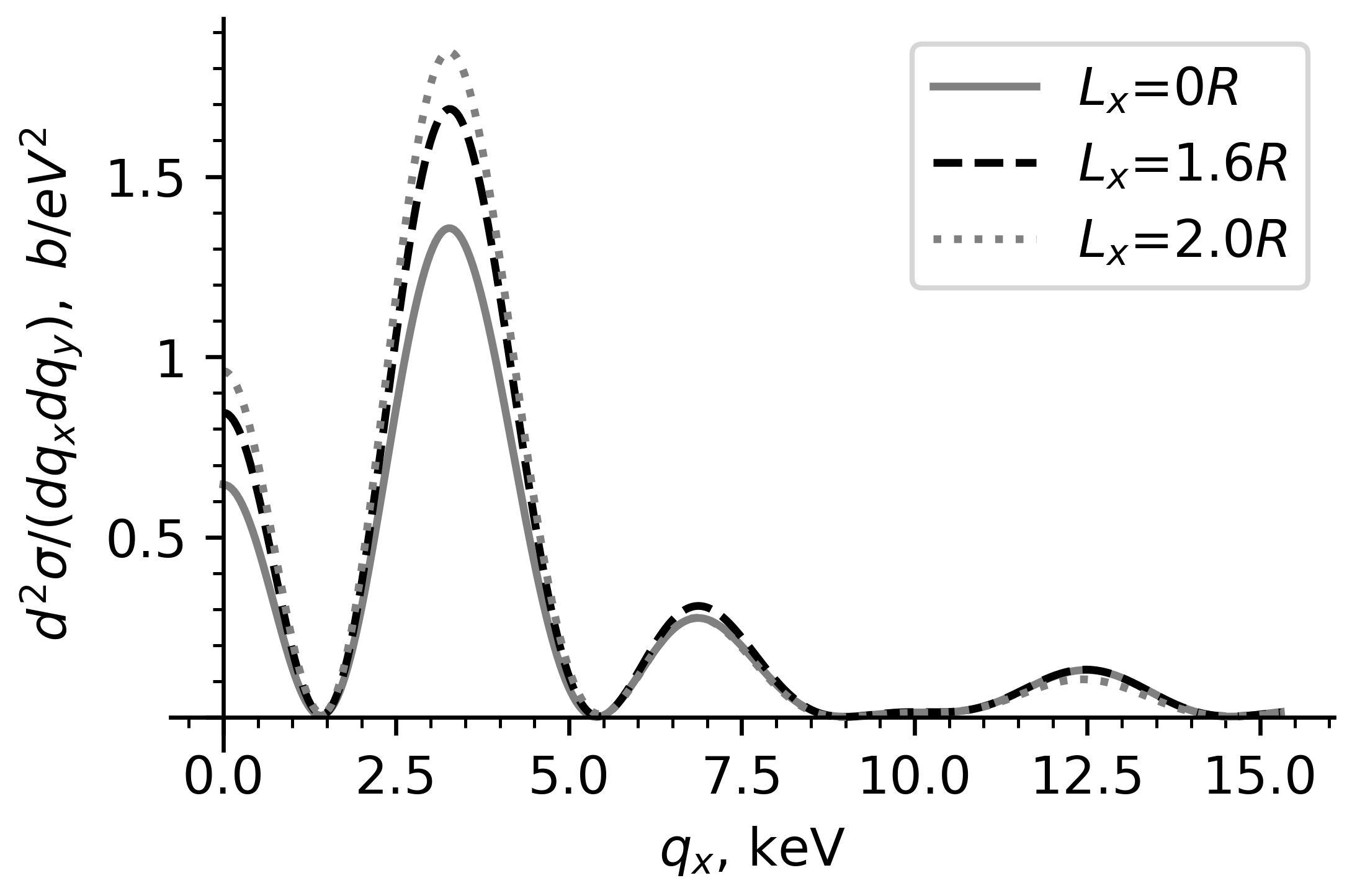}
\caption{$q_y=8.5$ keV}
\label{fig:cs_c2_5}
\end{subfigure}
\vfill
\begin{subfigure}{0.32\linewidth}
\includegraphics[width=\textwidth]{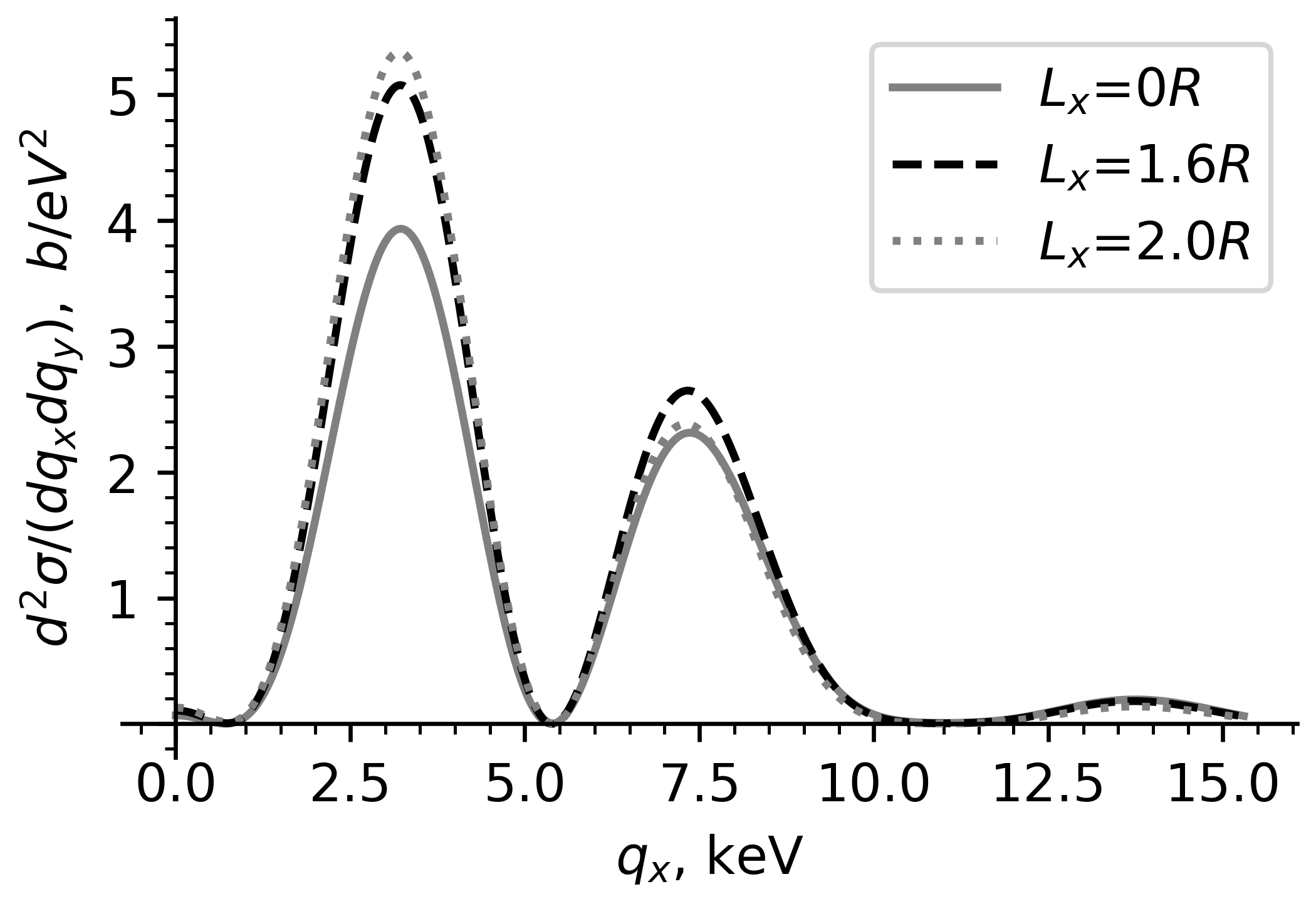}
\caption{$q_y=10.2$ keV}
\label{fig:cs_c2_6}
\end{subfigure}
\begin{subfigure}{0.32\linewidth}
\includegraphics[width=\textwidth]{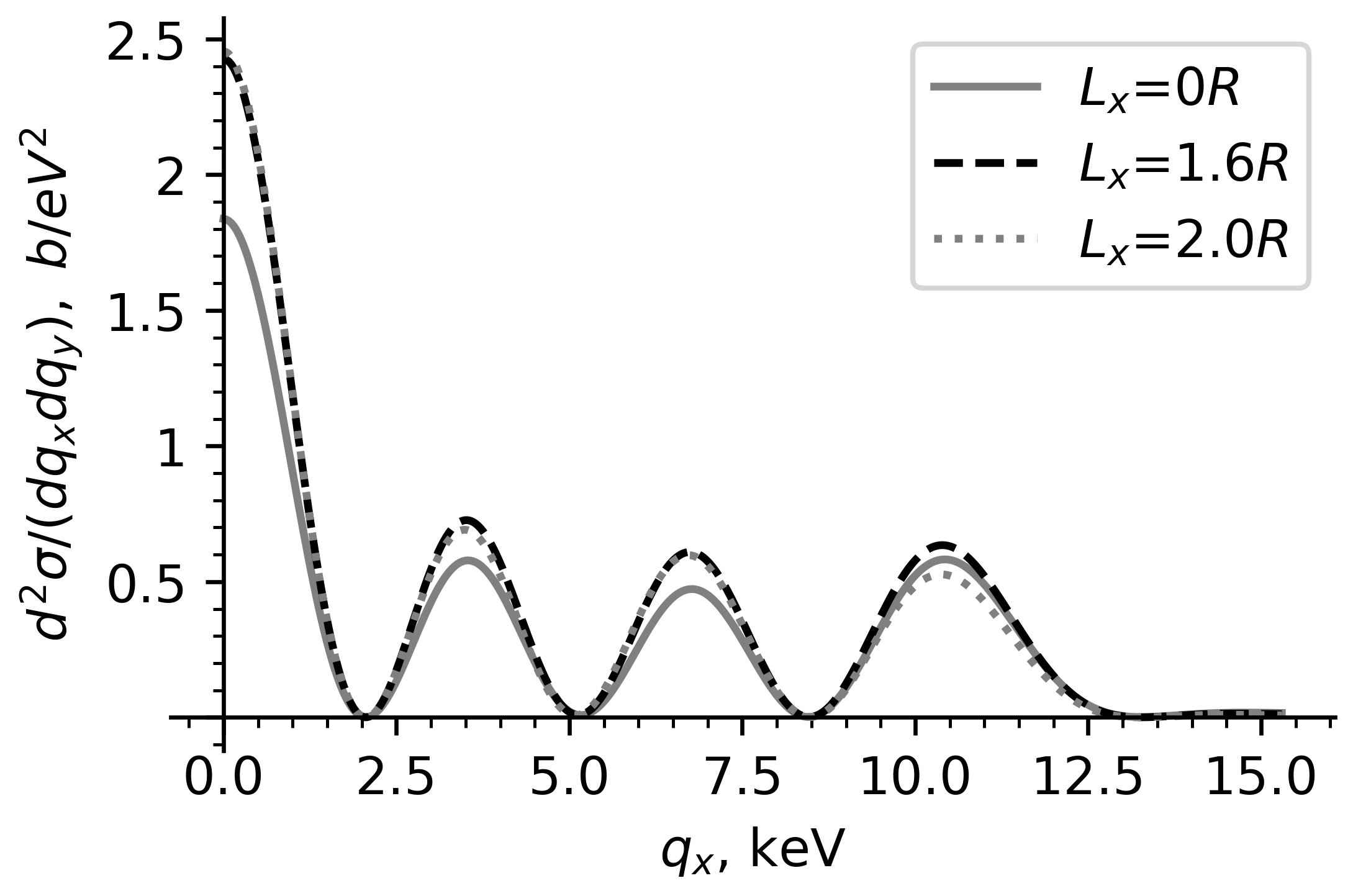}
\caption{$q_y=11.9$ keV}
\label{fig:cs_c2_7}
\end{subfigure}
\begin{subfigure}{0.32\linewidth}
\includegraphics[width=\textwidth]{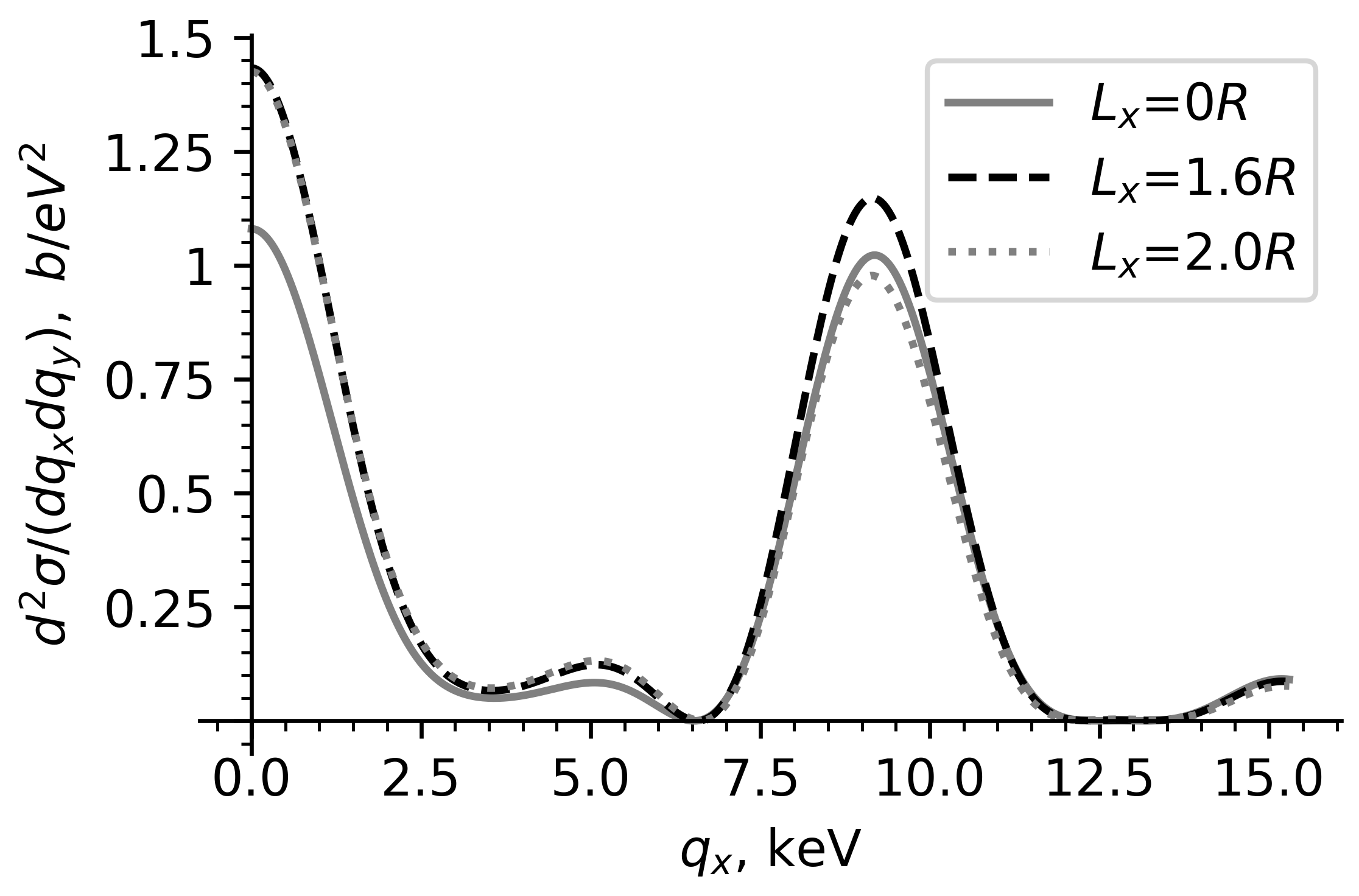}
\caption{$q_y=13.6$ keV}
\label{fig:cs_c2_8}
\end{subfigure}
\caption{Comparison of "numerical" differential cross sections of a fast charged particle scattering on a straight and tilted nanotubes with $L_x=\{1.6R, 2.0R\}$, $L_x=0R$ corresponds to the straight nanotube}
\label{fig_cs_c2}
\end{figure}

\begin{figure}[!h]
\begin{subfigure}{0.32\linewidth}
\includegraphics[width=\textwidth]{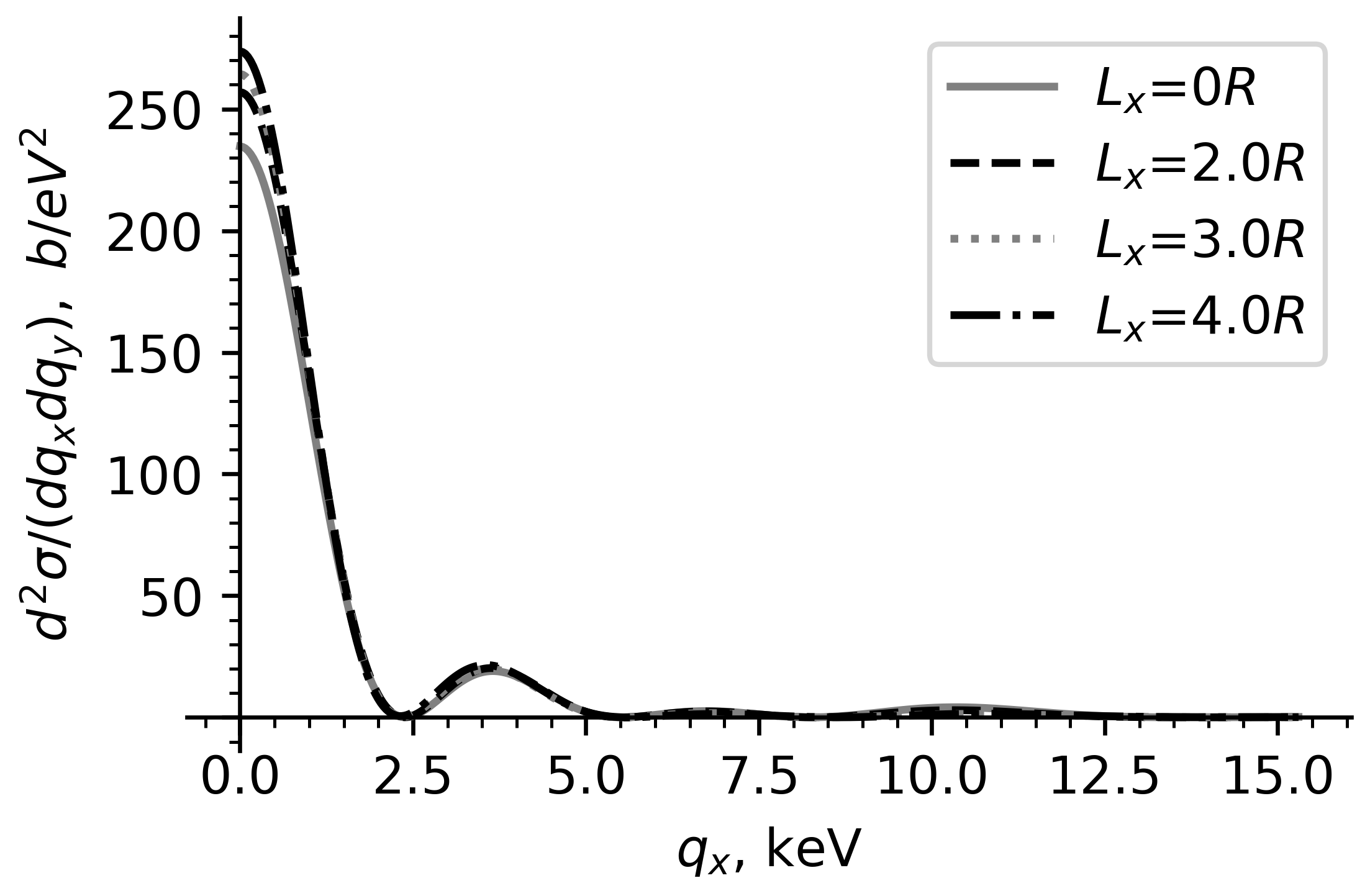}
\caption{$q_y=0.0$ keV}
\label{fig:cs_c3_0}
\end{subfigure}
\begin{subfigure}{0.32\linewidth}
\includegraphics[width=\textwidth]{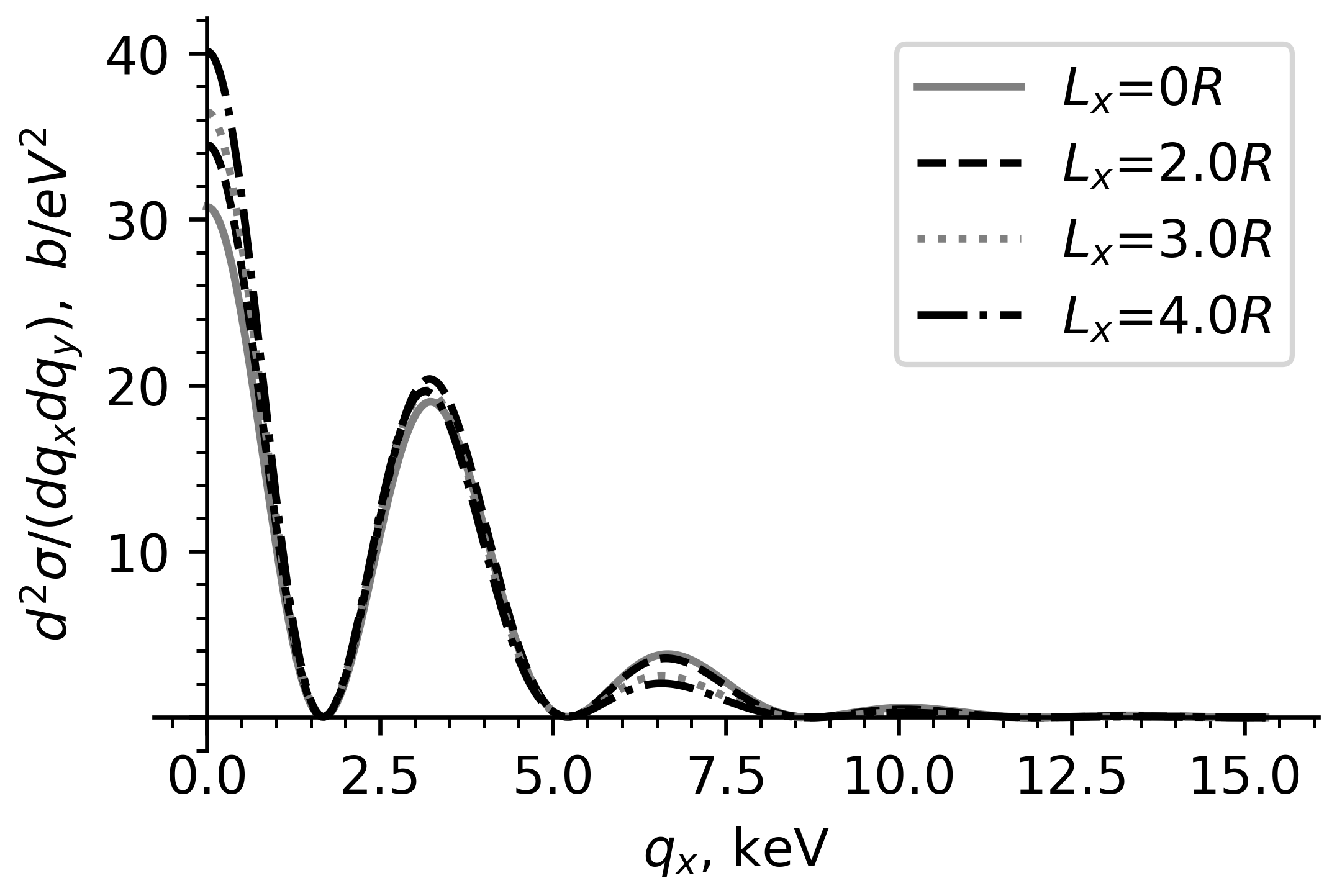}
\caption{$q_y=1.7$ keV}
\label{fig:cs_c3_1}
\end{subfigure}
\begin{subfigure}{0.32\linewidth}
\includegraphics[width=\textwidth]{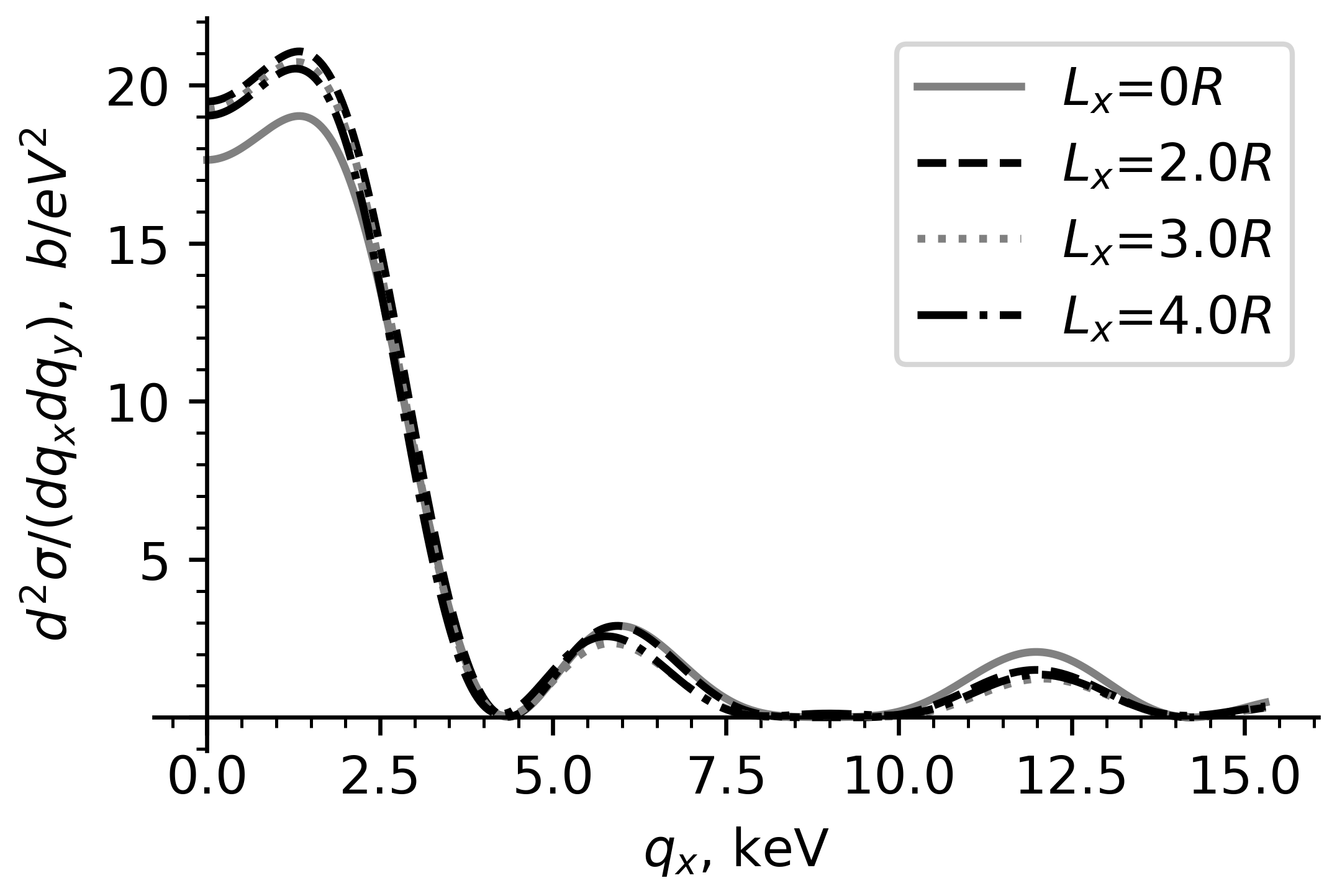}
\caption{$q_y=3.4$ keV}
\label{fig:cs_c3_2}
\end{subfigure}
\vfill
\begin{subfigure}{0.32\linewidth}
\includegraphics[width=\textwidth]{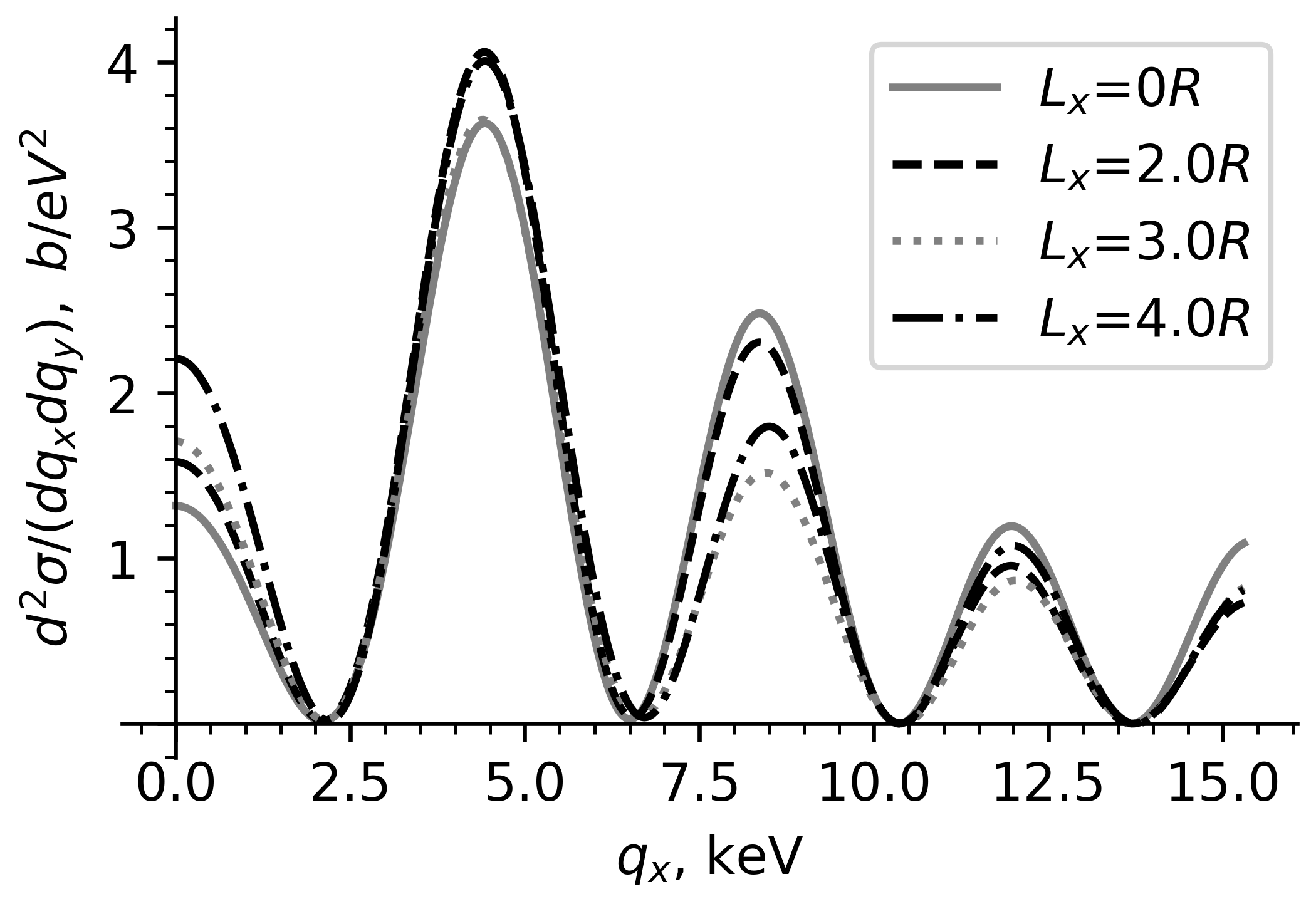}
\caption{$q_y=5.1$ keV}
\label{fig:cs_c3_3}
\end{subfigure}
\begin{subfigure}{0.32\linewidth}
\includegraphics[width=\textwidth]{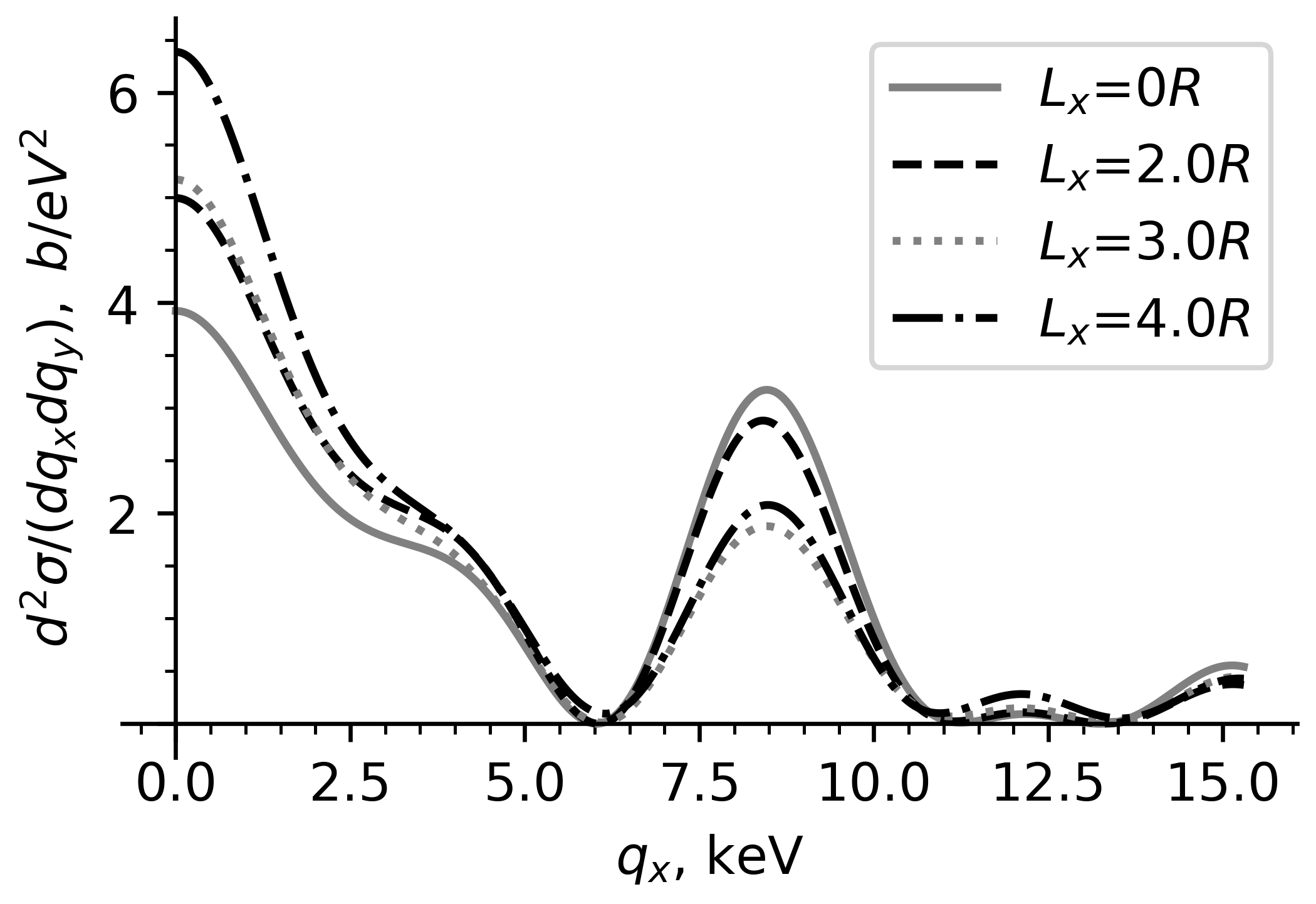}
\caption{$q_y=6.8$ keV}
\label{fig:cs_c3_4}
\end{subfigure}
\begin{subfigure}{0.32\linewidth}
\includegraphics[width=\textwidth]{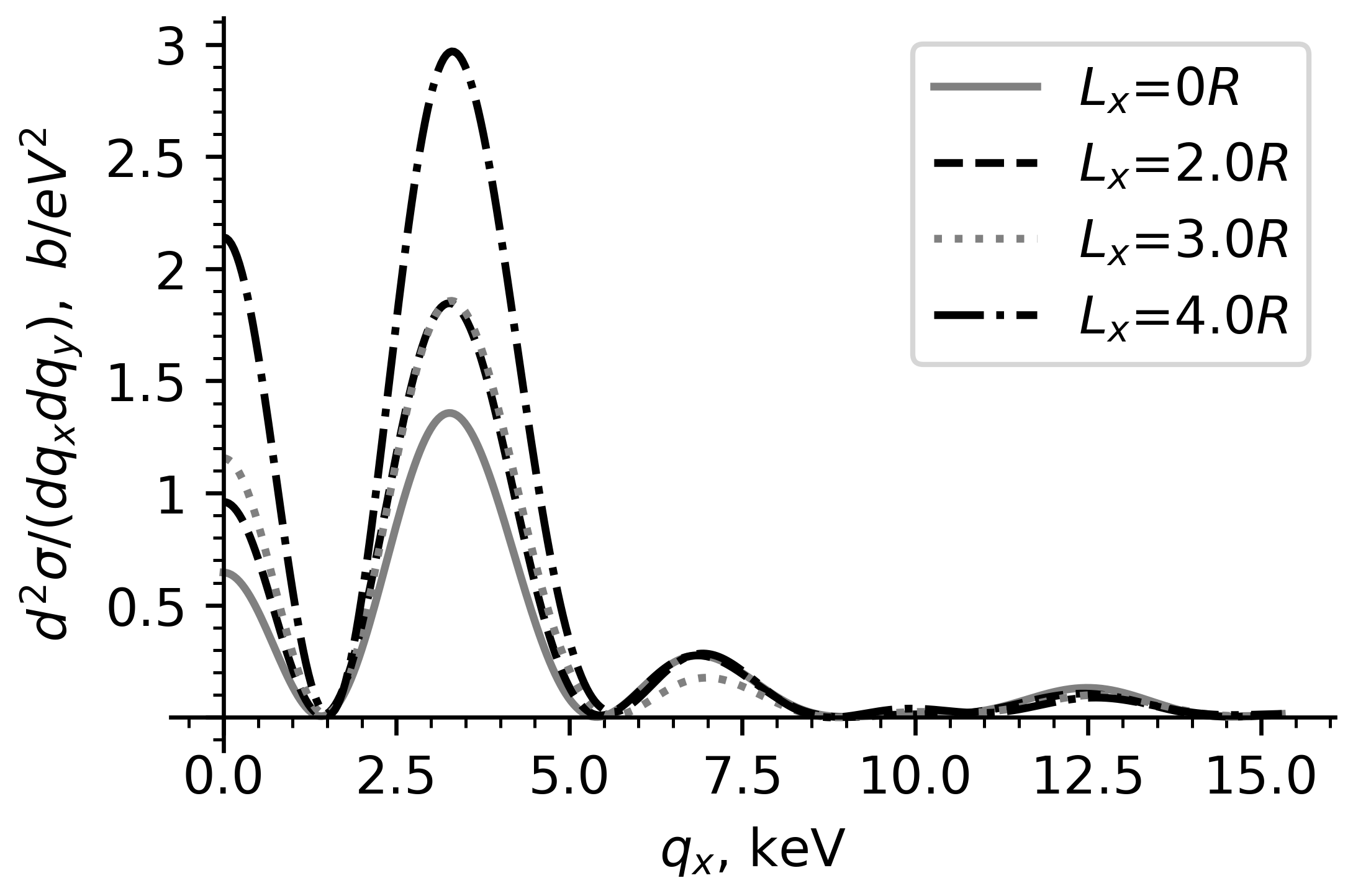}
\caption{$q_y=8.5$ keV}
\label{fig:cs_c3_5}
\end{subfigure}
\vfill
\begin{subfigure}{0.32\linewidth}
\includegraphics[width=\textwidth]{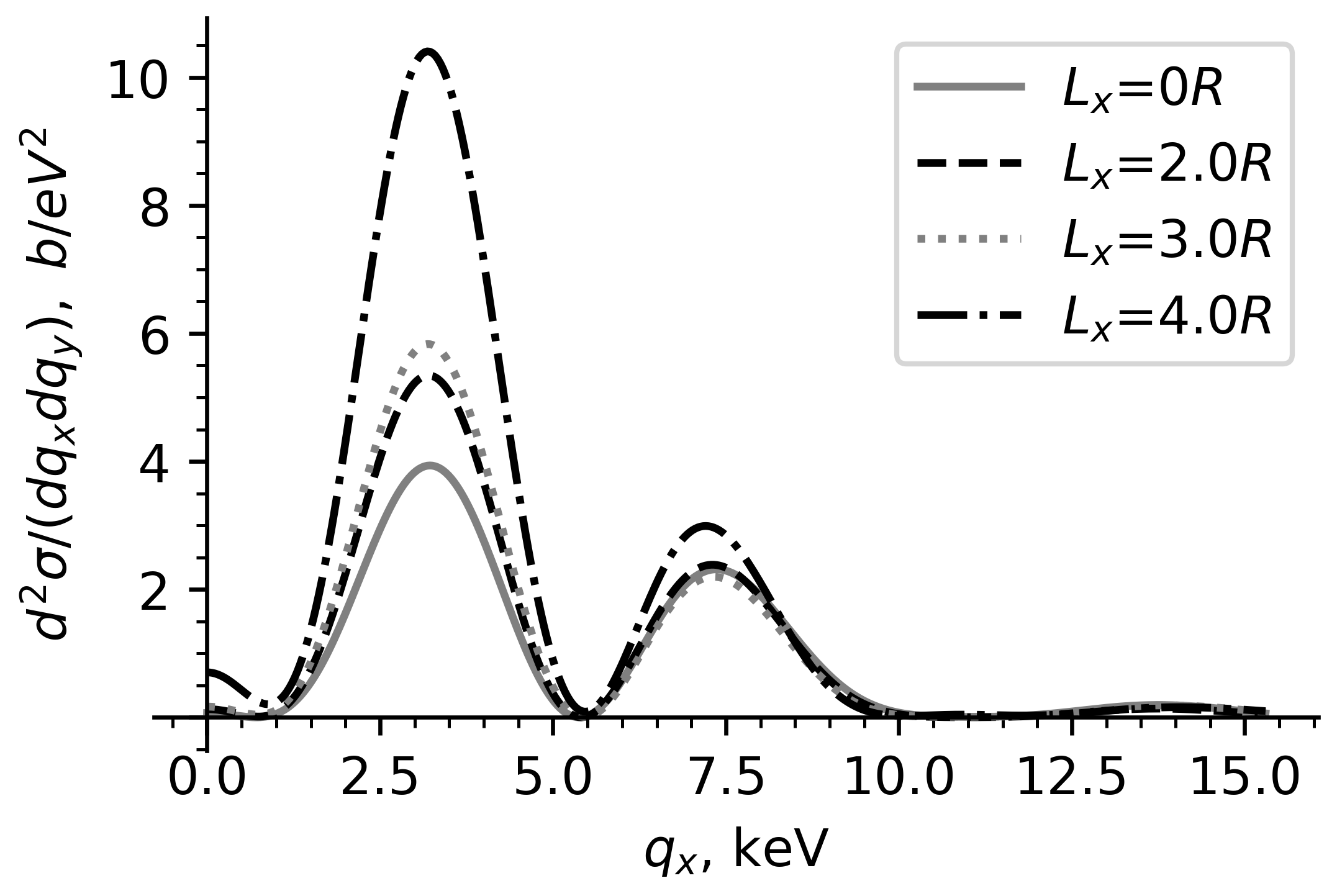}
\caption{$q_y=10.2$ keV}
\label{fig:cs_c3_6}
\end{subfigure}
\begin{subfigure}{0.32\linewidth}
\includegraphics[width=\textwidth]{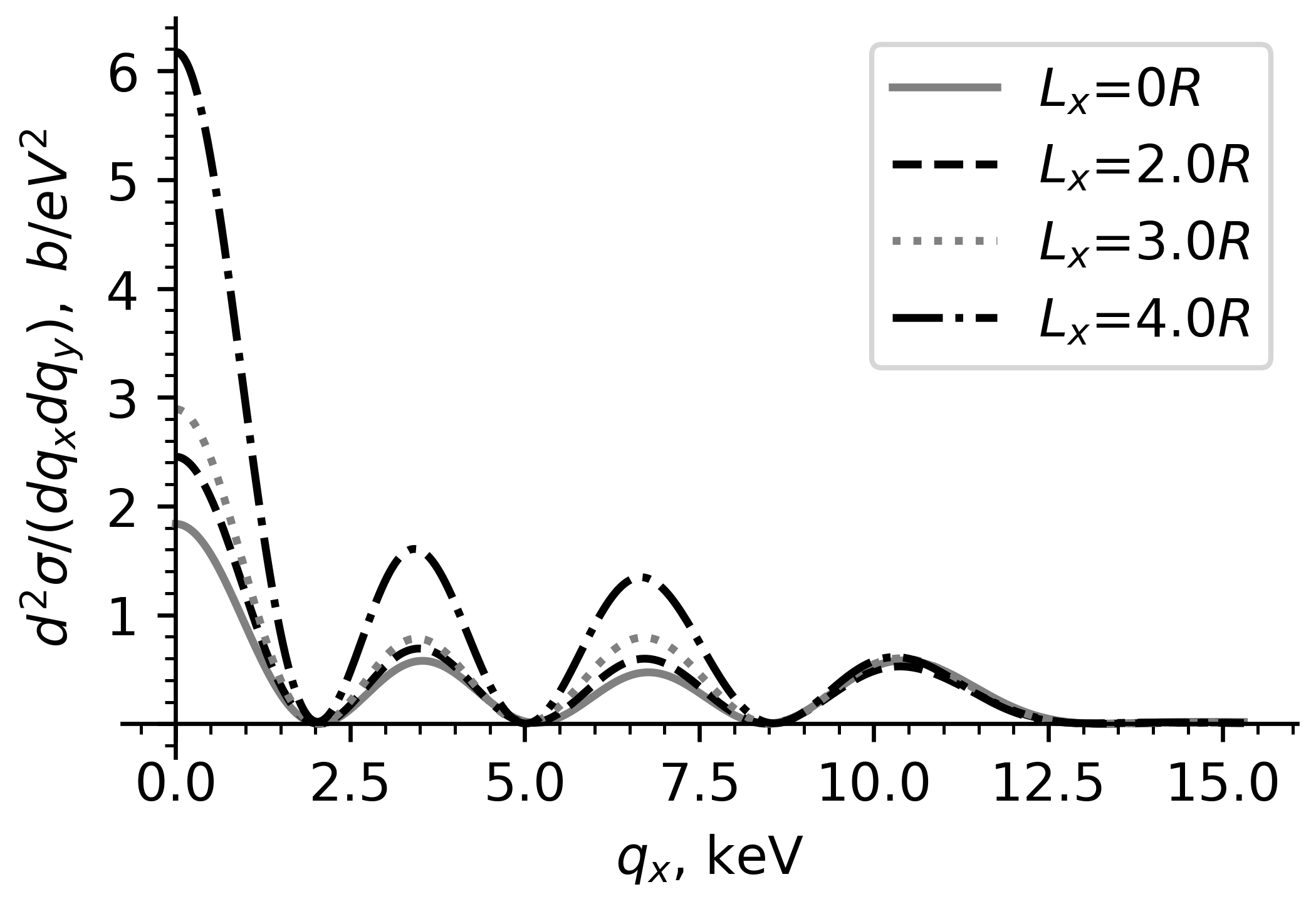}
\caption{$q_y=11.9$ keV}
\label{fig:cs_c3_7}
\end{subfigure}
\begin{subfigure}{0.32\linewidth}
\includegraphics[width=\textwidth]{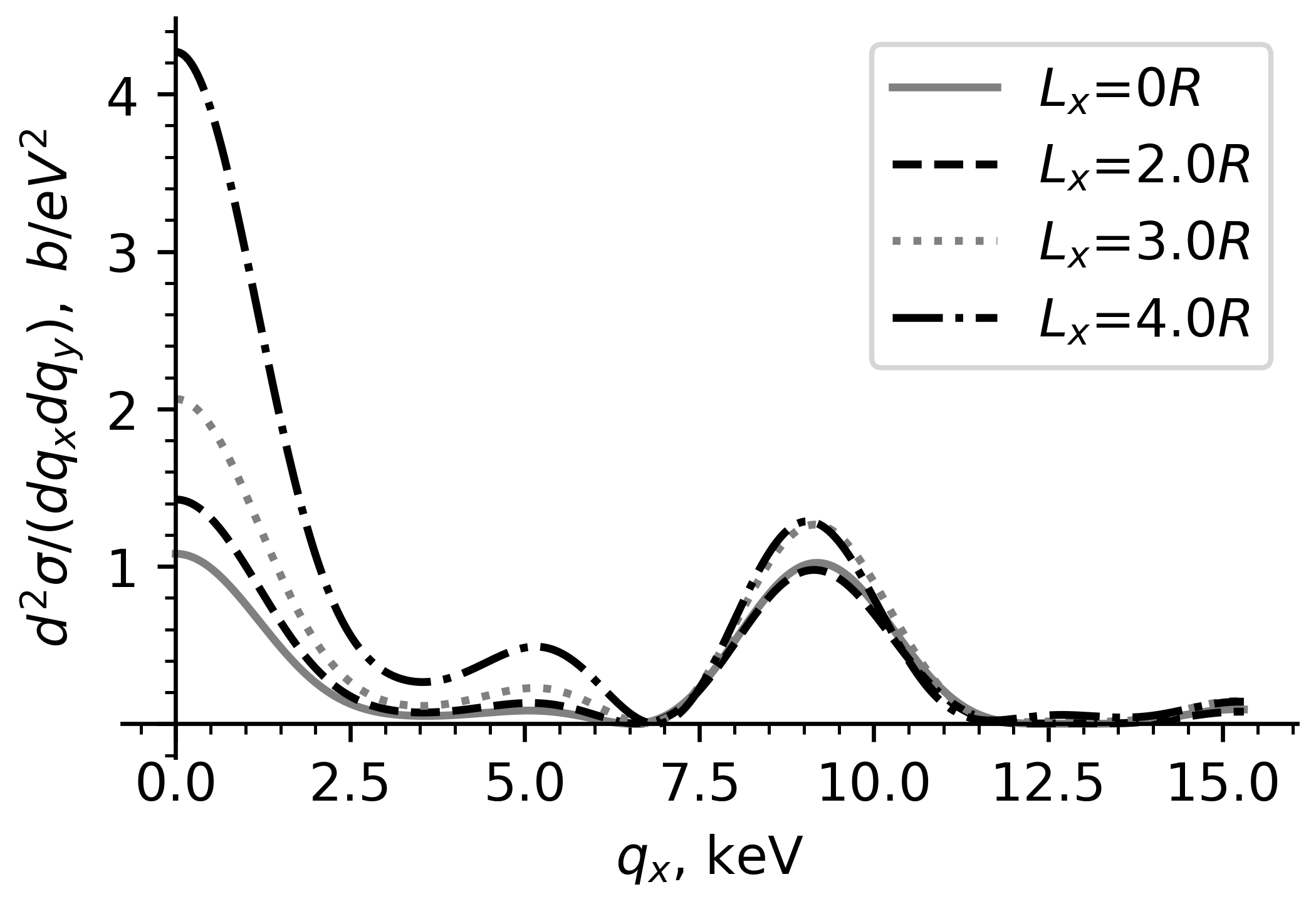}
\caption{$q_y=13.6$ keV}
\label{fig:cs_c3_8}
\end{subfigure}
\caption{Comparison of "numerical" differential cross sections of a fast charged particle scattering on a straight and tilted nanotubes with $L_x=\{2.0R, 3.0R, 4.0R\}$, $L_x=0R$ corresponds to the straight nanotube}
\label{fig_cs_c3}
\end{figure}

Also we compared "numerical" differential cross section for scattering on nanotubes tilted under different angles with respect to $z$-axis (for various $L_x$ parameters). The comparison is presented in Figs. \ref{fig_cs_c1}-\ref{fig_cs_c3}. Analyzing these figures we can discuss resolution necessary to distinguish different orientations of nanotubes. From Fig. \ref{fig_cs_c1}, we see that differential cross sections for $L_x \leq R$ are almost indistinguishable from corresponding cross sections of scattering on a straight nanotube. Figs. \ref{fig_cs_c2}-\ref{fig_cs_c3} illustrate how larger $L_x$ leads to greater difference between the corresponding cross section and the cross section of scattering on a straight nanotube.

The eikonal approximation that we used for this problem has wider applicability region compared with frequently used Born approximation. The applicability condition for a straight nanotube can be derived based on \cite{AIA96} as
\begin{eqnarray}\label{eq27}
\frac{|\tilde{\chi}_0^{(N_s)}|_{max} L_z}{pR^2} \ll 1.
\end{eqnarray}
where $\tilde{\chi}_0^{(N_s)}=\sum_{j=1}^{N_s}  \tilde{\chi}_{(j)}^{(s)} $.

The maximum absolute value of $\tilde{\chi}_0^{(N_s)}$ -- $|\tilde{\chi}_0^{(N_s)}|_{max}$ does not exceed $2A$ for the spacing between atomic strings which occurs in a carbon zigzag nanotube. For this spacings the value $|\tilde{\chi}_0^{(N_s)}|_{max}$ does not depend on the number of strings in the nanotube, only on $N_z$: $A=2Z \alpha N_z$. From \eqref{eq0_1}, the length of the nanotube can be estimated as $L_z \approx 1.5 N_z a_g$ for large enough $N_z$. Then, \eqref{eq27} corresponds to the following applicability condition:
\begin{eqnarray}\label{eq28}
\varepsilon[GeV] \gg 244 (L_z[\mu m])^2.
\end{eqnarray}
For considered in this paper length $L_z \approx 24$ nm, the required energy of the incident particle is $\varepsilon \gg 0.14$ GeV. The necessary energy for applicability of the eikonal approximation increases quadratically with increasing length of the nanotube.   

The Figs. \ref{fig_cs_l20}-\ref{fig_cs_c3} appear on the pages that follow. 

\FloatBarrier
\begingroup

\section*{Acknowledgements}
\sloppy
{Author thanks God and His Blessed Mother for saving us and our Ukraine. The work was partially supported by the National Academy of Sciences of Ukraine (project 0125U002866). The author is deeply thankful to her late research supervisor, an academic of NAS of Ukraine N.F.~Shul'ga, who gave a lot of necessary knowledge to her. Author also acknowledges the fruitful discussions with S.P. Fomin, I.V. Kyryllin, S.V.~Trofymenko.  
\par}

\endgroup

\end{document}